\documentclass[3p,times]{elsarticle}

\usepackage{ecrc}
\usepackage{hyperref}
\usepackage{soul}
\usepackage{xcolor} % for the \textcolor command

\usepackage{amsmath, amsfonts}
\usepackage{upgreek}
\usepackage{subfigure}
\usepackage{siunitx}
\usepackage{hyperref}
\usepackage[utf8]{inputenc} %needed for special characters like ß or äöü
\usepackage[T1]{fontenc} %for correct hyphenation particularly if special characters are being used
\usepackage[capitalise]{cleveref} %for cref command
\usepackage{multirow}
\usepackage{setspace}

\soulregister\cite7
\soulregister\cref7

\usepackage[margins,adjustmargins]{changes}
\usepackage{makecell}
\usepackage{titlesec}

\volume{00}

\firstpage{1}

\journalname{Optics Communications}

\jid{procs}

\jnltitlelogo{Optics Communications}

\CopyrightLine{2023}{Published by Elsevier Ltd.}

\usepackage{amssymb}
\usepackage[figuresright]{rotating}

\begin{document}
	%\linenumbers

\begin{frontmatter}



\dochead{}
%% e.g. \dochead{17th International Conference on Dynamical Processes in Excited States of Solids}

\title{Method Comparison for Simulating Non-Gaussian Beams and Diffraction for Precision Interferometry}

\author[1,2,3]{Mengyuan Zhao\corref{cor1}}
\ead{zhaomengyuan17@mails.ucas.ac.cn}

\author[2,3]{Yazheng Tao}
\author[3,4]{Kevin Weber}  
\author[3,4]{Tim Haase}
\author[3,4]{Sönke Schuster}
\author[2,3,5]{Zhenxiang Hao}
\author[3,4]{Gudrun Wanner\corref{cor1}}
\ead{gudrun.wanner@aei.mpg.de}

\address[1]{Key Laboratory of Electronics and Information Technology for Space System, National Space Science Center, Chinese Academy of Sciences, No.1 Nanertiao, Zhongguancun, Haidian district, Beijing 100190, China}
\address[2]{University of Chinese Academy of Sciences, No.19(A) Yuquan Road, Shijingshan District, Beijing 100049, China}
\address[3]{ Max Planck Institute for Gravitational Physics (Albert Einstein Institute), Callinstr. 38, 30167 Hannover, Germany}
\address[4]{ Institute for Gravitational Physics of the Leibniz Universit\"at Hannover, Callinstr. 38, 30167 Hannover, Germany}
\address[5]{Institute of Theoretical Physics, Chinese Academy of Sciences, No.55 Zhongguancun East Road, Haidian District, Beijing 100190, China}

\cortext[cor1]{Corresponding author}

\begin{abstract}
In the context of simulating precision laser interferometers, we compare via several examples two wavefront decomposition methods: the Mode Expansion Method (MEM) and the Gaussian beam decomposition (GBD) for their precision and applicability. 
To judge the performance of these methods, we define different types of errors and study their properties. We specify how the two methods can be fairly compared and based on that, the quality of the MEM and GBD are compared in several examples. We test here cases for which analytic results are available, i.e., non-clipped circular and general astigmatic Gaussian beams, as well as clipped circular Gaussian beams, in the near-, far-, and extreme far-field of millions of kilometers occurring in space-gravitational wave detectors. Additionally, we compare the methods for aberrated wavefronts and the interaction with optical components by testing reflections from differently curved mirrors. 
We find that both methods can be generally used for decomposing non-Gaussian beams. However, which method is more accurate depends on the optical system and simulation settings. In the given examples, the MEM more accurately describes non-clipped Gaussian beams, while for clipped Gaussian beams and the interaction with surfaces, the GBD is more precise. 
\end{abstract}

\begin{keyword}
simulation \sep diffraction \sep interferometry



\end{keyword}

\end{frontmatter}

% \linenumbers


	\section{Introduction}

% Diffraction is common, diffraction integrals for comparison
In classic optics textbooks, e.g. \cite{hecht2001optics,ghatak1989optics,bea1991fundamentals1,siegman1986lasers}, diffraction is defined as the phenomenon occurring when a wave is obstructed while propagating. This is, for example, the case when a Gaussian beam is clipped by an aperture. The phenomenon has been known for centuries, and there are various methods for the computation of diffracted wavefronts and their propagation. The most classic approach is the evaluation of diffraction integrals such as the Fresnel-Kirchhoff diffraction formula or the Fraunhofer diffraction equation, for instance, described in \cite{born2013principles}.

% Non-gaussian, i.e. aberrated wavefronts
The propagation of diffracted light closely relates to the propagation of arbitrary wavefronts for which there exists no analytic propagation equation. Such arbitrary wavefronts include all clipped and diffracting beams, but likewise, for instance, aberrated wavefronts. Therefore, the very same methods are used for diffracted light and aberrated wavefronts.

% weekness of diffraction integrals for our application (+ strengths for other fields?) 
Even though we have known diffraction and aberration for a long time, there seems to be no suitable method at hand which allows propagating diffracting beams with high precision through complex optical setups where the beam repeatedly reflects and refracts at various tilted or even curved surfaces. 
Given our particular context of space-based interferometry for gravitational wave detection, i.e. the gravitational wave detectors LISA (Laser Interferometer Space Antenna) \cite{2006LISA}, and Taiji \cite{2017The}, our simulation methods are required to provide at least picometer resolution. This precision needs to be achieved for the propagation of a clipped beam through a telescope, followed by an optical bench with in the order of 50-100 components, where the beam might be clipped at different points while propagating through the three-dimensional optical layout\cite{amaro2017laser}.

% no diffraction integrals
Diffraction integrals are not ideal for this type of application since they are built for free-space propagation and need non-trivial adaptation for propagation through the described complex three-dimensional layouts. Yet, here are alternative approaches, which allow a comparably simple propagation of diffracting wavefronts through such setups.

%solution: MEM + GBD	
These alternatives are based on a decomposition into fundamental or higher-order Hermite- or Laguerre-Gaussian beams. Once the diffracting beam is decomposed, it can be easily propagated using well-known and fast algorithms (e.g. described in \cite{bea1991fundamentals2}) which are simple ray tracing for the beam axis and the ray transfer matrix formalism \cite{siegman1986lasers} for the propagation of the Gaussian $q$-parameter for the wavefront propagation.

These decomposition methods are well established but used under a variety of different names. The decomposition into higher-order Hermite- or Laguerre-Gaussian modes which all share the very same beam axis and beam parameters was firstly proposed in \cite{goubau1961}. Like \cite{Tanaka:88,NS2003Modeling,2007Modeling}, we refer to this method as Mode Expansion Method (MEM). It is also known as modal decomposition \cite{2010Interferometer} or truncated orthogonal series expansion \cite{2010Optimum}. If Laguerre-Gaussian modes are used for the decomposition, it is referred to Laguerre-Gauss expansions \cite{Borghi1996Optimization}, Laguerre–Gaussian series expansion method \cite{2014Simulated}, Laguerre–Gaussian mode decomposition \cite{xiao2019laguerre}, or if Hermite-Gaussian modes are used, truncated Hermite–Gauss series expansion \cite{Yongxin2006Truncated}.

Another decomposition method, with the concept of decomposing an arbitrary wavefield into Gaussian beams, was proposed by Popov in 1982 for acoustics\cite{popov1982new}. A similar idea was conceived by Graynolds in 1981\cite{2014Fat}, when he began developing a ray tracing code that eventually became the commercial software ASAP and published his paper on the subject in 1985\cite{1985Propagation,2014Fat}.
In this original description, the fundamental Gaussians were all parallel, had all the same waist size, and the waist was positioned in the decomposition plane. 
We refer to this original version of the method proposed by Graynolds as the Gaussian Beam Decomposition (GBD) and use this throughout this paper.
This original method was adapted over time, for instance, with non-parallel grid beams or grid beams with initial wavefront curvature, and implemented in several common commercial software tools, including ASAP\cite{narayananl2004gaussian}, FRED, and Code~V\cite{ashcraft2020open}. It is, unfortunately, proprietary and unknown, in which form or adaptation the method was implemented in the different software tools, but this shows, that the method is well established. 
The method of decomposing wavefront into Gaussian beams and its adaptations are often also referred to as Gaussian beam summation \cite{white1987some,leye2016gaussian}, Gaussian beam superposition\cite{SPIES2000155}, Gaussian beamlet decomposition \cite{ashcraft2020open}, Gaussian beamlet summation \cite{Alonso:02}, and particularly known by the name Beam Synthesis Propagation (BSP) in Code~V \cite{kong2013design}. 
% GBD adaptations
The GBD method has undergone further development in recent years, with one notable example being the Stable Aggregates of Flexible Elements (SAFE), which provides a field estimate that comprises a series of Gaussian contributions. Each member is linked to a ray and also contains the information about the adjacent rays \cite{Alonso:13}.
	Code~V's BSP decomposes the wavefront into Gaussian beams emitted from a single point but in various directions, as opposed to decomposing the wavefront into Gaussian beams on a grid. 
	Additionally, there are some adaptations made without changing the name. An improved GBD technique was suggested by Tanushev et al. to decompose high frequency wave fields into a sparse set of Gaussian beams. The selection principle used to determine the Gaussian beam parameters aimed to minimize the energy difference between the original wave field and the superimposed Gaussian beams \cite{tanushev2009gaussian}. In order to compute the scalar diffraction field of a two-dimensional field specified on a curved surface, \c{S}ahin et al. proposed an improved Gaussian Beam Decomposition (GBD) method. The three-dimensional field is expressed as a summation of Gaussian beams, each propagating in a different direction with waist positions located at discrete points on the curved surface, obtained through regular sampling \cite{Sahin:13}. Worku et al. introduced a revised Gaussian Beam Decomposition (GBD) method that enables the computation of vectorial field propagation through high numerical aperture (NA) objectives, where the decomposed Gaussian beams in their study are polarized \cite{worku2017}. Worku et al. \cite{2018Decomposition} presented a modified GBD which decomposes arbitrary fields with smooth wavefronts into fundamental Gaussian beams with initial curvatures. Finally, half or quarter Gaussian beams were suggested to be applied in the GBD to optimize the simulation of sharp beam edges after passing through a hard aperture with an arbitrary shape \cite{2019Propagation}. 

% weakness of MEM+GBD
Despite the number of publications using these methods, the number of publications describing and comparing the MEM and GBD in detail is currently low. Optimal settings for the methods are, therefore, often unknown, and the limitations of these methods are unclear.

% current knowledge about MEM: settings + review	
For the MEM, Borghi et al.\ \cite{Borghi1996Optimization} derived the optimal decomposition parameters using Laguerre-Gaussian (LG) modes for circular symmetric fields of \SI{1}{mm} radius, and particularly for top-hat beams. Yan Rong et al.\ \cite{YANRong2006Application} extended Borghi’s optimal rule to an arbitrary radius of aperture. 
Liu et al.\ \cite{Yongxin2006Truncated} presented the optimal decomposed beam waist for plane waves clipped by an arbitrary radius of aperture with Hermite-Gaussian (HG) modes. 

% current knowledge about GBD
Regarding the GBD, the publication status seems fairly sparse. About 30 years after the proposal of the basic concept of decomposing arbitrary electric fields into fundamental Gaussian beams, in Graynolds' overview article \cite{2014Fat}, he provided a revisit on the GBD, described the method's history and development over time and gave detailed implementation steps for the field decomposition, tracing, and computation of the resulting field in the target plane.  In \cite{2015Modeling}, various examples of modeling complex optical phenomena with the GBD were demonstrated, including interference and diffraction. However, these papers do not include a discussion of parameter settings, i.e. what waist size or waist location should be chosen for the grid beams, what overlap the grid beams should have or what type of grid would be ideal.

% main goal of this paper
As previously mentioned, recent publications have further advanced the development of the GBD method. However, they again do neither address the question of parameter settings, nor do they compare the performance of the GBD with the MEM. Within this paper, we specify our experience values for parameter settings in the GBD, when it is used for simulating simple cases such as non-clipped and clipped Gaussian beams, for which we have analytical results available to compare with. We then directly compare the performance of the MEM and the original GBD method as introduced by Greynolds. This comparison is performed for strongly varying propagation distances, ranging from the common case of a few millimeters in the very near field, up to millions of kilometers. With this exceptionally large propagation distance, we test the applicability of the methods for the context of space interferometry and particularly space gravitational wave detectors like LISA and Taiji, for which the properties of the electric field needs to be characterized for distances up to 3 million kilometers. Additionally, we qualitatively test the MEM and GBD in decomposing and propagating aberrated wavefronts, for which we do not have an analytic result to compare with. Finally, we test the propagation through an optical setup by reflecting the decomposed fields from curved mirrors with different curvatures.

% information on the paraxial approximation
It should be noted that the MEM and GBD are based on the paraxial approximation and are only suitable for specific scenarios. This includes cases where the Gaussian beam waist is significantly larger than the wavelength. The MEM and GBD underly the paraxial approximation, because higher order or fundamental Gaussian beams are derived based on the paraxial approximation, this extends to their propagation using the ray transfer matrix method as well. As the examples used in this context meet the assumptions of the paraxial approximation, the analytical methods are considered reliable and used as references.

% organization of this paper
After this introduction, we start this paper in \cref{se:2} with an introduction of the MEM and GBD, specify the free parameters in both decomposition methods and define different types of errors to evaluate the precision of electric field estimates obtained by each method. We discovered that while the MEM performs well in accurately resolving far field wavefronts even with small mode orders, it is not able to ideally resolve high-frequency spatial oscillations in the near-field. The MEM's accuracy naturally improves with higher mode orders, it also improves with the propagation distance, which is typically within the ranges of interest. Similarly, we observed that the GBD's typical settings are inadequate for accurately resolving high-frequency spatial oscillations in the near field, while comparably smoother far fields are resolved with higher accuracy. The GBD's accuracy improves with increasing grid sizes.

We then discuss in \cref{se:3} what mode order and grid size we chose to achieve a fair comparison of the methods. 
	Afterwards we directly compare the performance of the MEM and the original GBD method as introduced by Greynolds in \cref{se:4}. 
	This comparison is performed firstly in free space in \cref{sec:GBs-free-space}, for strongly varying propagation distances, ranging from the common case of a few millimeters in the very near field, up to millions of kilometers. With this exceptionally large propagation distance, we test the applicability of the methods for the context of space interferometry and particularly space gravitational wave detectors like LISA and Taiji, for which the properties of the electric field needs to be characterized for distances up to 3 million kilometers. Our demonstration indicated that both techniques can accurately resolve the field at extremely far distances without requiring re-decomposition at an intermediate point. A comparison between the direct methods revealed that the MEM outperforms the GBD when decomposing and propagating non-clipped circular and general astigmatic Gaussian beams. On the other hand, for clipped circular Gaussian beams, the GBD exhibited greater precision. However, these results may be dependent on various factors, such as the software and its implementation of both methods, the computer used, the operating system, and compilers.
	While in space gravitational wave detectors, the wavefronts will be aberrated, for instance, due to the telescope, and it also propagates through complex optical setups, therefore, we qualitatively test the MEM and GBD in decomposing and propagating aberrated wavefronts in \cref{se:5}, for which we do not have an analytic result to compare with. Based on our results, it can be inferred that both methods are generally suitable for decomposing and propagating aberrated wavefronts.
	Afterwards, we test the propagation through an optical setup by reflecting the decomposed fields from curved mirrors with different curvatures in \cref{se:6}. Our findings verified that when propagating through an optical setup that involves interactions with surfaces, the GBD method outperforms the MEM. Finally, we summarize the work and findings throughout this paper in \cref{se:7}.

%IfoCAD
All the simulations described throughout this paper have been performed using the software library IfoCAD \cite{ifocad}, which has both an implementation of the MEM and GBD. For all simulations performed for this paper, circular symmetric Gaussian beams were used for the decompositions. However, the IfoCAD algorithm allows these beams to become simple or general astigmatic during propagation through the local setup. For this, the methods described in \cite{kochkina2013simulating} are being used. The propagation of all modes and Gaussian beams follows the methods discussed in \cite{bea1991fundamentals2}.

\section{Wavefront Decomposition Methods} \label{se:2} 
Within this section, we introduce the MEM and GBD methods in detail and individually test their performance using a simple exemplary case. We start with the MEM in \cref{MEM_sec} and continue with the GBD in \cref{GBD_sec}. Both subsections are structured the same way: In \cref{MEM_definition} and \cref{sec:introGBD} we specify the mathematical properties and implementation of the methods. In \cref{MEM_error} and \cref{GBD_error}, we define three kinds of different error definitions for estimating the decomposition precision. In \cref{MEM_settings} and \cref{se:GBD_settings}, we discuss how the MEM and GBD parameters should be chosen for minimal error. This is known only for the MEM for typical setups, while we can state only our experience values for the GBD. Finally, we individually test in \cref{se:exampleMEM} and \cref{GBDexample} the performance of each method on examples. 

\subsection{Properties and individual test of the Mode Expansion Method}	\label{MEM_sec}
\subsubsection{MEM: method description} \label{MEM_definition}

The MEM is a well-known method defined, for instance, in \cite{siegman1986lasers}, which describes the decomposition of an arbitrary wavefront into higher-order Laguerre-Gaussian (LG) modes or Hermite-Gaussian (HG) modes. LG modes are radially symmetric and therefore defined in cylindrical coordinates, while HG modes are defined in rectangular coordinates due to their axial symmetry. A conversion between both types of modes is known and described in detail, for instance, in \cite{Kimel1993, ONEIL200035}. Throughout this paper, we therefore focus fully on decompositions using HG modes which are defined as

\begin{align}
	\begin{split}
		{\rm HG}_{mn}(x,y,z;w_{0d}) =& \frac{ c_{mn}}{w(z)}H_m\left( \frac{\sqrt{2}x}{w(z)}\right) H_n\left( \frac{\sqrt{2}y}{w(z)}\right) \\
		&\cdot\exp \left( -\frac{x^2+y^2}{w^2(z)}\right) \exp \left( -ik\frac{x^2+y^2}{2R(z)} + i(m+n+1)\zeta(z)\right)\exp \left( -ikz\right)\,,
	\end{split}
	\label{eq:HG-definition}
\end{align}
where $w_{0d}$ is the waist of the fundamental mode HG$_{00}$ which is used as a parameter for all HG modes, and the beam radius $w(z)$, the radius of curvature $R(z)$ and the Gouy phase $\zeta(z)$ have the same definitions as for fundamental Gaussian beams. The coefficients $c_{mn}$ are normalization constants:
\begin{equation}
	c_{mn} = \sqrt{\frac{2}{\pi}}\frac{1}{\sqrt{m! 2^m}} \frac{1}{\sqrt{n! 2^n}}  \,.
	\label{eq:cmn}
\end{equation}

The function $H_m\left( \bullet \right)$ is the $m^{\rm th}$ Hermite polynomial given by \cite{siegman1986lasers}
\begin{equation}
	H_m(x) = (-1)^me^{x^{2}}\frac{d^m}{dx^m}e^{-x^{2}} \,.
	\label{eq:hermite}
\end{equation}
An important property of these HG modes is that they are orthonormal and complete and thereby form a basis \cite{siegman1986lasers}:
\begin{equation}
	\iint {\rm HG}_{mn}^*(x,y,z;w_{0d}) 
	{\rm HG}_{kl}  (x,y,z;w_{0d})
	\, dx dy= \delta_{mk} \delta_{nl} \,,
	\label{eq:HG-orthogonality}
\end{equation}
where HG$^{*}_{mn}(x,y;w_{0d})$ is the complex conjugated Hermite Gaussian mode, and $w_{0d}$ is the waist of the fundamental mode HG$_{00}$ which is used as parameter for all HG modes. The Kronecker delta function $\delta_{mk}$ equals 1 for $m = k$ and equals 0 if $m \neq k$. This implies that any wavefront $E(x,y)$ can be decomposed into a superposition $E_{\rm \infty}(x,y)$ of these modes \cite{siegman1986lasers}:	%
\begin{equation}
	E_{\rm \infty}(x,y) :=\sum_{m=0}^{+\infty}\sum_{n=0}^{+\infty}a_{mn}{\rm HG}_{mn}(x,y;w_{0d})\,,
	\label{es}
\end{equation}
where $E_{\rm \infty}(x,y)$ is a mathematically exact representation of $E(x,y)$: 
\begin{equation}
	E_{\rm \infty}(x,y) \equiv E(x,y) \,,
\end{equation}
where $\equiv$ indicates the functions are equivalent for every point $(x,y)$. The complex coefficients $a_{mn}$ with $|a_{mn}|^2 = P_{mn}$ are usually referred to as mode overlap and describe how much beam power $P_{mn}$ is stored in each mode. They can be calculated by the inner product \cite{siegman1986lasers}
\begin{equation}
	a_{mn}={\iint {\rm HG}^{*}_{mn}(x,y;w_{0d})E(x,y)dxdy }\,.
	\label{eq:amn}
\end{equation}

In real computations, it is not possible to use either an infinite number of modes in the decomposition (\cref{es}), or an infinite overlap integral (\cref{eq:amn}) for the determination of the mode overlap. Replacing the infinite surface integral in \cref{eq:amn} with a finite one is uncritical provided the surface is chosen to be sufficiently large because electric fields of interest are usually fading out towards higher radial distances. Therefore, by choosing appropriately large integration boundaries, the introduced error becomes negligible.
However, the error made by working with a finite mode order $N$ is often non-negligible. Consequently, the decomposed field is no longer an exact representation of the input field $E(x,y)$, but only an approximation:	
\begin{equation}
	E(x,y) \approx {E}^N(x,y,w_{0d})=\sum_{m=0}^{N}\sum_{n=0}^{N-m}a_{mn}{\rm HG}_{mn}(x,y;w_{0d}) \,.
	\label{es1}
\end{equation}
Here, we refer to $N$ as the maximum mode order of the MEM. 
In \cref{es1}, we use a triangular summation of the modes by summing $n$ only up to $N-m$ rather than $N$ \cite{ONEIL200035}.
This ensures, that within any decomposition, all polynomials up to the given order $N$ are considered, and no polynomial orders larger than $N$ are included. Consequently, in any MEM with mode order $N$, there are $(N+1)(N+2)/2$ HG modes superimposed. For radially symmetric fields $E(x,y)$, the overlap $a_{mn}$ is set to zero if either the index $m$ or $n$ is odd. 
This means that for any mode order $N$ the number $\nu$ of modes used in the MEM is given by 
\begin{equation}
	\nu = \begin{cases}
		(N+1)(N+2)/2  & \text{if the $E(x,y)$ is non-symmetric } \\
		(N/2+1)(N/2+2)/2  & \text{if $N$ is even and $E(x,y)$ is symmetric } \\ 
		((N-1)/2+1)((N-1)/2+2)/2  & \text{if $N$ is odd and $E(x,y)$ is symmetric. } 
	\end{cases}
	\label{eq:no_of_modes_in_MEM}
\end{equation}

%%%%%%%%%%%%%%%%%%%%%%%%%%%%%%%
\subsubsection{Error definitions for the MEM}\label{MEM_error}
The finite mode order decomposition given in \cref{es1} is not exact and will, therefore, have an error. It can be learned that for any given mode order, the decomposition error of the MEM depends on the mode order and the waist size $w_{0d}$ chosen for the modes in the decomposition. Using the norm 
\begin{equation}
	\left\| f(x,y)\right\| ^2 = \iint_{\mathbb{R}^2} \left| f(x,y)\right|^2 dxdy
	\label{eq:2ndnorm} \,,
\end{equation}
the \emph{normalized mean squared error (NMSE)} is given by

\begin{align}
	\varepsilon^\text{NMSE}(N, w_{0d}) &:= 
	\frac{\left\|  {E}^N(x,y,w_{0d})-E(x,y)\right\| ^2}{\left\| E(x,y)\right\| ^2} \label{eq:ierror_def} \\
	&
	=\frac{\left\|  E(x,y)\right\| ^2}{\left\| E(x,y)\right\| ^2} 
	- 2 \Re \frac{{E}^N(x,y,w_{0d}) E^*(x,y)}{\left\| E(x,y)\right\| ^2} 
	+ \frac{\left\|  {E}^N(x,y,w_{0d})\right\| ^2}{\left\| E(x,y)\right\| ^2}
	\nonumber \\
	&= 
	1 - 2 \Re \left( \sum_{m=0}^{N}\sum_{n=0}^{N-m}a_{mn} \iint_{\mathbb{R}^2} E^*(x,y){\rm HG}_{mn}(x,y;w_{0d}) dxdy \right) /{\left\| E(x,y)\right\| ^2}  \nonumber\\ 
	&+\left( \sum_{m=0}^{N}\sum_{n=0}^{N-m} a_{mn}^2 \iint_{\mathbb{R}^2}{\rm HG^2}_{mn}(x,y;w_{0d}) dxdy\right) 
	/{\left\| E(x,y)\right\| ^2}\nonumber \\
	&= 1 - \frac{\sum_{m=0}^{N}\sum_{n=0}^{N-m}a_{mn}^2}{P} \,,
	\label{eq:ierror}
\end{align}
using ${\left\| E(x,y)\right\| ^2 = P}$, where $P$ is the power of the initial beam. For any input field $E(x,y)$, this normalized mean squared error $\varepsilon^\text{NMSE}(N, w_{0d})$ depends solely on the mode order $N$ and waist size $w_{0d}$ chosen during the decomposition, and it has the property of being propagation distance independent (cf. \cref{eq:amn}, and \cite{Borghi1996Optimization}).

The NMSE is defined via infinite surface integrals, which are replaced by numerical integrals over finite surfaces in optical simulations. This means, that in simulations a \emph{discretized NMSE (DNMSE)} $\varepsilon^\text{DNMSE}_\text{o}(N_R,R,z)$ is evaluated. For radial surfaces and assuming radially symmetric beams, this is given by 
\begin{align}
	\varepsilon^\text{DNMSE}_\circ(N_R,R,z) &:= \frac{\sum_{i = 0}^{N_R}{2 \pi \left| {E}^N(r_i,z,w_{0d})-E(r_i,z)\right| ^2} r_i \Delta r}{\left\| E(x,y,z)\right\| ^2}
	\nonumber \\
	&= 
	\frac{\sum_{i = 0}^{N_R}{2 \pi \left| {E}^N(r_i,z,w_{0d})-E(r_i,z)\right| ^2} r_i \Delta r}{P}\,,
	\label{eq:11a}
\end{align}
and for non-radially symmetric beams on rectangular surfaces by 
\begin{equation}
	\varepsilon^\text{DNMSE}_\Box (N_X,N_Y,X,Y,z) := \sum_{i=1}^{N_X} \sum_{j=1}^{N_Y} \frac{\left | E^N(x_i,y_j,z,w_{0d})-E(x_i,y_j,z) \right |^{2} \Delta x \Delta y }{P} \,.
	\label{eq:11a1}
\end{equation}
Here, $r = \sqrt{x^2+y^2}$ denotes the radial distance, $N_R, N_X, N_Y$ are the numbers of sampling points, and $\Delta r$, $\Delta x$ and $\Delta y$ are the step sizes in the different directions, such that the maximal distances are $R=N_R \Delta r$, $X = N_X \Delta X$, $Y= N_Y \Delta Y$. Due to the assumed radial symmetric beams, we substituted $\iint_{\mathbb{R}^2} dxdy$ by $\sum 2 \pi r\Delta r$ for \cref{eq:11a}. Only the numerator is discretized in the DNMSE because the NMSE is normalized by the initial beam's total power $P$, which is usually known.

The discretized NMSE is a numerical representation of the NMSE, and thereby propagation distance independent provided that enough sampling points are chosen and the radial distance $R$ is sufficiently large. However, this implies that with non-ideal settings, such as too few sampling points or a too small radial range, the error is indeed propagation distance dependent, which we highlight in \cref{eq:11a} by the explicitly stated $z$-dependency.

One disadvantage of all shown errors is that they give only an integrated information and no distribution over a plane. We, therefore, define a  sampling point dependent error $\varepsilon^\text{rel}(x_i,y_i,z)$ which we name \emph{relative error:}

\begin{equation}
	\varepsilon^\text{rel}(x_i,y_i,z) := \frac{\left | {E}^N(x_i,y_i,z,w_{0d}) - E(x_i,y_i,z)\right |}{\left | E(x_i,y_i,z) \right |}\,. 
	\label{eq:rerrorsk}
\end{equation}

For radially symmetric beams, we sample along the $x$-axis by setting $y = 0$. Therefore, the reduced 1D version of \cref{eq:rerrorsk} can be written as:
\begin{equation}
	\varepsilon^\text{rel}(r_i,z)=
	\varepsilon^\text{rel}(x_i, 0, z)= \frac{\left | {E}^N(x_i,0,z,w_{0d}) - E(x_i,0,z)  \right |}{\left | E(x_i,0,z) \right |}\,. 
	\label{eq:rerrorsk1d}
\end{equation}

Throughout this paper, we use these relative errors to visualize the performance of the MEM, as well as for a qualitative comparison of the MEM with the GBD. To quantify the resulting information and judge the total error in the finite surface of interest, we define the \emph{summed relative error}, %in 2D and 1D respectively, by
for rectangular target surfaces and no assumed symmetry, or circular symmetric beams on a circular surface
\begin{align}
	\varepsilon_{\sum}^\text{rel}(N_X,N_Y,X,Y,z) &= \sum_{i=1}^{N_X}\sum_{j=1}^{N_Y}  \frac{\left |{E}^N(x_i,y_j,z,w_{0d}) - E(x_i,y_j,z) \right |}{\left | E(x_i,y_j,z) \right |} \Delta x\Delta y
	\nonumber \\
	&= \sum_{i=1}^{N_X}\sum_{j=1}^{N_Y}  \varepsilon^\text{rel}(x_i,y_j,z) \Delta x\Delta y\,,
	\label{eq:rerror} \\
	%
	%		\varepsilon_{\sum}^\text{rel}(N_R, R,z) &= \sum_{i=1}^{N_X}  \frac{\left |{E}^N(x_i,0,z,w_{0d}) - E(x_i,0,z) \right |}{\left | E(x_i,0,z) \right |}\Delta x = \sum_{i=1}^{N_R}  \varepsilon^\text{rel}(r_i,z)\Delta r\,,
	%		\label{eq:rerror1d}
	\varepsilon_{\sum}^\text{rel}(N_R,R,z) &:= {\sum_{i=0}^{N_R}}\frac{2\pi \left | E^N(r_i,z,w_{0d})-E(r_i,z) \right | r_i \Delta r }{\left | E(r_i,z) \right | } \,.
	\label{eq:src}
\end{align}
This summed error definition is, for typical settings in optical simulations,  fairly independent of the chosen number of sampling points since \cref{eq:rerror} and \cref{eq:src} are representations of a discretized integral and therefore represent the surface under by the given function. However, the summed relative error is, unfortunately, not normalized. \\% We also define a circular summed relative error, for radially symmetric beams
We use all introduced error types throughout this paper to study the performance of the MEM and to compare it with the GBD. 

\subsubsection{MEM settings} \label{MEM_settings}
When a wavefront is decomposed using the MEM, there are only two parameters that need to be chosen: the waist $w_{0d}$ of the modes used in the decomposition, and the maximum mode order $N$. For any maximum mode order, the choice of $w_{0d}$ directly affects the magnitude of the mode overlap $a_{mn}$ and hence the resulting error $	\varepsilon^\text{NMSE}(N, w_{0d})$. One can then for instance choose the decomposition waist $w_{0d}$ such that the mode overlap of a specific mode is maximal (e.g. used in \cite{2007Modeling}) or, alternatively, such that the error made in the decomposition is minimal. Throughout this paper, we use the latter criterion, minimizing the error $\varepsilon^\text{NMSE}(N, w_{0d})$, following the examples of \cite{Borghi1996Optimization,Yongxin2006Truncated,YANRong2006Application}. 

%	 Therefore, for any given mode order $N$, the choice of the modes' waist size $w_{0d}$ affects the residual errors $\varepsilon^N $ and $ \hat\varepsilon^N(X, Y, z)$ and thereby the precision of the decomposition.
Which waist is optimal for the decomposition depends on the properties of the initial wavefront $E(x,y,z)$. For an arbitrary wavefront, the optimal decomposition waist is therefore unknown.  However, for the common special case of circular symmetric wavefronts originating from clipping at an aperture of radius $R_{\rm a}$ the optimal waist was found to be \cite{Borghi1996Optimization,YANRong2006Application,Yongxin2006Truncated,item_1660139}: 
\begin{equation}
	w_{0d}=R_{\rm a}\sqrt{\frac{2}{N}} \label{borghi}\,.
\end{equation}
This choice results in the minimum NMSE $\varepsilon^\text{NMSE}(N, w_{0d})$ in \cref{eq:ierror} for a given mode order $N$ \cite{Borghi1996Optimization,YANRong2006Application,Yongxin2006Truncated,item_1660139}.

\subsubsection{Example: MEM performance for a clipped Gaussian beam} \label{se:exampleMEM}

% headline: the example is on clipped Gaussians
We will now demonstrate the performance of the MEM in an example, for which the electric field is analytically known. In this example we investigate a clipped Gaussian beam. We assume that a Gaussian beam impinges orthogonally and perfectly aligned to a circular aperture with radius $R_a = \SI{0.5}{mm}$.  The waist of the incident Gaussian beam has a radius of $w_0= \SI{2}{mm}$ which is located in the aperture plane. The resulting circular symmetric clipped Gaussian beam is decomposed by an MEM with varying mode orders: $N= 10,20,...50$. For every mode order, the waist size $w_{0d}$ of the modes is calculated using \cref{borghi}.
% headline: then Fresnel numbers, Campbell and Tanaka
We compute the electric fields amplitude and phase in various propagation distances and compare the results with the numerical evaluation of an analytic formula developed by Campbell in \cite{1987Fresnel} for clipped Gaussian beams in the Fresnel region. Analogously, we use the analytical method of Tanaka et al.\ cf.\cite[Eq.(1) $\sim$ Eq.(6)]{1985Field} for the Fraunhofer region. The distinction between the Fresnel and Fraunhofer region, is done using the Fresnel number $F$ given by
\begin{equation}
	F:=\frac{R_{a}^{2}}{\lambda d}\,,
	\label{eq:fresnelnumber}
\end{equation}
where $\lambda$ denotes the wavelength of the beam, $d$ the propagation distance after the clipping aperture, and $R_a$ the radius of the aperture. The near field refers to propagation distances, which make $F$ larger than 1. If the Fresnel number is smaller than  1, the beam has propagated to the far field.
% headline: we have an example of clipped Gaussians and the parameters.
In this example, we use the propagation distances $d$ of $\SI{5}{mm}, \SI{20}{mm}, \SI{100}{mm}$, i.e. $F$ is $46.9925, 11.7481, 2.3496$ in the near field, and  $d=\SI{1000}{mm}$ with $F = 0.2350$ in the far-field.   
The number of sampling points for requesting the complex electric field is set to 3001. 
For convenience, all parameters of this example are listed in \cref{mem_example_setting}.
\begin{table}[htbp]
	\caption{Parameters list of the MEM example.}
	\footnotesize
	\resizebox{\textwidth}{!}
	{
		\begin{tabular}{@{}lll}
			\hline
			\textbf{parameters} & \textbf{description} & \textbf{value}  \\ 
			\hline
			$\lambda$          & wavelength  & \SI{1064}{\nm} \\
			$P_0$          & beam power  & \SI{1}{W} \\  
			$w_0$         & beam waist  &  \SI{2}{\mm}     \\ 
			$z_0$        & distance from the waist           &   \SI{0}{mm}      \\ 
			$R_a$          & aperture radius       &  \SI{0.5}{\mm}     \\ 
			$N$          & mode order of the MEM    &   10, 20, 30, 40, 50     \\ 
			$w_{0d}$          & waist of the modes used in the MEM         &  \SI{0.2236}{\mm}, \SI{0.1581}{\mm}, \SI{ 0.1291}{\mm}, \SI{0.1118}{\mm}, \SI{0.1}{\mm}            \\
			$d$ & propagation distance & \SI{5}{mm},\SI{20}{mm},\SI{100}{mm},\SI{1000}{mm}\\
			$X$ & number of sampling points & 3001 \\
			\hline
		\end{tabular}
	}
	\label{mem_example_setting}
\end{table}
\normalsize
%

\begin{figure}[htbp]
	\begin{center}
		\includegraphics[]{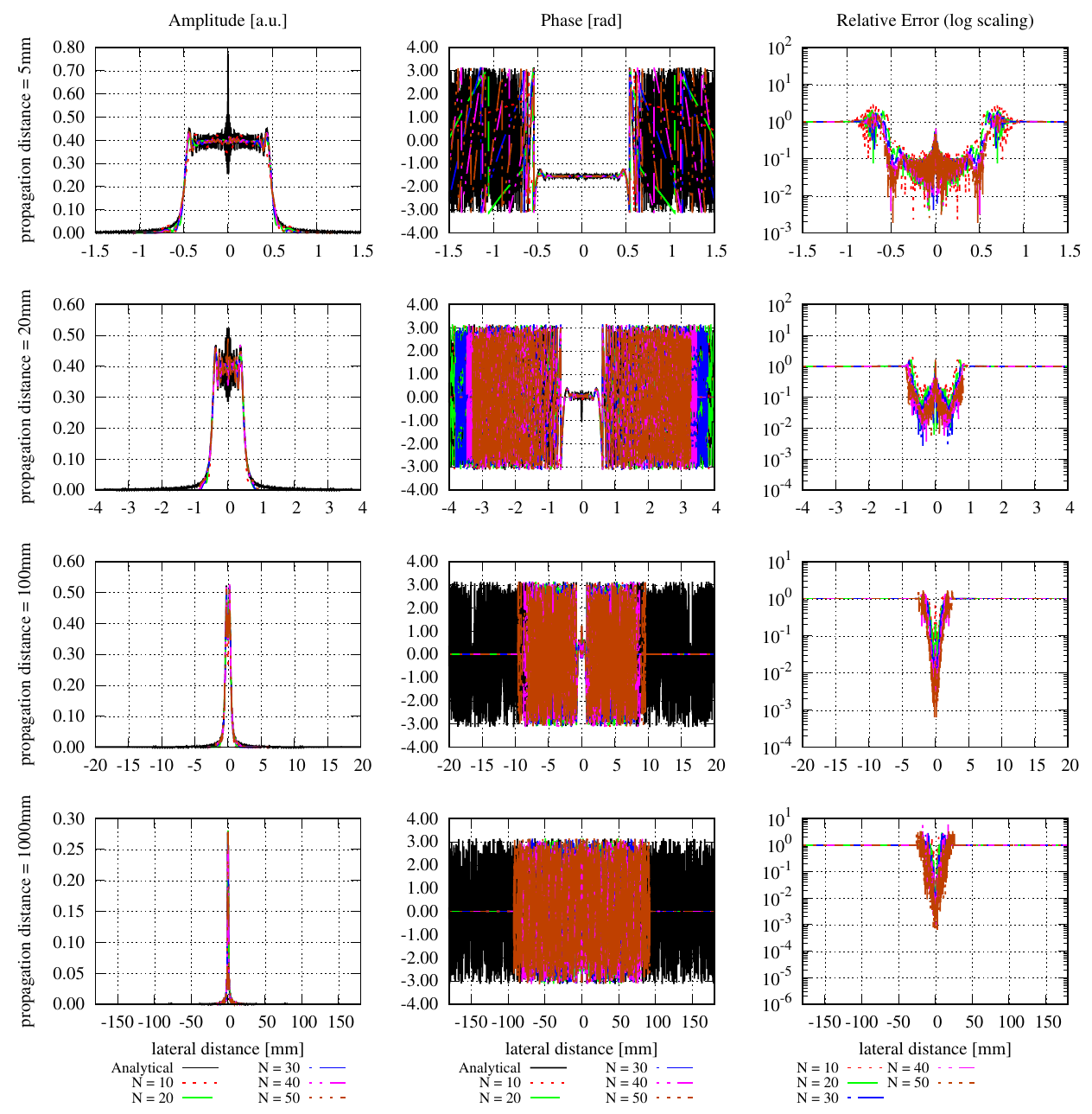}
	\end{center}
	\caption{Performance of a MEM with different maximum mode orders $N$ = 10, 20, 30, 40 and 50, for an incoming circular Gaussian beam with a \SI{2}{mm} waist being clipped by a \SI{0.5}{mm} radius circular aperture. Shown are the amplitude (absolute value), phase and relative error $\varepsilon^\text{rel}(N_R,R,z)$ at different propagation distances $z=$ \SI{5}{mm}, \SI{20}{mm}, \SI{100}{mm} (near field) and \SI{1000}{mm} (far field) after the clipping aperture. The analytical methods for the near and far field are Campbell \cite{1987Fresnel} and Tanaka et al.\ \cite{1985Field}, respectively. Lateral distances for each propagation distance are chosen large enough to cover all the power. For these large lateral ranges, the MEM is effectively failing, generating zero amplitude and phases from lateral ranges that are about 3 times the spot size of the highest mode in the decomposition.}
	\label{error_modes}
\end{figure}

\begin{figure}[htbp]
	\begin{center}
		\includegraphics[]{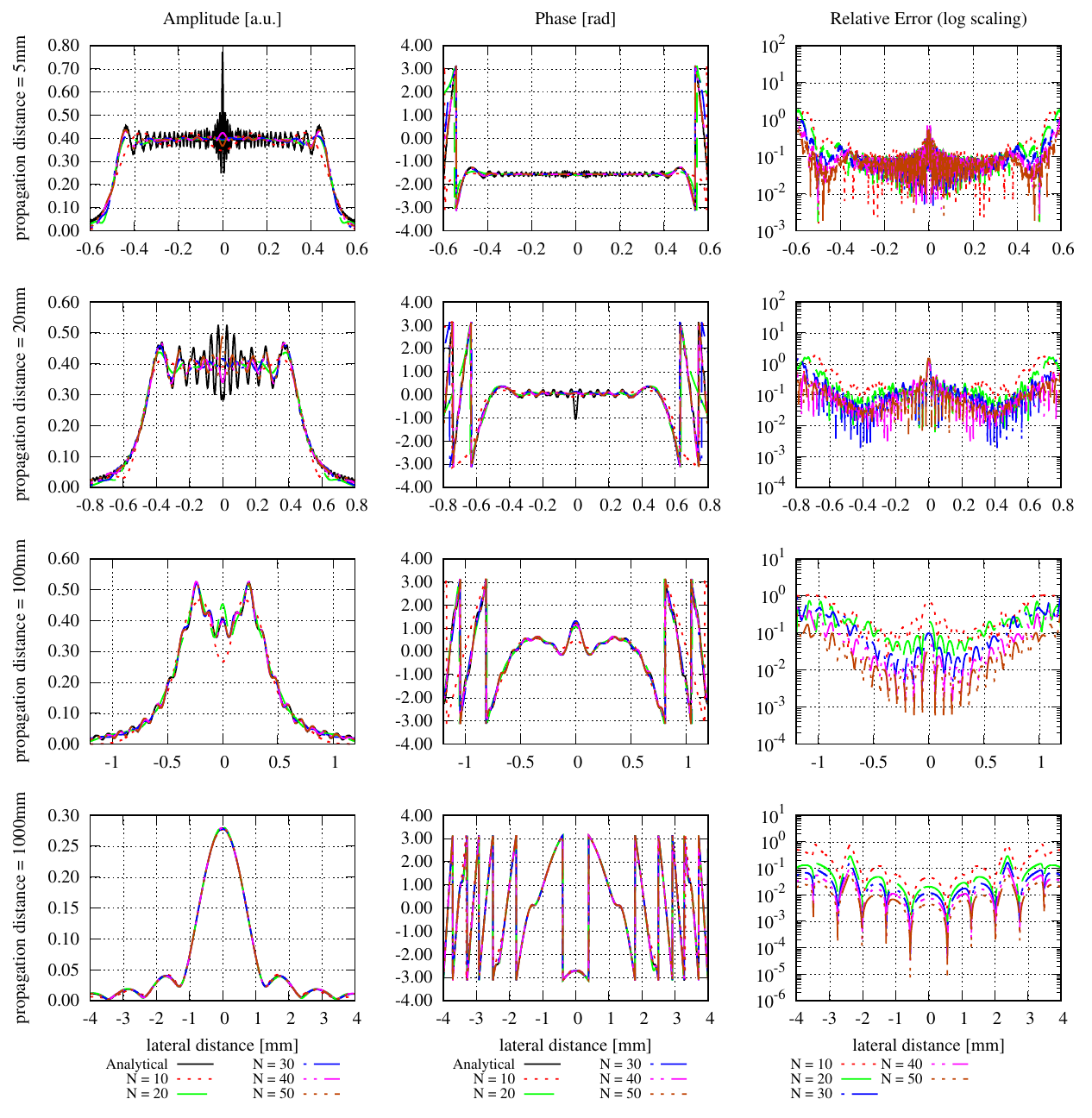}
	\end{center}
	\caption{MEM performance for the very same test case shown in \cref{error_modes} but for smaller lateral ranges. Shown is the performance of a MEM with different maximum mode orders $N$ = 10, 20, 30, 40 and 50, for an incoming circular Gaussian beam with a \SI{2}{mm} waist being clipped by a \SI{0.5}{mm} radius circular aperture. Shown are the amplitude (absolute value), phase and relative error $\varepsilon^\text{rel}(N_R,R,z)$ at different propagation distances $z=$ \SI{5}{mm}, \SI{20}{mm}, \SI{100}{mm} (near field) and \SI{1000}{mm} (far field) after the clipping aperture. The analytical methods for the near and far field are Campbell \cite{1987Fresnel} and Tanaka et al.\ \cite{1985Field}, respectively. One can see that the further the beam propagates, or the higher the mode order is, the better is the performance of the MEM.}
	\label{error_modes_small_range}
\end{figure}

%headline: showing the results, fig.1 for larger lateral ranges, fig.2 for smaller, and table 1 shows all errors
The amplitude, phase, and relative error distribution calculated via \cref{eq:rerrorsk1d} are shown in \cref{error_modes} for different propagation distances after the clipping aperture with different mode orders.
The lateral ranges chosen for this figure are unusually large. The corresponding results for a smaller lateral range are, therefore, shown in \cref{error_modes_small_range}.
All introduced error types have been calculated for both choices of lateral ranges and listed in \cref{sumofremem}.
\begin{table}[htbp]
	\caption{The MEM errors, including the NMSE, discretized NMSE and the summed relative error, defined in \cref{eq:ierror},  \cref{eq:11a} and \cref{eq:src} respectively, are calculated for increasing mode orders at different propagation distances. The NMSE and discretized NMSE, which are propagation distance independent, when the lateral ranges $R$ are large enough, are numerically equivalent. For smaller lateral ranges $R$, the discretized NMSE is propagation distance dependent. The summed relative error decreases  with increasing mode orders for any propagation distance, and for a given mode order it increases (but not consistently) with increasing propagation distance. The number of sampling points $X$ is 3001.}
	\footnotesize 
	\resizebox{\textwidth}{!}
	{
		%\centering
		\begin{tabular}{@{}lllllll}
			\hline
			\textbf{\makecell[c]{propagation \\distance \\(mm)}} & \textbf{\makecell[c]{mode \\order}}  & \textbf{\makecell[c]{NMSE\\$\varepsilon^\text{NMSE}(N, w_{0d})$}} & \textbf{\makecell[c]{DNMSE\\(for larger lateral ranges) \\$\varepsilon^\text{DNMSE}_\circ(N_R,R,z)$}}&\textbf{\makecell[c]{DNMSE\\(for smaller lateral ranges) \\$\varepsilon^\text{DNMSE}_\circ(N_R,R,z)$}} &\textbf{\makecell[c]{the summed relative error\\(for larger lateral ranges) \\$	\varepsilon_{\sum}^\text{rel}(N_R,R,z)$}} & \textbf{\makecell[c]{the summed relative error\\(for smaller lateral ranges) \\$	\varepsilon_{\sum}^\text{rel}(N_R,R,z)$}} \\ 
			\hline
			\multirow{5}{*}{5}&	10 & 0.0527&0.0519 &0.0436&13.5852&	0.9643 \\
			&	20 &0.0275&0.0268 &	0.0210&12.8045
			&0.7307\\  
			&	30 &0.0186&0.0178 &0.0116&12.5164&0.4855 \\ 
			&	40 &0.0139&0.0132 &0.0077&12.3217
			&0.3436\\ 
			&	50 &0.0112&0.0105 &0.0057&12.1582&0.2465\\ \hline
			\multirow{5}{*}{20}&	10 & 0.0527&0.0517 &0.0444&99.0521&2.5357\\ 
			&	20 &0.0275&0.0265 &0.0192&	98.2983&1.7465\\ 
			&	30 &0.0186&0.0175 &0.0101& 97.6844	&1.1216\\ 
			&	40 &0.0139&0.0130 &0.0057&97.0994&0.5350\\ 
			&	50 &0.0112&0.0102 &0.0042&96.5324&0.4630\\ \hline
			\multirow{5}{*}{100}&	10 &0.0527&0.0517 & 0.0323 & 2510.17&5.1104\\ 
			&	20 &0.0275&0.0265 &0.0083&2506.91&2.4619\\ 
			&	30 &0.0186&0.0175 & 0.0035&2503.34
			&2.2752\\ 
			&	40 &0.0139&0.0130 &$8.8482\times10^{-4}$&2499.37 &1.2239\\ 
			&	50 &0.0112&0.0103 &$1.9470\times10^{-4}$	&2494.99&0.5380\\ \hline
			\multirow{5}{*}{1000}&	10  &0.0527&0.0516 & 0.0051 &203568 &28.4408 \\ 
			&	20 &0.0275&0.0264 &$6.0815\times10^{-4}$&203402&8.4849	\\ 
			&	30 &0.0186&0.0174 &$1.8625\times10^{-4}$&203138&4.5648\\ 
			&	40 &0.0139&0.0129 &$6.9439\times10^{-5}$&202895&2.7671	\\
			&	50 &0.0112&0.0102 &$2.5676\times10^{-5}$&202455&1.6778	\\
			\hline
		\end{tabular}
	}
	\label{sumofremem}
\end{table}	
\normalsize
%

% headline: reasons for choosing these different lateral ranges
The large lateral ranges used for \cref{error_modes} were chosen such that they are large enough to make the DNMSE propagation distance independent. In this case, the deviation between the input beam power and the MEM beam power was less than 2\%. The incident beam power $P$ is calculated simply from the Gaussian beam power passing through the aperture radius $R_a$:
\begin{equation}
	P= P_0 [1-\exp\left(-2R_a^2/w_{0}^2\right)] \,,
	\label{clipped_power}
\end{equation}
with $R_a$ = \SI{0.5}{mm}, $w_{0}$ = \SI{2}{mm} and $P_0$ being the full power of the Gaussian beam prior to clipping. The resulting normalized power ($P/P_0$) of the clipped Gaussian beam is 0.1175. The power of the MEM beams were computed by the numerical sum $\sum_{i = 0}^{R}2 \pi\left| E(r_i)\right|^2 r_i \Delta r$, which is a numerical representation of the denominator of \cref{eq:11a}. 	
This procedure resulted in a slight variation of the MEM beam power in the different propagation distances. The deviations between the input beam power and the MEM beam power are 1.15879\%, 1.15949\%, 1.17214\% and 1.65811\% for propagation distances \SI{5}{mm}, \SI{20}{mm}, \SI{100}{mm}, \SI{1000}{mm}, respectively.
Ideally, the lateral range would be chosen from the spot size of the clipped beam in the various propagation distances. Yet, particularly for clipped and diffracted beams, there is not one uniquely defined spot size, but rather a number of different concurring options, which are often not analytically known. Although a detailed discussion on spot sizes of clipped beams is beyond the scope of this paper, we want to compare the beams spot size with the chosen lateral ranges.
In the near field behind the aperture, i.e. with Fresnel number $F \gg1$, the spot size of the clipped beam is still roughly equal the aperture radius and so in our example ($F =  46.9925$) the spot size is approximately \SI{0.5}{mm}. Therefore, in row one of \cref{error_modes}, the lateral range we are showing is \SI{1.5}{mm}, which is approximately 3 times the spot size. For the propagation distance \SI{1000}{mm}, which is in the far-field, we can use \cite[Eq.(8)]{Drege:00} to estimate that the spot size of the clipped beam to be \SI{0.8751}{mm}. This is consistent with the spot size of a Gaussian beam with \SI{0.5}{mm} waist at propagation distance \SI{1000}{mm}, which is \SI{0.8419}{mm}, and thereby slightly smaller than the clipped beam, as expected.
Yet, in our computation, 3 times \SI{0.8751}{mm} did by far not fulfill the propagation distance independent DNMSE, such that we had to extend the lateral range to \SI{180}{mm} instead.

These large lateral distances with a constant MEM beam power result in the expected propagation distance independency of the discretized normalized mean square error $\varepsilon^\text{DNMSE}$, as shown in the fourth column of \cref{sumofremem}, except of some minor variations.
We can see from \cref{error_modes} that for propagation distances of \SI{20}{mm} and larger, there exists a maximum lateral range, for which the phase is correctly approximated by the MEM (see the zero lines for larger lateral distance). This is due to the finite size of the modes used in the decomposition. 
For a propagation distance $z$, the spot size of the higher order modes along $x$ and $y$ is \cite{1980spot}

\begin{align}
	w_{x,mn}(z) = \sqrt{2m+1} w(z)= \sqrt{2m+1} w_{0d}\sqrt{1+(z/z_r)^2} \\
	w_{y,mn} (z)= \sqrt{2n+1} w(z)= \sqrt{2n+1} w_{0d}\sqrt{1+(z/z_r)^2}
\end{align}
with $w(z)$ being the spot size of the fundamental mode HG$_{00}$ used in the decomposition, and the Rayleigh $z_r = \pi w_{0d}^2/\lambda$. Using these equations, we can estimate the spot size of the highest mode used in the MEM. 
The MEM can then only resolve fields in the range of maximally three times this spot size of the highest used mode.
We can show this on the example of a propagation distance of \SI{1000}{mm} (lowest row in  \cref{error_modes}) and a mode order of 50. For this, we find $w_{x,50\,0}= w_{y,{0\,50}} = \SI{32}{mm}$, resulting in a maximal resolvable range of approximately \SI{96}{mm}, which fits precisely the observation in the phase graph.

It might be expected that a higher mode order automatically implies that a larger lateral range can be resolved. However, that is not necessarily the case, as can be seen in \cref{error_modes} for a propagation distance of \SI{20}{mm}. Here, this inverts: the higher the mode order, the smaller the resolvable lateral range. This is a consequence of using \cref{borghi} to compute the optimal waist size, which decreases with increasing mode order.

Outside the maximal resolvable lateral range, the MEM is failing and generates zero amplitudes and phases. Consequently, the relative error is approximately 1 outside the maximal resolvable lateral range. This is clearly visible in \cref{error_modes}. However, the large lateral range is not a choice usually taken in simulations since the spot properties are barely visible in these lateral ranges. Instead, simulations are usually performed with smaller lateral ranges in the target plane, such as shown in \cref{error_modes_small_range}. Here, the lateral ranges cover only 0.4, 0.2, 0.06 and 0.022 times the ranges shown in \cref{error_modes}, respectively. 
For instance, for the propagation distance \SI{5}{mm}, the lateral distance shown in \cref{error_modes_small_range} is \SI{0.6}{mm}, compared to the calculated spot size of \SI{0.5}{mm}. For the propagation distance of \SI{1000}{mm}, the computed spot size is \SI{0.8751}{mm}, in comparison of the \SI{4}{mm} lateral distance shown in \cref{error_modes_small_range}.
These lateral changes were chosen simply for good visibility of the amplitude and phase profiles without any hard criterion. 

% discussion of the figure and the performance of the MEM
From the first row of \cref{error_modes_small_range}, one sees that the MEM with the given settings and mode orders up to $N=50$ insufficiently resolves the high-frequency spatial oscillation in the very near field behind the aperture. However, the further the beam propagates, the better is the performance of the MEM, such that after \SI{1000}{mm} (i.e. at $F=0.235$, the wavefront is well represented even with a mode order of 10. So while the NMSE is propagation distance independent and therefore constant for any choice of $N$ we see from the left and center columns of \cref{error_modes_small_range} how the precision of the MEM increases with increasing propagation distance. This means that the error radially transmits outwards and might therefore be in a radial distance of no interest in the application.
This also shows that it is not always necessary to choose high mode orders, particularly for far field simulations. Instead, the mode order should be chosen as a compromise between different criteria. The primary criterion is the increasing computational effort with increasing mode order. A second criterion is that the evaluation of the sum of Hermite-Gaussian modes with high polynomial orders is a typical mathematical challenge, resulting in numerical errors for high mode orders. Finally,  
the optimal decomposed beam waist calculated according to \cite{Borghi1996Optimization} and \cref{borghi} decreases with increasing mode order and needs to be sufficiently large to not violate the paraxial approximation. Consequently, the mode order should be chosen carefully under consideration of the intended precision and the costs and risks if the mode order is chosen too high.

% dicussion the errors and their properties
We can now compare the different errors for the case of the large lateral ranges. In that case, the DNMSE (fourth column of \cref{sumofremem}) is propagation distance independent and deviates from the analytically computed NMSE (third column) only slightly, with a maximum deviation of 9.32\% (at \SI{1000}{mm} with mode order 50). It can be seen that both the NMSE and DNMSE decrease with increasing mode orders for all propagation distances, as expected. In contrast to the DNMSE, the summed relative error $\varepsilon_{\sum}^\text{rel}$ shown in column 6 is not propagation distance independent, but increases for any mode order with the propagation distance. However, for any propagation distance, the summed relative error decreases with the increasing mode order. 
For the smaller lateral ranges, the DNMSE (fifth column of \cref{sumofremem}) is propagation distance dependent, as indicated before, but decreases with increasing mode orders for any given propagation distance, just like the NMSE. It generally decreases also with increasing propagation distances for a given mode order, but not strictly monotonously. Similarly, the summed relative error shown in column 7 is also propagation distance dependent and decreases with increasing mode orders for any propagation distance. For any given mode order, it increases as the beam propagates, due to the increasing step sizes $\Delta r$.
%\RemarkM{This modification comes from the results where we use the new error definition.}

%Key findings on errors
Concerning the various error definitions, we find that the DNMSE is not propagation distance independent in typical simulation scenarios because the lateral range is then chosen too small. %DNMSE key finding
The relative error we have introduced here is a useful quantity that allows qualitatively judging the performance of the MEM directly from a graph. It allows, for instance, to directly see in \cref{error_modes_small_range}), that the accuracy of the MEM increases with increasing mode order. The same finding is also found by the DNMSE or NMSE, but only in numbers that cannot be visualized comparably. In cases, where the relative errors cannot be clearly distinguished from the graph, like e.g. in the first row of \cref{error_modes_small_range}, the summed relative error can help quantify physical dependencies (like the performance change with mode order or propagation distance).
%\hl{However, an obvious disadvantage of the summed relative error is that the step size $\Delta r$ increase with increasing propagation distances, when the lateral ranges increase and the number of sampling points remains the same, and because it is not normalized, sometimes its value is quite large.}
%\RemarkM{This modification comes from the results where we use the new error definition.}

In conclusion, we find that all defined types of errors have their individual strengths and weaknesses, such that a comparison of the performance of the MEM with the different error types can be helpful. Concerning the MEM itself, we find that it is not ideally resolving the high spatial oscillations of the diffracted beam in the near field but describes the beam accurately in the far field even if only low mode orders are used.

\subsection{Properties and individual test of the Gaussian Beam Decomposition} \label{GBD_sec}

\subsubsection{GBD: method description} \label{sec:introGBD} 
\begin{figure*}[htbp]
	\begin{center}
		\includegraphics[width=0.49\linewidth]{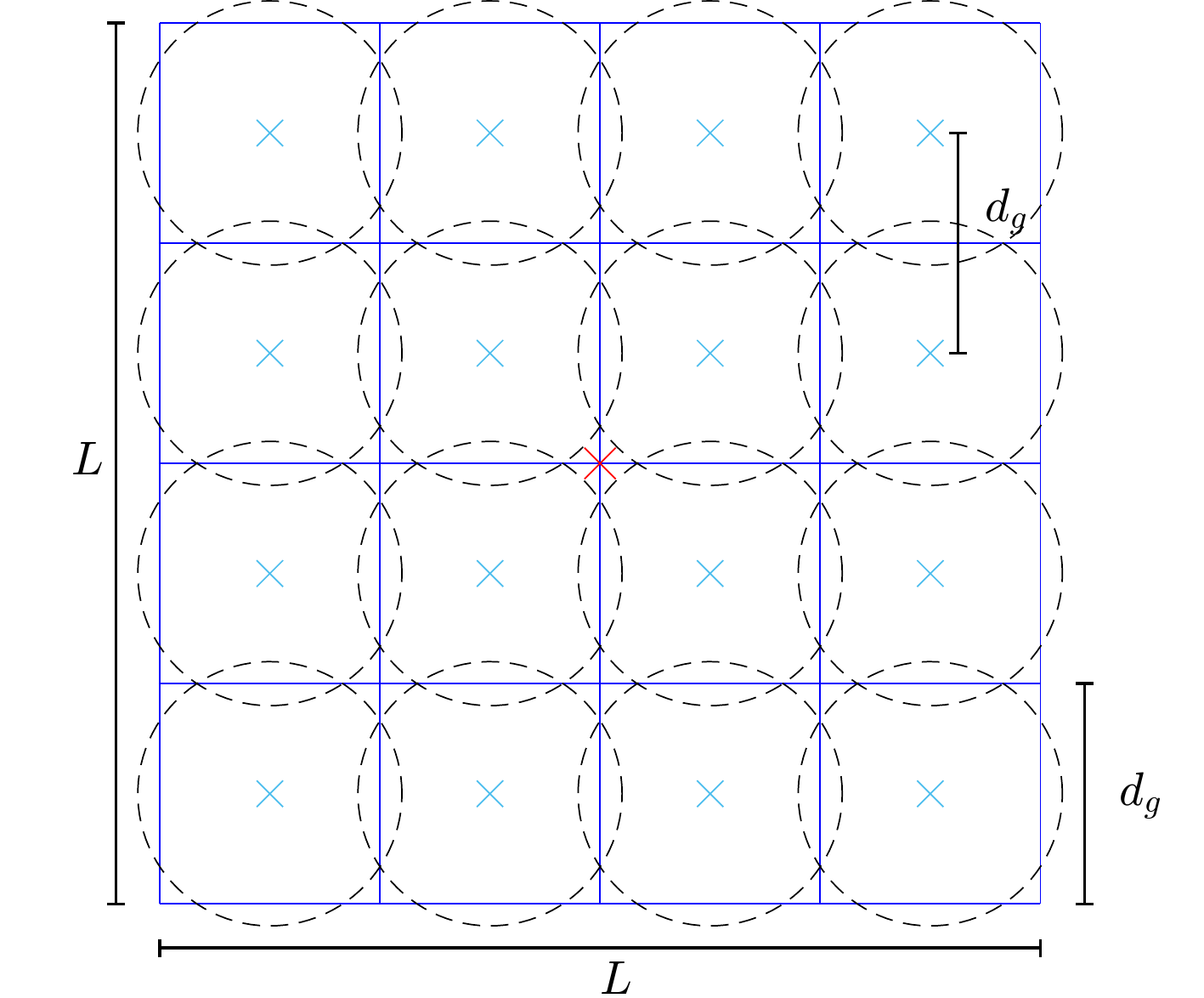}
		\includegraphics[width=0.49\linewidth]{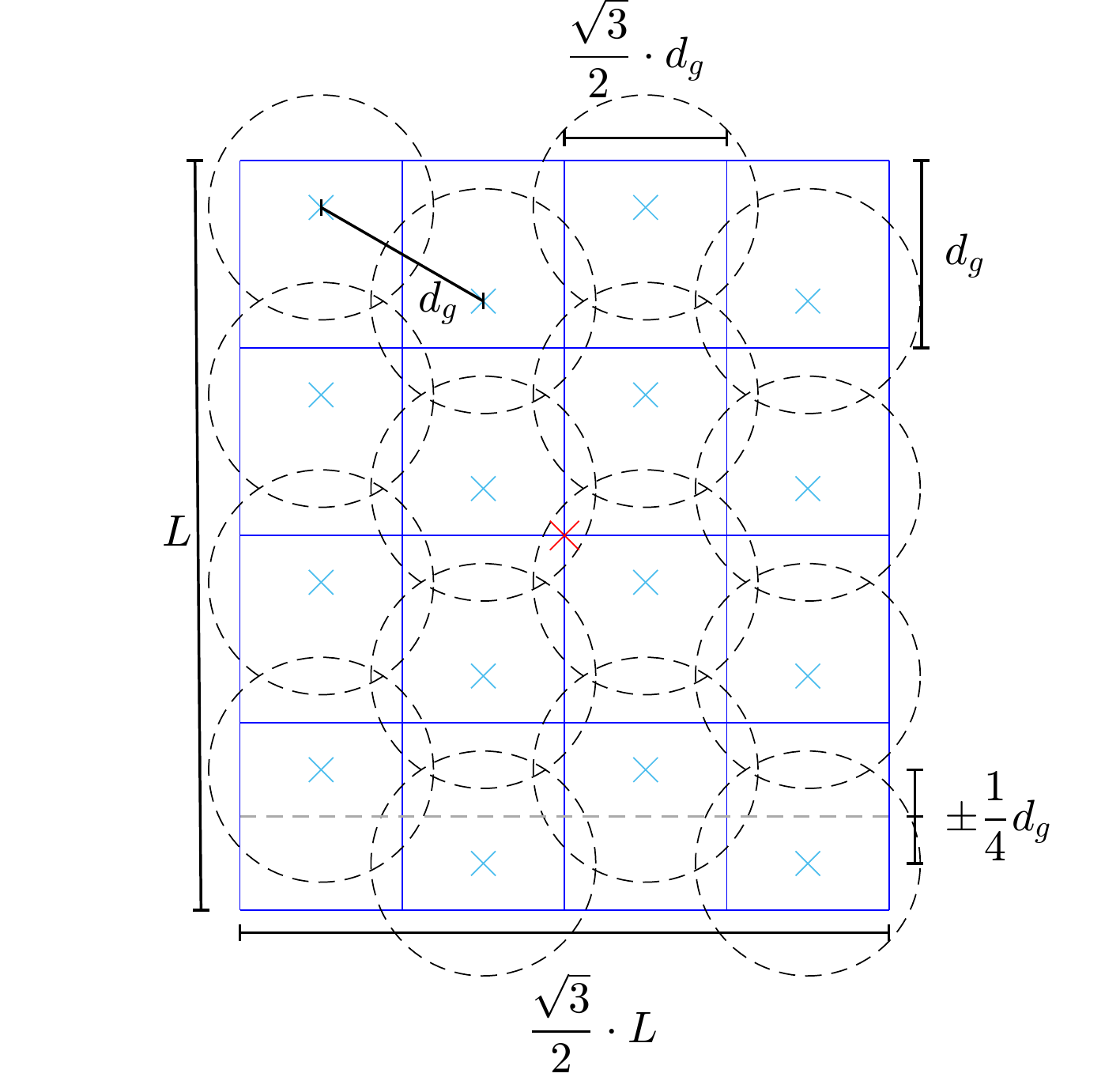}
		\includegraphics[width=0.49\linewidth]{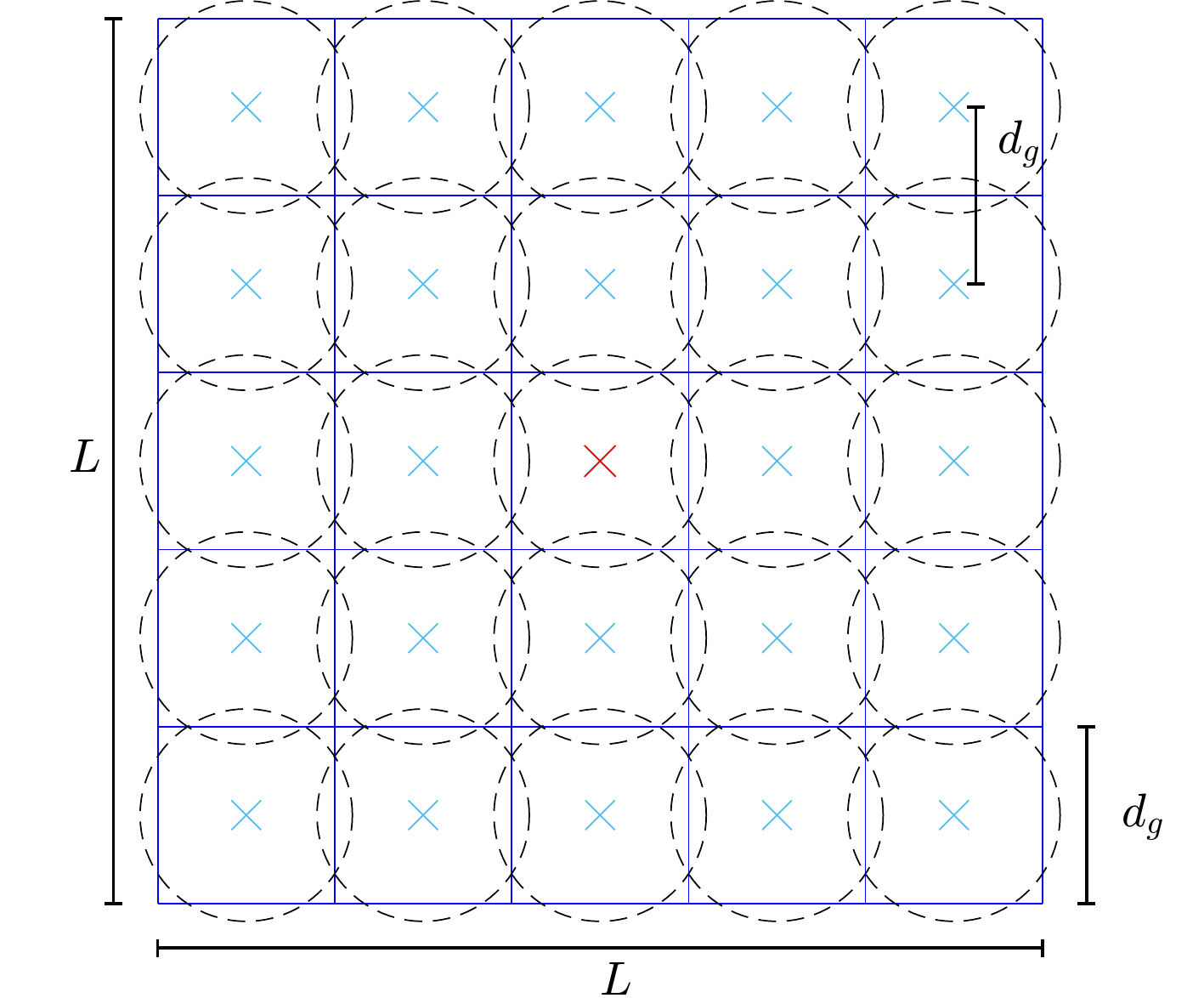}
		\includegraphics[width=0.49\linewidth]{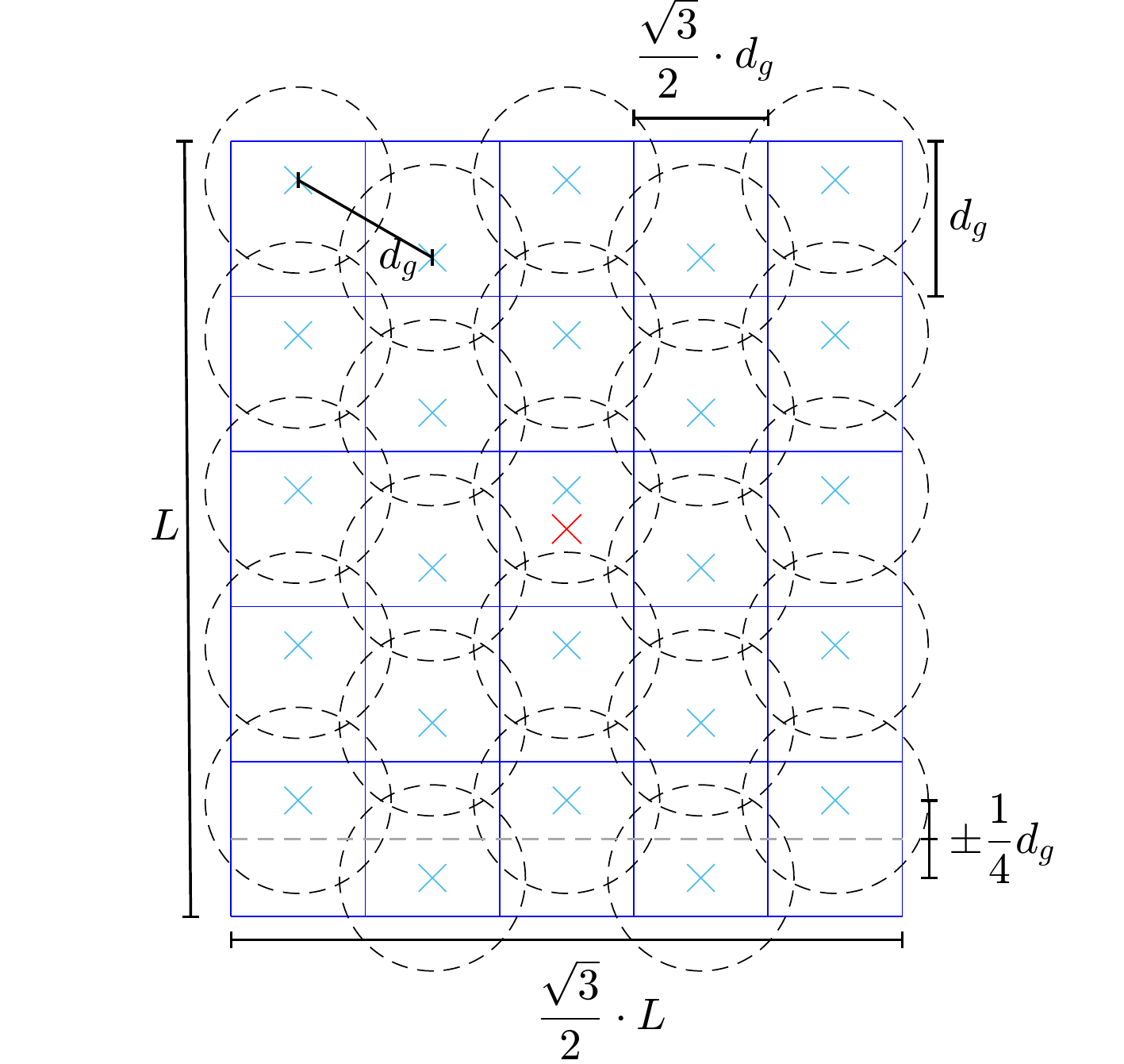}
	\end{center}
	\caption{
		Illustration of the GBD method. Shown on the left is the square grid structure, shown on the right is the hexagonal grid structure,  both for even and odd $g$, respectively. One can see the (virtual) grid in blue, with the center marked in red. The origin points of the individual grid beams are marked in cyan, their waists with radius $w_{0g}$ are shown as the dashed circles. The waist scaling factor $f_{\rm ws}$ was set to \num{1.2} in the diagrams.}
	\label{GBDi}	
\end{figure*}

\begin{figure*}[htbp]
	\begin{center}
		\includegraphics[scale=0.8]{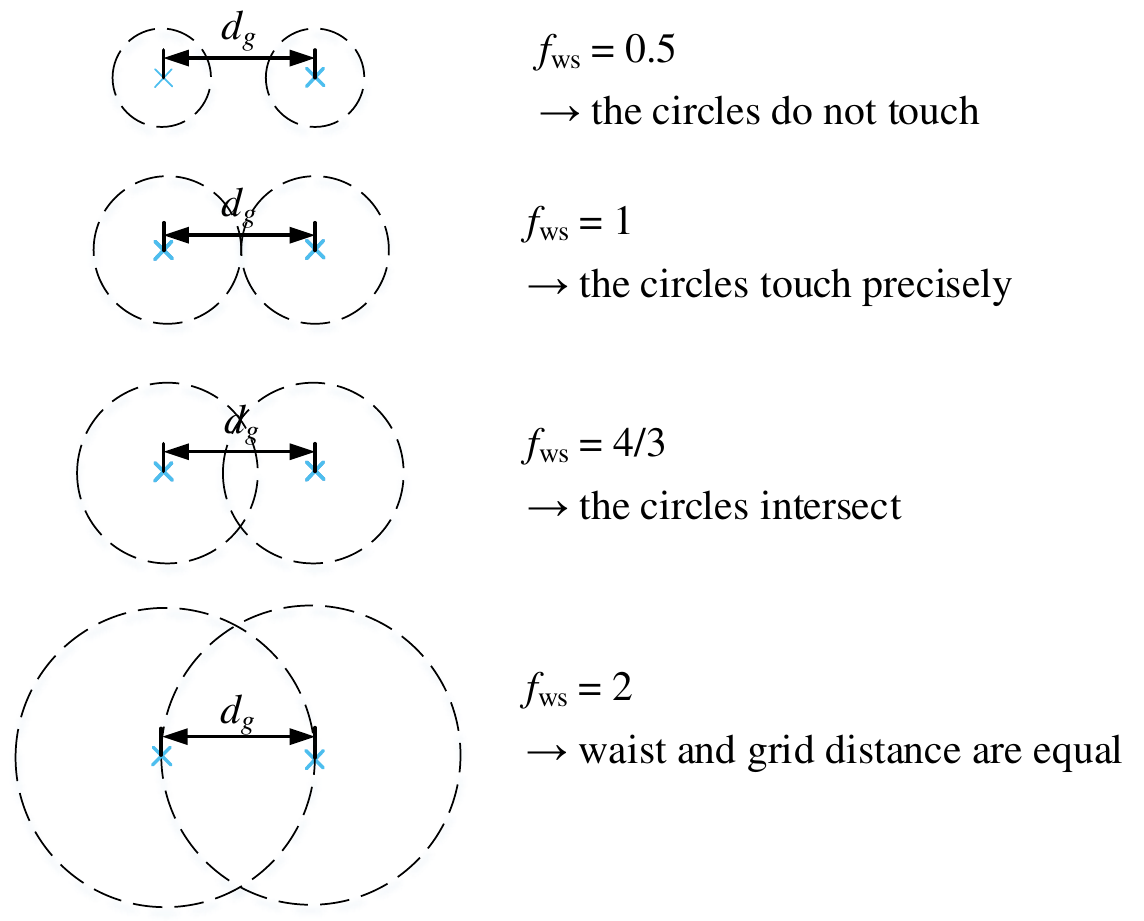}
	\end{center}
	\caption{%
		Illustration of the waist scaling factor $f_{\rm ws}$. If $f_{\rm ws} = \num{0.5}$, the grid beams do not touch; If $f_{\rm ws} = \num{1}$, the grid beams precisely touch; if $f_{\rm ws} = \frac{4}{3}$, the grid beams intersect; if $f_{\rm ws} = \num{2}$, the grid beam waist radius and grid distance are equal.%
	}
	\label{fws_illustration}	
\end{figure*}

Similar to the MEM, the GBD is a wavefront decomposition method. However, it decomposes any wavefront into fundamental Gaussian beams on a grid, as illustrated in \cref{GBDi}. 
There are two supported shapes for the decomposition grid at the moment: square or hexagonal. Both grid shapes are depicted in \cref{GBDi}. The quantities defining the grid are the edge length $L$, called window size, and the number of fundamental Gaussian beams along each dimension $g$. The lattice constant $d_g$, called grid distance, is defined as $d_g = \frac{L}{g}$. The images show the grid of fundamental Gaussian beams, depicted here by dashed circles that denote their waist radius $w_{0g}$. The waist radius is defined as
{%
	\begin{equation}
		w_{0g} = f_{\rm ws} \cdot \frac{d_g}{2} = f_{\rm ws} \cdot \frac{L}{2g}\,,
		\label{fws}
	\end{equation}%
}
where $f_{\rm ws}$ is the so called waist scaling factor. For $f_{\rm ws} = 1$, the waists exacly touch each other. For larger $f_{\rm ws}$, the overlap of the fundamental Gaussian beams increases, for smaller $f_{\rm ws}$, it decreases, as shown in \cref{fws_illustration}.
The number of grid beams along each dimension is denoted by $g$ with $g\times g = G$ being the total number of grid beams placed within the window in \cref{GBDi}. We refer to this total number of grid beams as grid size. The definition of the window size and waist scaling factor shown here is the same as those in the IfoCAD, however, other software could have different definitions.

The hexagonal grid can be directly constructed from the square grid without changing the number of points or underlying math. Therefore, the algorithm to compute the GBD can be left unchanged when switching between grid geometries. To construct the hexagonal grid, columns with an even index are shifted up by $\frac{d_g}{4}$ relative to the square grid position, columns with an odd index are shifted down by the same amount to create a hexagonal point structure. A rescaling by the factor of $\frac{\sqrt{3}}{2}$ along the horizontal direction is required to create an equidistant separation between the nearest neighbours of points, thus forming equilateral triangles. The mapping can be described by the function
\begin{equation}
	\begin{split}
		x_{ij}' &= \frac{\sqrt 3}{2}\left(x_{ij} - x_0\right) + x_0 \\
		y_{ij}' &=
		\left\{
		\begin{array}{ll}
			y_{ij} + \frac{d_g}{4} \text{ if }i\text{ is even,} \\
			y_{ij} - \frac{d_g}{4} \text{ if }i\text{ is odd}
		\end{array}
		\right.
	\end{split}
\end{equation}
where $x_{ij}'$ and $y_{ij}'$ are the coordinates of the $ij$-th grid point calculated from the coordinates of the square grid point $x_{ij}$ and $y_{ij}$ and $x_0$ is the x-coordinate of the grid center. However, this transformation shrinks the window-size in $x$-direction due to the rescaling, resulting in a new window-size of $\frac{\sqrt{3}}{2}L \times L$. The waist size $w_{0g}$ is the same as for the square grid, because the nearest neighbour distance remains $d_g$ in both cases.

Within this paper, we will use mostly the square grid, and only use the hexagonal grid in one particular case in Section \ref{GBDexample}. Therefore, the mathematical description below will focus on the square grid, because the basic theory remains the same for both grids.

The goal of the GBD is, to represent the wavefront as a superposition of the fundamental Gaussian grid beams weighted by coefficients:
\begin{equation}
	E(x, y) \approx \sum_{i=0}^g \sum_{j=0}^g b_{ij} E_{ij}(x - x_{ij,0}, y - y_{ij,0}),
\end{equation}%
where $E$ is the continuous wavefront to be decomposed, $E_{ij}$ are the electric fields of the grid beams with unity intensity, $(x_{ij,0}, y_{ij,0})$ are their origin points and $b_{ij}$ are the complex weighing coefficients. To determine the coefficients, the above equation is evaluated at a discrete set of sampling points $(x_k, y_l)$, where $k$ and $l$ describe the location in the grid: one is the column, the other is the row. The resulting linear equation system is then solved for $b_{ij}$. There must be at least as many sampling points as there are coefficients, which is $G$, one sampling point per grid beam. For better precision, you can also choose more sampling points. But because only the minimal number of required sampling points was used in the paper's simulations, we'll focus our explanations on this case.%

Both pairs of indices are compressed into a single sequential index to be able to write the linear equation system in matrix form. The $E(x_k, y_l)$ and the $b_{ij}$ can be written as column vectors, $\vec{W_{\rm s}}$ and $\vec{b}$ respectively. The $E_{ij}(x_k - x_{ij,0}, y_l - y_{ij,0}) := m_{ijkl}$ can be interpreted as a matrix $M$ with each row corresponding to a sampling point containing the electric fields of each grid beam at this sampling point. These matrix entries describe, how strongly each grid beam influences the value of the superimposed electric field in the sampling point. Therefore, the GBD can be expressed as
\begin{equation}
	\vec{W_{\rm s}} = M\vec{b}, \label{eq:GBDlinear} \,
\end{equation}
which can be solved by well understood methods such as the ${\rm QR}$ decomposition, which is adopted in IfoCAD. The following equations show in detail, how $\vec{W_{\rm s}}$, $\vec{b}$ and $M$ are composed for an equal number of grid beams and sampling points of $G$.
\begin{equation}
	\begin{array}{ll}
		\vec{W_{\rm s}} = 
		\begin{pmatrix}
			E(x_1, y_1)\\
			\vdots \\
			E(x_1, y_g)\\
			\vdots \\
			E(x_k, y_l) \\
			\vdots \\
			E(x_g, y_1) \\
			\vdots \\
			E(x_g, y_g) \\
		\end{pmatrix}, \
		\vec{b} = 
		\begin{pmatrix}
			b_{11}\\
			\vdots \\
			b_{1g}\\
			\vdots \\
			b_{ij}\\
			\vdots \\
			b_{g1} \\
			\vdots \\
			b_{gg} \\
		\end{pmatrix} \,.
	\end{array}
\end{equation}
\begin{equation}
	M =
	\begin{bmatrix}
		m_{1111} & \cdots & m_{ij11} & \cdots & m_{gg11} \\
		\vdots & \cdots & \vdots & \cdots & \vdots \\
		m_{11kl} & \cdots & m_{ijkl} & \cdots & m_{ggkl} \\
		\vdots & \cdots & \vdots & \cdots & \vdots \\
		m_{11gg} & \cdots & m_{ijgg} & \cdots & m_{gggg}
	\end{bmatrix}
\end{equation}
The GBD can be computationally expensive, if large grid sizes are chosen. For example, if $G = 1000\times 1000$, $10^6$ beams will be superimposed, so both $\vec W_s$ and $\vec b$ have $10^6$ entries, which makes $M$ of size $10^{6}\times 10^6$. Another approximation is employed to reduce the complexity of the problem further. Gaussian beam intensities drop off rapidly with increasing distances from the center. At points a few waist sizes apart, their contribution is near zero. Therefore, if the distance between the sampling point and the grid beam origin is larger then $3w_{0g}$, the beam's contribution to the electric field at the sampling point is neglegible. The corresponding element in $M$ can be set to zero. Consequently, $M$ becomes a sparse matrix, and a software implementation making use of this can reduce both memory consumption and computational effort. Nonetheless, the high dimensionality of the equations should be kept in mind, when choosing the grid size of a GBD. The mathematical form of the sparsification can be expressed by
\begin{equation}
	m_{ijkl}=
	\begin{cases}
		m_{ijkl} & \text{if } d_{ijkl} \leq 3w_{0g} \\ 
		0 & \text{else} \,,
	\end{cases}	
\end{equation}
where $d_{ijkl}$ is the distance between grid beam origin $(x_{ij,0}, y_{ij,0})$ and sampling point $(x_k, y_l)$.

\subsubsection{Finite GBDs and their error} \label{GBD_error}
To judge the quality and performance of the GBD to define a comparable set of errors like for the MEM. A NMSE for the GBD equivalent to \cref{eq:ierror_def} could be defined but not analytically evaluated like in the MEM (\cref{eq:ierror}). Whether this error would be propagation distance independent is, therefore, not clear. However, we can still define the discretized NMSE comparable with the MEM error defined in \cref{eq:11a} and \cref{eq:11a1} as follows
\begin{equation}
	\varepsilon^\text{DNMSE}_\circ(N_R,R,z) := \frac{\sum_{i = 0}^{N_R} 2 \pi \left |M\vec{b}(r_i,z,L,w_{0g},f_{{\rm ws}})-\vec{W_{\rm s}} \right |^2r_i \Delta r}{P}\,. 
	\label{eq:sregbd}
\end{equation}
\begin{equation}
	\varepsilon^\text{DNMSE}_\Box(N_X,N_Y,X,Y,z) := \frac{\sum_{i=1}^{N_X}\sum_{j=1}^{N_Y}   \left |M\vec{b}(x_i,y_j,z,L,w_{0g},f_{{\rm ws}})-\vec{W_{\rm s}} \right |^2\Delta x\Delta y}{P}\,. 
	\label{eq:sregbd1}
\end{equation}
Similarly, we define the 2D and 1D version of the relative error, and the summed relative error for the GBD according to \cref{eq:rerrorsk,eq:rerrorsk1d,eq:rerror,eq:src}:
\begin{equation}
	\varepsilon^\text{rel}(x_i,y_i,z):= \frac{\left |M\vec{b}(x_i,y_i,z,L,w_{0g},f_{{\rm ws}})-\vec{W_{\rm s}} \right |}{\left | \vec{W_{\rm s}} \right |}\,,
	\label{eq:rerrorskgbd}
\end{equation}
\begin{equation}
	\varepsilon^\text{rel}(r_i,z)= \frac{\left |M\vec{b}(x_i,0,z,L,w_{0g},f_{{\rm ws}})-\vec{W_{\rm s}}\right |}{\left | \vec{W_{\rm s}} \right |}\,,
	\label{eq:rerrorskgbd1d}
\end{equation}
\begin{align}
	\varepsilon_{\sum}^\text{rel}(N_X,N_Y,X,Y,z):&=\sum_{i=1}^{N_X}\sum_{j=1}^{N_Y} \frac{ \left |M\vec{b}(x_i,y_j,z,L,w_{0g},f_{{\rm ws}}) -\vec{W_{\rm s}}\right |}{\left | \vec{W_{\rm s}} \right |} \Delta x\Delta y
	\nonumber \\
	&= \sum_{i=1}^{N_X}\sum_{j=1}^{N_Y} 	\varepsilon^\text{rel}(x_i,y_j,z)\Delta x\Delta y\,,
	\label{eq:rerrorgbd}
\end{align}
%	\begin{align}
	%		\varepsilon_{\sum}^\text{rel}(N_X,X,z):=\sum_{i=1}^{N_X} \frac{ \left | M\vec{b}(x_i,0,z,L,w_{0g},f_{{\rm ws}}) -\vec{W_{\rm s}} \right | }{\left | \vec{W_{\rm s}} \right |} \Delta x = \sum_{i=1}^{N_R} 	\varepsilon^\text{rel}(r_i,z)\Delta r\,, 
	%		\label{eq:rerrorgbd1d}
	%	\end{align}
\begin{equation}
	\varepsilon_{\sum}^\text{rel}(N_R,R,z):=\sum_{i=1}^{N_R} \frac{ 2\pi \left | M\vec{b}(r_i,z,L,w_{0g},f_{{\rm ws}}) -\vec{W_{\rm s}} \right |r_i\Delta r }{\left | \vec{W_{\rm s}} \right |}  \,, 
	\label{eq:rerrorgbd1dr}
\end{equation}

We have now defined the same type of errors for the GBD as for the MEM and can use these for comparison. However, we do not know any major characteristics of the given errors when applied to a GBD. We, therefore, study and discuss their characteristics on the examples given throughout this paper.

\subsubsection{GBD settings} \label{se:GBD_settings}
For the MEM, it is known that for the stated set of applications, the relation defined in \cref{borghi} can be used to achieve minimal error in the decomposition. For the GBD, we could not find any comparable information. It is, therefore, not clear how the parameters of the GBD should be chosen for minimal error. We can therefore only state fairly general information and the typical settings we choose in our simulations.

For the current implementation of the GBD in IfoCAD,  the parameters which can be chosen explicitly, are the waist scaling factor $f_{{\rm ws}}$, the number $g$ of grid beams along each primary axis of the square grid, and the window size $L$. The waist radius $w_{0g}$ of the grid beams and grid distance $d_g$ are then determined using \cref{fws}. Therefore, there are three parameters that influence the precision of the decomposition, from which the one dimensional number $g$ of grid beams roughly compares with the mode order $N$ of the MEM. One intuitively expects for a fixed windows size: the larger the number of grid beams $g$, the higher the precision, though this is valid only within a certain range. We show this property in \cref{GBDexample} below. Furthermore, we investigate in \cref{faircomparison} how to choose the grid length $g$ and mode order $N$ if both methods are being directly compared. 
We, therefore, understand the number $g$ of grid beams as the primary handle for the precision of the GBD. The remaining two parameters (the waist scaling factor $f_{{\rm ws}}$ and the window size $L$) are secondary handles, and we discuss their settings below. 

Within this paper, we have two different types of examples: either a non-clipped Gaussian beam is being decomposed, or a wavefront that is clipped by an aperture. In the first example, the window size needs to be at least three times larger than the waist size, or else the beam would be clipped by the window during the decomposition, resulting in unintended and unphysical diffraction. On the other hand, the window should not be chosen too large to avoid an unnecessarily high number of grid beams with zero amplitudes and no influence on the final result.
Comparable arguments hold for the second type of examples. Here, the window size needs to be sufficiently larger than the aperture. If the window was chosen to be smaller than the aperture size, the beam would obviously be clipped by the window, not the aperture. If the window size was chosen to equal the aperture size or be only slightly larger than it, the GBD would not be able to resolve the step function in the electric field, resulting in a GBD-beam with a considerable residual electric field amplitude outside the window.  Only if the window is sufficiently oversized (compared to the aperture) the grid beams can resolve the step function in the electric field amplitude that originates from the clipping aperture and thereby accurately decompose the entire wavefront of interest. Within the examples of this paper, the window size is chosen between 1 to 1.5 times the diameter of the aperture. The window size equal to the diameter of the aperture is only used for the examples of non-clipped Gaussian beams (see \cref{sec:GBs-free-space}), as in this case, the diameter of the aperture is already sufficiently large and does not clip the beam.
Please note, the window size is defined as a full width, comparing rather to the diameter of the aperture.

The waist scaling factor should be chosen such that the grid beams have a non-negligible overlap. If the waist scaling factor was chosen too small, the GBD could accurately resolve the incident electric field in the grid points, but due to the lacking overlap of the grid beams, the GBD-beam would effectively have `holes' between the grid points. On the other hand, if the waist scaling factor is chosen very large, a high number of grid beams contribute to the electric field in every sampling point, thereby significantly increasing the computational effort. Within this paper, the range of $f_{\rm ws}$ chosen in all examples is between $3/2$ and $10/3$. 

Unfortunately, we do not know an analytic relation between grid size, window size and waist scaling factor that forms an ideal choice for typical decompositions. The waist scaling factor and window sizes chosen within this paper are also not strictly optimized for the given examples, but simply follow the given logic.

\subsubsection{Example: GBD performance for a clipped Gaussian beam} \label{GBDexample}
In this subsection, we show two examples on the performance of the GBD. In both cases, we decompose a Gaussian, which is clipped by a circular aperture. We assume normal incidence and the Gaussian beam to be optimally centred on the aperture. In the first example, we investigate the performance of the GBD with increasing grid size and compare and test the different error definitions. The second example illustrates the behavior of square and hexagonal grid shapes for the same grid size.

\subparagraph{Example 1: comparing different grid sizes}
For the MEM, it is known from analytic equations that the precision of the decomposition increases monotonously with increasing mode order $N$. This is then also observed in simulations, as long as numerical errors are sufficiently small. 
For the GBD, one might likewise want to assume that for a fixed window size, the precision of the GBD increases with an increasing number of grid beams. However, we showed in \cref{sec:introGBD} and \cref{fws} that the waist $w_{0g}$ of the grid beams scales inversely with the number $g$ of grid beams. This means the more grid beams are used, the smaller the grid beams' waist will get, provided the waist scaling factor is not adapted. Therefore, an increasing number of grid beams can quickly result in a violation of the paraxial approximation. It can, therefore, not be generally expected that an increasing number of grid beams is expected to increase the precision of the decomposition.  \\
In this example, we test the hypothesis of an increase in precision with an increasing number of grid beams. Intentionally, we work with a fixed window size $L$ and a fixed waist scaling factor $f_\text{ws}$ and increase the grid size $G$ up to values that cause the waist sizes to be in the order of the wavelength, thereby violating the paraxial approximation assumption. With this, we test in one simple example how the precision changes with the grid size, and we test slightly beyond settings that would normally be chosen.

The parameter settings of this example are listed in \cref{gbd_example_setting}. 
\begin{table}[tbp]
	\caption{Parameters list of the GBD example.  }
	\footnotesize
	\resizebox{\textwidth}{!}
	{
		\begin{tabular}{@{}lll}
			\hline
			\textbf{parameters} & \textbf{description} & \textbf{value}  \\ 
			\hline
			$\lambda$          & wavelength  & \SI{1064}{\nm} \\
			$P_0$          & beam power  & \SI{1}{W} \\   
			$w_0$         & beam waist  &  \SI{2}{\mm}     \\ 
			$z_0$        & distance from the waist           &   0      \\ 
			$R_a$          & aperture radius       &  \SI{0.5}{\mm}     \\ 
			$G$          &grid size of the GBD    &   $100\times100$, $200\times200$, $500\times500$, $1000\times1000$     \\ 
			$L$          & window size of the GBD             &   \SI{1.5}{\mm}     \\ 
			$f_{\rm ws}$        &waist scaling factor of the GBD             &  1.5      \\ 
			$w_{0g}$        &grid beam waist of the GBD             &  11.2\,$\upmu$m, 5.6\,$\upmu$m, 2.2\,$\upmu$m, 1.1\,$\upmu$m        \\ 
			grid shape         &grid shape of the GBD             &   square     \\ 
			$d$ & propagation distance & \SI{5}{mm}, \SI{20}{mm}, \SI{100}{mm}, \SI{1000}{mm}\\
			$X$ & the number of sampling points & 3001 \\ 
			\hline
			
		\end{tabular}
	}
	\label{gbd_example_setting}
\end{table}
\normalsize
In this example, the beam parameter, aperture size, shape and alignment, the propagation distances and sampling points are all chosen to be the same as in the MEM example in \cref{se:exampleMEM}. The grid sizes are $100\times100$, $200\times200$, $500\times500$ and $1000\times1000$ respectively, using a square grid with a window size of \SI{1.5}{mm} and a waist scaling factor $f_{\rm ws}=1.5$ and the grid beam waist $w_{0g}$ are calculated by \cref{fws}.
As shown in \cref{gbd_example_setting}, the resulting waist sizes are critically small, up to a clear violation of the paraxial approximation in the case of $1000\times1000$ grid beams. 

\begin{figure*}[htbp]
	\begin{center}
		\includegraphics[]{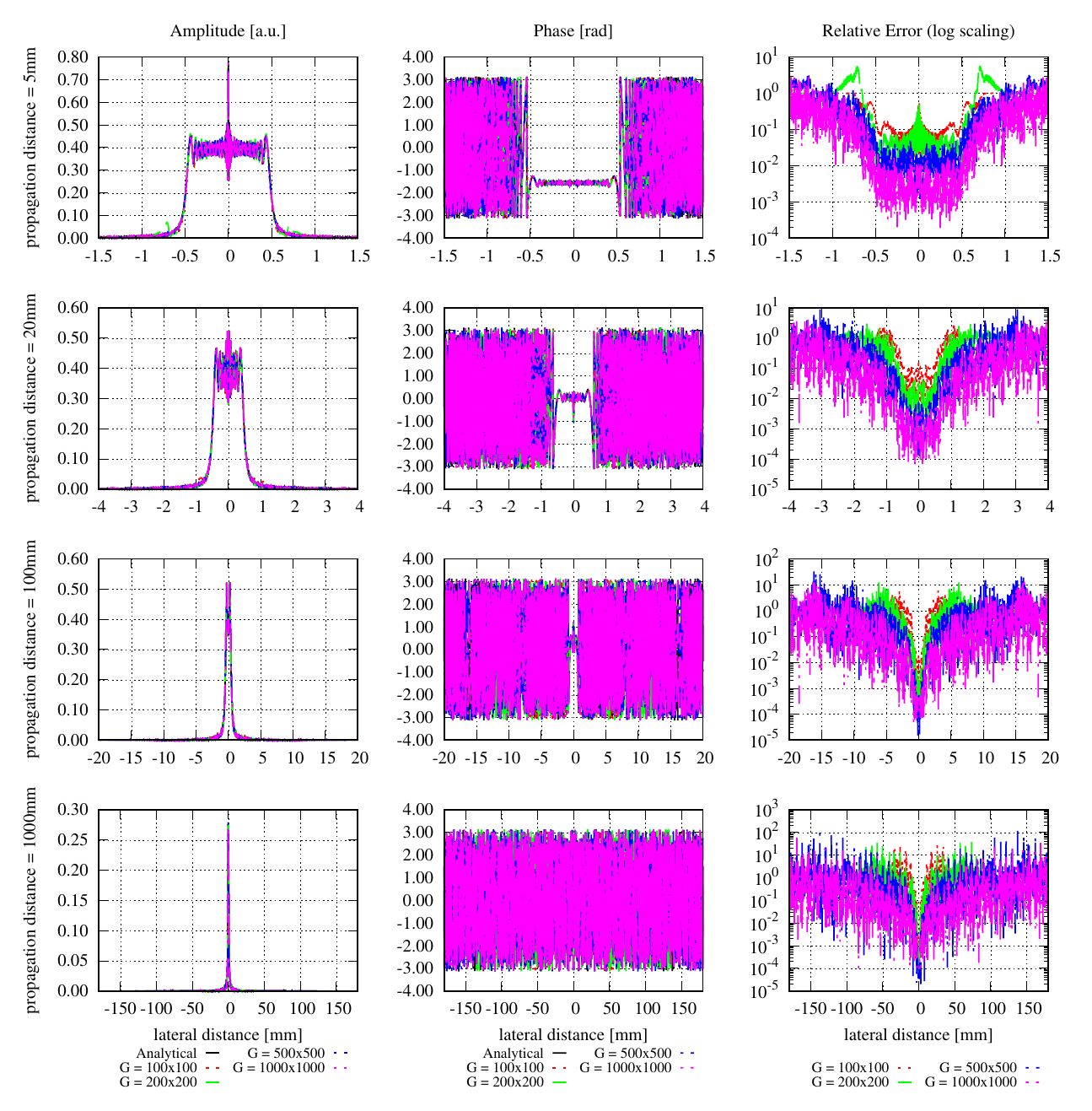}
	\end{center}
	\caption{Performance of a GBD for large lateral ranges with different grid sizes G = $100\times100$,  $200\times200$,  $500\times500$, and  $1000\times1000$ for an incoming circular Gaussian beam with a \SI{2}{mm} waist being clipped by a \SI{0.5}{mm} radius circular aperture. Shown are the amplitude (absolute value), phase and relative error $\varepsilon^\text{rel}(N_R,R,z)$ at different propagation distances $z=$ \SI{5}{mm}, \SI{20}{mm}, \SI{100}{mm} (near field) and \SI{1000}{mm} (far field) after the clipping aperture. The analytical methods for the near and far field are Campbell \cite{1987Fresnel} and Tanaka et al.\ \cite{1985Field}, respectively. Lateral distances for each propagation distance are chosen large enough to cover all the power. For any propagation distance, the performance of the GBD becomes better with the increasing grid size.}
	\label{grid_error}
\end{figure*}

\begin{figure*}[htbp]
	\begin{center}
		\includegraphics[]{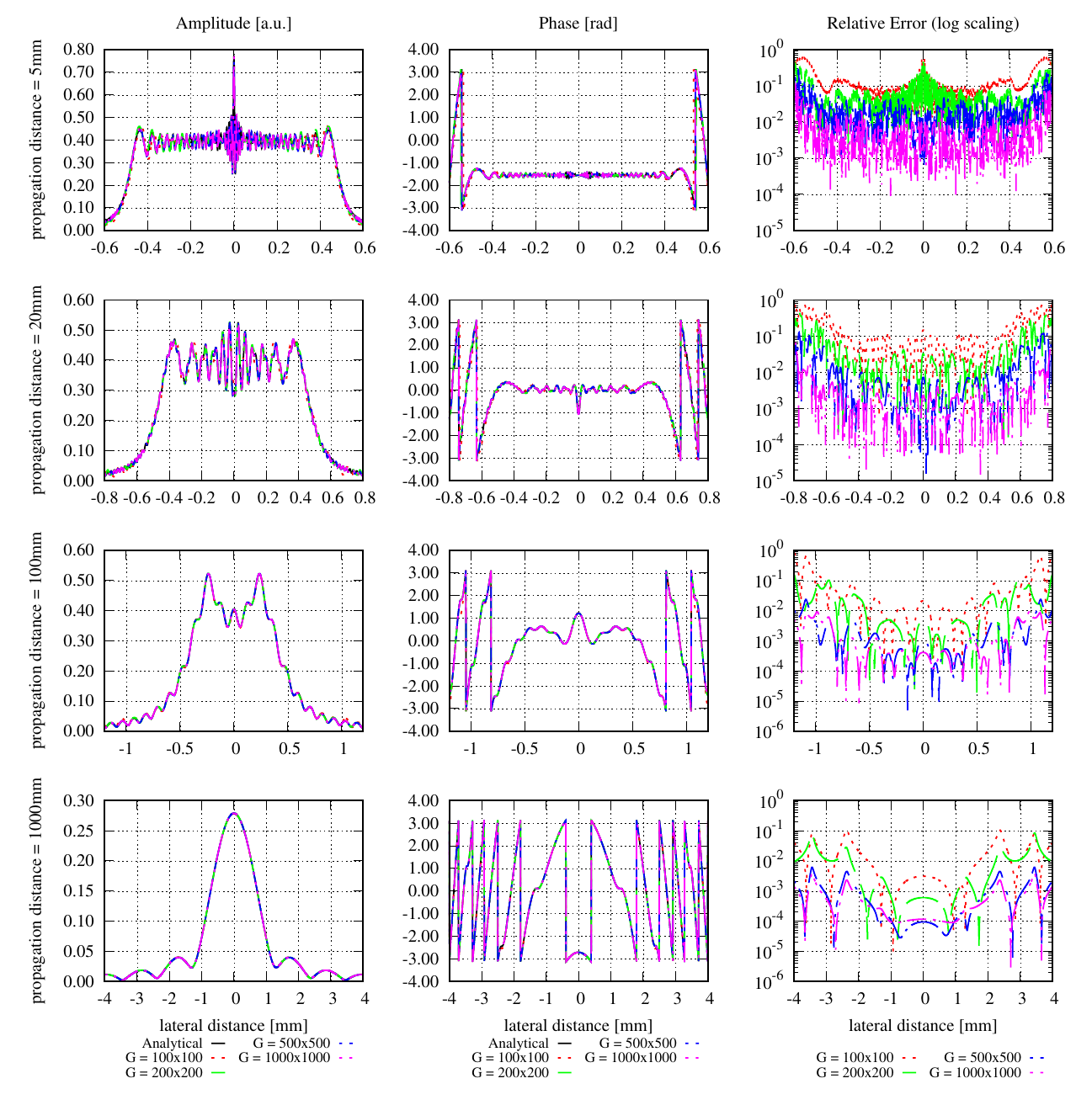}
	\end{center}
	\caption{Same as \cref{grid_error}, but for smaller lateral ranges. Shown is the performance of a GBD with different grid sizes for an incoming circular Gaussian beam with a \SI{2}{mm} waist being clipped by a \SI{0.5}{mm} radius circular aperture. Shown are the amplitude (absolute value), phase and relative error $\varepsilon^\text{rel}(N_R,R,z)$ at different propagation distances $z=$ \SI{5}{mm}, \SI{20}{mm}, \SI{100}{mm} (near field) and \SI{1000}{mm} (far field) after the clipping aperture. The analytical methods for the near and far field are Campbell \cite{1987Fresnel} and Tanaka et al.\ \cite{1985Field}, respectively. One sees that the further the beam propagates, or the higher the grid size is, the better is the performance of the GBD. }
	\label{grid_error_small_range}
\end{figure*}

The amplitude, phase, and relative error are plotted for a large and small lateral range in \cref{grid_error} and \cref{grid_error_small_range}, respectively. The corresponding errors are summarized in  \cref{sumofregbd}.
Since the electric field of interest is the very same as in the MEM example, we use the very same lateral ranges as in \cref{se:exampleMEM}.
\begin{table}[htbp]
	\caption{The GBD errors, including the discretized NMSE and the summed relative error, defined in \cref{eq:sregbd} and \cref{eq:rerrorgbd1dr} respectively, are calculated for increasing grid sizes at different propagation distances. The discretized NMSE for both lateral ranges are propagation distance dependent. The summed relative errors for smaller lateral ranges decrease with increasing grid size at any propagation distance. The number of sampling points $X$ is 3001. }
	\footnotesize 
	\resizebox{\textwidth}{!}
	{
		\begin{tabular}{@{}llllll}
			\hline
			\textbf{\makecell[c]{propagation distance \\(mm)}} & \textbf{grid size}  & \textbf{\makecell[c]{DNMSE \\(for larger ranges)\\$	\varepsilon^\text{DNMSE}_\circ(N_R,R,z)$}} &\textbf{\makecell[c]{DNMSE \\(for smaller ranges)\\$\varepsilon^\text{DNMSE}_\circ(N_R,R,z)$}}  & \textbf{\makecell[c]{the summed relative error \\(for larger ranges)\\$\varepsilon_{\sum}^\text{rel}(N_R,R,z)$}} & \textbf{\makecell[c]{the summed relative error \\(for smaller ranges)\\$\varepsilon_{\sum}^\text{rel}(N_R,R,z)$}} \\ 
			\hline
			\multirow{4}{*}{5}&	$100\times100$ & 0.0158 & 0.0134&11.7783&0.4706\\
			&	$200\times200$ & 0.0196& 2.5690$\times10^{-3}$&15.6724& 0.1893\\  
			&	$500\times500$ & 1.9541$\times10^{-3}$ &5.8425$\times10^{-4}$&12.0022&0.0886\\  
			&	$1000\times1000$ & 3.6873$\times10^{-4}$ &5.8403$\times10^{-5}$&5.7221	&  0.0304\\ \hline
			\multirow{4}{*}{20}&	$100\times100$ & 0.0126 & 6.7781$\times10^{-3}$& 96.2485&0.9669\\ 
			&	$200\times200$ &5.3759$\times10^{-3}$ &8.5460$\times10^{-4}$&97.6313
			&0.3810\\ 
			&	$500\times500$& 1.1952$\times10^{-4}$& 6.9181$\times10^{-5}$&81.7791&0.1112\\ 
			&$1000\times1000$ &1.9347$\times10^{-5}$& 5.9805$\times10^{-6}$& 40.5455 &0.0253\\ \hline
			\multirow{4}{*}{100}&	$100\times100$ &9.7793$\times10^{-3}$ &8.3938$\times10^{-4}$&2488.71&1.1545\\ 
			&	$200\times200$ & 5.2608$\times10^{-3}$&9.5334$\times10^{-5}$&2589.92&0.2732\\ 
			&		$500\times500$ & 1.2207$\times10^{-3}$ & 1.5016$\times10^{-6}$&2890.48&0.0396 \\ 
			&$1000\times1000$ & 1.9361$\times10^{-4}$& 6.8164$\times10^{-7}$& 1449.3	& 0.0251\\ \hline
			\multirow{4}{*}{1000}&	$100\times100$ & 9.8745$\times10^{-3}$ & 4.6425$\times10^{-5}$ &208054&2.3840\\ 
			&	$200\times200$ & 5.1443$\times10^{-3}$ &8.2525$\times10^{-6}$ &231535&1.5544\\ 
			&		$500\times500$ & 1.1037$\times10^{-3}$&1.4047$\times10^{-7}$ &332595&0.1414\\ 
			&$1000\times1000$& 1.4711$\times10^{-4}$ &7.5511$\times10^{-8}$&137490&0.0780
			\\ 
			\hline
		\end{tabular}
	}
	
	\label{sumofregbd}	
\end{table}
\normalsize

% general description on the graphs and the table
Like in the MEM example shown in \cref{error_modes}, the GBD results with the large lateral ranges shown in \cref{grid_error} are not very descriptive because the lateral ranges are simply too large to judge the quality of the decomposition. However, \cref{grid_error_small_range} shows that the GBD describes the beam well, particularly in the far field, even if only $100\times100$ grid beams are used. 
Additionally, it can be seen that the further the beam propagates, or the higher the grid size, the better the performance of the GBD. This can also be seen from both the DNMSE (in column 4) and the summed relative error (in column 6), as they decrease with increasing grid sizes, although it is not strictly monotonous for increasing propagation distances with a given grid size in column 6. This is particularly interesting here, given that the high grid sizes imply unadvisably small waist sizes were used. The observed increasing precision of the GBD with these high grid sizes was, therefore, not naturally given. Additionally, the summed relative error in column 6 also changes nonmonotonically as the diffracted beam propagates.

% discussion of errors in detail, first for larger lateral, then for smaller ranges
From column 3 in \cref{sumofregbd}, it can be seen that the DNMSE of GBD is propagation distance dependent not only for small lateral ranges but also in the case of large lateral ranges. This is unlike in the case of the MEM. 
Additionally, it can be seen that the DNMSE decrease with increasing grid sizes for any given propagation distance. This holds again for both choices of lateral ranges.  However, it is not consistently given that for any choice of grid size, there is a strictly monotonous decrease of the DNMSE with increasing propagation distance. This is again different from the behavior of the MEM. However, it is currently unclear whether this originates from the method itself or its implementation.
Likewise, also the summed relative error $\varepsilon_{\sum}^\text{rel}$ does not show a strictly monotonous decrease with increasing grid size.  For example, the summed relative error in column 5 at a propagation distance of \SI{5}{mm} increases in the step from a grid size of $100\times100$ to $200\times200$, and decreases for any further increase of the grid size. 
Finally, the summed relative error in column 5 is increasing with the propagation distance, unlike the DNMSE, which was mostly decreasing with the propagation distance. The reason for these observations could again originate from the method itself or be numerical precision or implementation problems, or the nearly zero-valued denominator in the error computation.
Despite the non-monotonous behavior of the DNMSE and the summed relative error, we still use both for the total performance evaluation of the GBD in order to allow a direct comparison with the MEM.

\begin{figure*}[htbp]
	\begin{center}
		\includegraphics[]{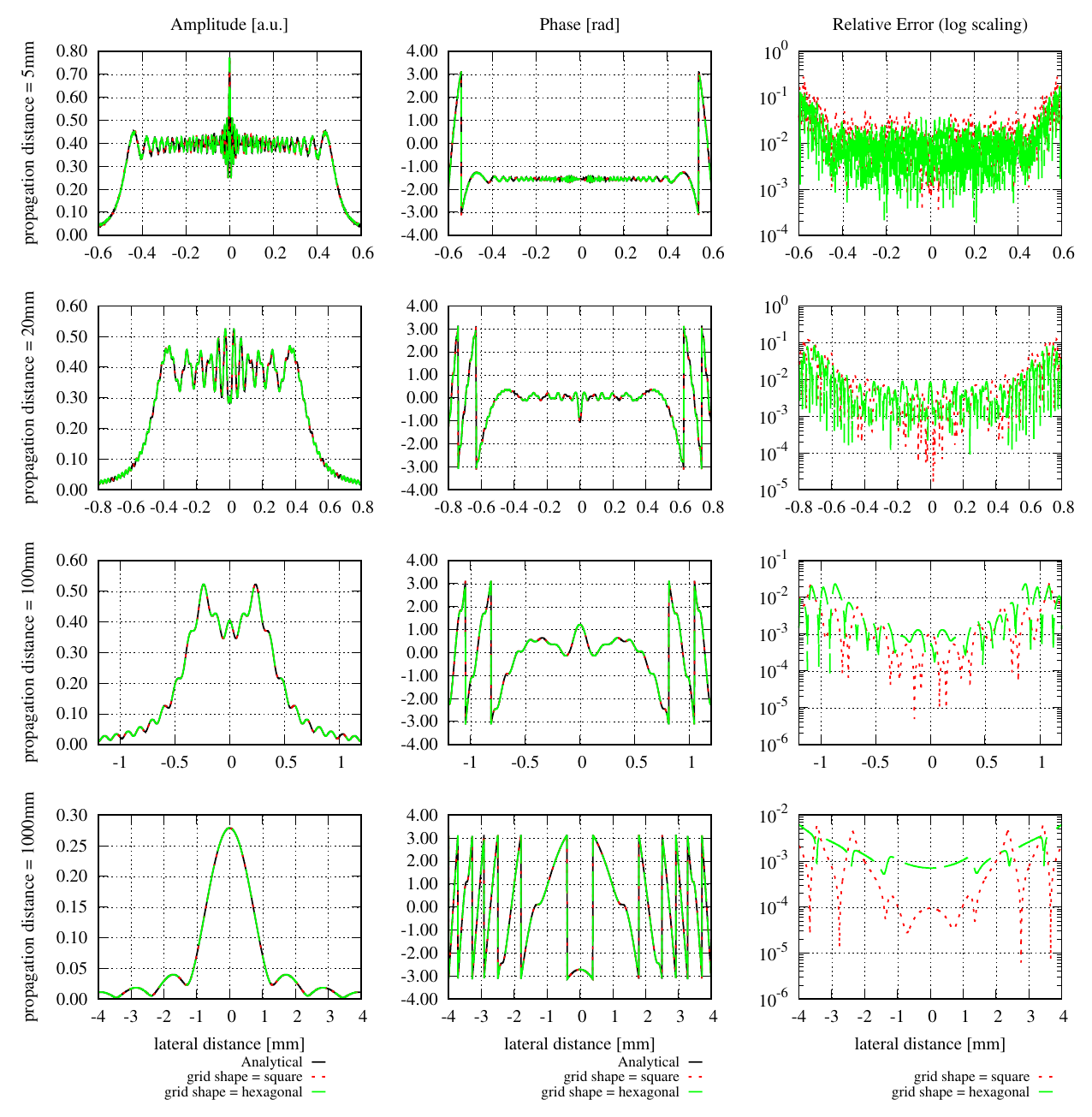}
	\end{center}
	\caption{Performace of a GBD with hexagonal and square grid shapes. Shown are the amplitude (absolute value), phase, and relative error distributions at different propagation distances from a circular aperture with \SI{0.5}{mm} radius. The incoming circular-symmetric Gaussian beam was centered to the aperture and had its \SI{2}{mm} waist located in the aperture plane. The analytical methods for the near and far field are Campbell \cite{1987Fresnel} and Tanaka et al.\ \cite{1985Field}, respectively. The number of sampling points $X$ is 3001.}
	\label{grid_shape}
\end{figure*}

\subparagraph{Example 2: comparing grid shapes}
The shape of the grid can affect the accuracy of the GBD. We, therefore, repeat the previous example with the very same settings and a grid size of $500\times500$ beams, but compare this time the performance of the GBD with a square and a hexagonal grid. As introduced in \cref{sec:introGBD}, the window size in the horizontal direction is rescaled by a factor of $\sqrt{3}/2$ for the hexagonal grid, which is \SI{1.5}{mm}$\times\sqrt{3}/2$ in this example. We use the same waist scaling factor $f_{\rm ws} = 1.5$ and the waist radius $w_{0g}$ for both square and hexagonal grid in IfoCAD. The resulting amplitude, phase and relative error of the square and hexagonal are plotted in \cref{grid_shape}. 
The corresponding DNMSE and the summed relative error are listed in \cref{sumofreshape}.
\begin{table}[htbp]
	\caption{The DNMSE and the summed relative error for different grid shapes at different propagation distances given grid size $500\times500$. It can be seen that the hexagonal grid caused smaller errors for short propagation distances of \SI{5}{mm} and \SI{20}{mm}, while the square grid generated more precise results at higher propagation distances. The number of sampling points $X$ is 3001.}
	\footnotesize
	%\centering
	%\resizebox{\textwidth}{!}
	{
		\begin{tabular}{@{}llll}
			\hline
			\textbf{\makecell[c]{propagation distance \\ (mm)}} & \textbf{grid shape} &
			\textbf{\makecell[c]{DNMSE \\$	\varepsilon^\text{DNMSE}_\circ(N_R,R,z)$}} &
			\textbf{\makecell[c]{the summed relative error \\$\varepsilon_{\sum}^\text{rel}(N_R,R,z)$}} \\ 
			\hline
			\multirow{2}{*}{5}&	square & 5.8425$\times10^{-4}$ &0.0886\\ 
			&	hexagonal &	2.4141$\times10^{-4}$ & 0.0603 \\ \hline
			\multirow{2}{*}{20}&	square & 6.9181$\times10^{-5}$ &0.1112\\ 
			&	hexagonal & 3.1151$\times10^{-5}$ &	0.0641 \\ \hline
			\multirow{2}{*}{100}&	square & 1.5016$\times10^{-6}$ &0.0396\\ 
			&	hexagonal & 3.8222$\times10^{-6}$&0.0583\\ \hline
			\multirow{2}{*}{1000}&	square & 1.4047$\times10^{-7}$ &0.1414	\\ 
			&	hexagonal& 2.4972$\times10^{-7}$ &0.1672\\ 
			\hline
		\end{tabular}
	}
	\label{sumofreshape}
	
\end{table}
\normalsize
%

%discussion
Both \cref{grid_shape} and \cref{sumofreshape} show that at propagation distances i.e. \SI{5}{mm} and \SI{20}{mm}, the simulations performed with a hexagonal grid gradually show slightly better results than the simulations using a square grid. However, at larger distances of \SI{100}{mm} or \SI{1000}{mm}, the square grid resulted in higher accuracy. However, this is only one example, and we cannot draw generalized conclusion from it.

\section{Fair Comparison} \label{se:3}
\label{faircomparison}
% why we discuss what in this section
So far, the MEM and GBD have been introduced, their settings discussed, and their performance was individually tested on an example. In the next step, we want to directly compare the two methods, for which we need to define criteria to evaluate which method performed better or whether they performed equally well. Particularly, we need to define which mode order of the MEM should be compared with what grid size of the GBD and why this is chosen. This is discussed in the following.

\subparagraph{Criteria for a fair comparison}
% definition of fair comparison
There are two aspects which should be considered in the comparison of the MEM and GBD: accuracy and computational efforts. We judge the accuracy by the introduced errors, and the computational effort by the run time of the simulation. We consider a fair comparison a case, where either both methods achieve the same accuracy, and then the run time is used to judge the performance, or, both methods are set to have approximately the same run time, and the performance is judged by the achieved accuracy. Within this study, we use the later criterion and choose as run time the elapsed real time (not the CPU time). 

% factors that impact the computational effort
From the basic principles of the two methods,the computational effort comprises three parts: the decomposition time, propagation time, and the superposition time. For the MEM, the decomposition time is the time spent on the integration to calculate the coefficients of the higher-order modes. For the GBD, it is the time span needed for the QR decomposition solving \cref{eq:GBDlinear}. Therefore, the decomposition time depends on the mode order and grid size, respectively, but also on the properties of the input field. 
The propagation time is very short in comparison because ray tracing methods, including the propagation of the Gaussian beam parameters with the ABCD matrix formalism, are highly efficient and computationally low demanding. For the MEM, all modes even share the same axis, such that only one ray needs to be traced, which then represents the beam axis of all modes. 
Finally, the superposition time depends on the number of sampling points in the target plane, as well as the number of modes or grid beams, respectively. The computational effort is therefore dominated by the decomposition time and the superposition time, and is naturally affected by other criteria such as the efficiency of the original implementation of the methods into the used software tool, the computational power of the used computer, and possibly even the used operation system.

In this study, we compare the performance of the MEM with GBD using IfoCAD (Version of 2022/10, git commit adf19a5b) to find mode orders and grid sizes which result in similar computational effort. All simulations shown here for testing the computational effort were performed on a MacBook Pro 2020 with \SI{8}{GB} RAM and a 2.3 GHz processor with 8 cores.

\subparagraph{Computational effort of the MEM and GBD}
%settings for finding the similar computational effort
In order to find settings that result in comparable computational effort for both methods, we performed a dedicated simulation where we varied the mode order and grid size for a fixed number of sampling points in the target plane. For this simulation, we chose again the case of a clipped Gaussian beam, with the same settings as in the previous examples:
a circular Gaussian beam with waist \SI{2}{mm} located in the aperture center incident onto this circular aperture with radius of \SI{0.5}{mm}. We set $N$ ranging from 10 to 100 with a step of 10 for the MEM and set $g$ ranging from 100 to 1000 with a step of 100. The target plane was \SI{5}{mm} away from the aperture and we used 3001 sampling points to compute the electric field for $x \in {-3,3}, y=0$. Here, the MEM and GBD were run with parallelization. We, therefore, distinguish two different times: the elapsed real time, which is the time the user needs to wait for a result, and the CPU time, which is the actual computation time and larger than the elapsed real time due to the used parallelization. We focus here on the elapsed real time and use this as the primary criterion. The computational effort of the MEM and GBD are summarized in \cref{mem_computational_effort} and \cref{gbd_computational_effort} contain therefore information on the elapsed time during the decomposition, propagation, superposition. The total time is given both as elapsed real-time (column 5) and CPU time (column 6).

\begin{table}[htbp]
	\caption{Computational effort of the MEM with increasing mode order. Shown are the decomposition, propagation, superposition, and the resulting total time. The number of sampling points in the target plane is 3001. All times are elapsed real-time, except for the last column, which shows total CPU time.}
	\label{mem_computational_effort}
	\footnotesize
	\resizebox{\textwidth}{!}
	{
		\begin{tabular}{@{}llllll}
			\hline
			\textbf{mode order} & \textbf{decomposition time} & \textbf{propagation time} &  \textbf{superposition time} & \textbf{total time} & \textbf{total CPU time}\\ 
			&  \textbf{ (ms)} &  \textbf{ (ms)} &  \textbf{(ms)} & \textbf{ (ms)} & \textbf{ (ms)}\\ 
			\hline
			10&567.139&0.002&8.262&575.403&2203.02\\
			20&2362.44&0.002&32.63&2395.07&13045.8\\
			30&4588.29&0.002&82.949&4671.24&29407.9\\
			40&7533.84&0.002&221.177&7755.02&52262.9\\
			50&12262&0.002&347.936&12609.9&86188.6\\
			60&17014.9&0.002&537.851&17552.8&119239\\
			70&25647.9&0.002&847.281&26495.1&188318\\
			80&33581.8&0.002&1180.16&34762&249430\\
			90&41533.5&0.002&1567.19&43100.7&309376\\
			100&55278.4&0.001&2069.17&57347.6&411494\\
			\hline
		\end{tabular}
	}
	
\end{table}
\normalsize
\begin{table}[htbp]
	\caption{Computational effort of the GBD with increasing grid size. Shown are the decomposition, propagation, superposition, and the resulting total time. The number of sampling points in the target plane is 3001. All times are elapsed real-time, except for the last column, which shows total CPU time.}
	\footnotesize
	\resizebox{\textwidth}{!}
	{
		\begin{tabular}{@{}llllll}
			\hline
			\textbf{grid size} & \textbf{decomposition time} & \textbf{propagation time} &  \textbf{superposition time} & \textbf{total time } & \textbf{total CPU time}\\ 
			&  \textbf{ (ms)} &  \textbf{ (ms)} &  \textbf{(ms)} & \textbf{ (ms)} & \textbf{ (ms)}\\ 
			\hline
			$100\times100$&124.121&1.818&383.069&509.008&1007.79 \\
			$200\times200$&486.003&8.089&2109.11&2603.2&10031.1  \\
			$300\times300$&1063.21&18.471&5592.88&6674.56&31341.2\\
			$400\times400$&1925.34&34.633&11483.6&13443.6&64804.6\\
			$500\times500$&3191.69&57.293&19316.7&22565.7&117748\\
			$600\times600$&4648.64&73.8&29109.8&33832.2&188760\\
			$700\times700$&6519.14&102.54&40750.7&47372.4&274034\\
			$800\times800$&8768.65&133.035&54117.1&63018.8&363051\\
			$900\times900$&11564.9&177.839&67911.2&79654&458815\\
			$1000\times1000$&14249.6&219.835&82782.7&97252.1&567600\\
			\hline
		\end{tabular}
	}
	\label{gbd_computational_effort}
\end{table}
\normalsize

For the MEM we see in \cref{mem_computational_effort}, that the decomposition time is by far the dominant time consumer in the given example, making out more than 95\% of the total elapsed real-time. For the GBD, this is different: \cref{gbd_computational_effort} shows that the superposition time is dominant over the decomposition time by a factor of more than 3. 
The propagation time is, as expected, insignificant. For the GBD with high grid sizes, some computational effort accumulates due to the large number of grid beams that need to be traced. For instance, for a grid size of $1000 \times 1000$, 1 million grid beams need to be propagated.
Finally, the different superposition times are noteworthy. We understand this as a consequence of the different numbers of beams that need to be computed and evaluated at every target grid point. For instance, a mode order of 100 implies that $51\times 52/2=1326$ modes are used in the MEM (cf.~\cref{eq:no_of_modes_in_MEM}).
In the case of the GBD, a grid size of $500\times500$ grid beams implies that in every sampling point of the target plane, 250,000 beams are being tested, whether they contribute to the electric field. Even though only a few electric fields are indeed being superimposed in the end, the test itself for the high number of grid beams costs considerable time in the current implementation.

If we compare now the total elapsed real time for different settings of the MEM and GBD, we see several pairings of grid size $G$ and mode order $N$, that can be used to achieve comparable computational effort. For instance, $\{N=10, G= 100\times100\},\{N=20, G= 200\times200\}, \{N=50, G= 400\times400\}$, and we chose  $\{N=50, G= 400\times400\}$ for all comparisons using 3001 sampling points within this paper. 

However, this choice depends on the number of sampling points used in the target plane since the computational effort of the GBD is dominated by the superposition time (which significantly depends on the number of sampling points in this plane), while the MEM is not. Therefore, the shown comparison should be repeated if a different number of sampling points is used. Within this paper, we use always the shown 3001 sampling points for all 2-dimensional cross sections of the electric field. However, we also show figures for the full cross-section ($x$ and $y$ for a fixed propagation distance $z$) and use $101\times101$ sampling points in these cases. Therefore, we repeat the above simulation for $x \in {-3,3}, y= \in {-3,3}$ with 101 points each axis, i.e. 10201 sampling points total. The elapsed real time and the CPU time, are summarized in \cref{mem_computational_effort_10201} and \cref{gbd_computational_effort_10201}. 
\begin{table}[htbp]
	\caption{Computational effort of the MEM with increasing mode order. Shown are the decomposition, propagation, superposition, and the resulting total time. The number of sampling points in the target plane is $101\times101$. All stated times are again elapsed real-time, except for the last column.}
	\footnotesize
	\resizebox{\textwidth}{!}
	{
		\begin{tabular}{@{}llllll}
			\hline
			\textbf{mode order} & \textbf{decomposition time} & \textbf{propagation time} &  \textbf{superposition time} & \textbf{total time} & \textbf{total CPU time}\\ 
			&  \textbf{ (ms)} &  \textbf{ (ms)} &  \textbf{(ms)} & \textbf{ (ms)} & \textbf{ (ms)}\\ 
			\hline
			10&589.91&0.002&28.607&618.519&2301.31\\
			20&2203.86&0.002&110.936&2314.8&13607.3\\
			30&4393.64&0.002&279.216&4672.86&30801.9\\
			40&7786.71&0.000999999&755.448&8542.15&54665.3\\
			50&12722.8&0.002&1176.24&13899.1&84440.1\\
			60&18766.7&0.002&1872.93&20639.6&126174\\
			70&23967.4&0.000999999&2770.08&26737.5&172244\\
			80&32632.7&0.000999999&3811.03&36443.7&230824\\
			90&43512&0.003&5176.18&48688.1&325858\\
			100&59277&0.002&6858.93&66135.9&446890\\
			\hline
		\end{tabular}
	}
	\label{mem_computational_effort_10201}
\end{table}
\normalsize
\begin{table}[htbp]
	\caption{Computational effort of the GBD with increasing grid size. Shown are the decomposition, propagation, superposition, and the resulting total time. The number of sampling points in the target plane is $101\times101$. All stated times are again elapsed real-time, except for the last column.}
	\footnotesize
	\resizebox{\textwidth}{!}
	{
		\begin{tabular}{@{}llllll}
			\hline
			\textbf{grid size} & \textbf{decomposition time} & \textbf{propagation time} &  \textbf{superposition time} & \textbf{total time } & \textbf{total CPU time}\\ 
			&  \textbf{ (ms)} &  \textbf{ (ms)} &  \textbf{(ms)} & \textbf{ (ms)} & \textbf{ (ms)}\\ 
			\hline
			$100\times100$&132.024&2.11&1229.4&1363.53&2610.24\\
			$200\times200$&477.872&8.187&6544.58&7030.64&28128\\
			$300\times300$&1039.3&18.413&17145.1&18202.8&85500.9\\
			$400\times400$&1837.53&32.663&34857.4&36727.6&183316\\
			$500\times500$&2991.38&54.024&58133.6&61179&328992\\
			$600\times600$&4441.29&74.153&86668.1&91183.6&527420\\
			$700\times700$&6134.74&100.717&120812&127047&787626\\
			$800\times800$&8512.32&133.028&172108&180754&1.08301$\times10^6$\\
			$900\times900$&10993.3&168.125&222619&233781&1.4554$\times10^6$\\
			$1000\times1000$&13971.6&206.211&279026&293204&1.85278$\times10^6$\\
			\hline
		\end{tabular}
	}
	\label{gbd_computational_effort_10201}
\end{table}
\normalsize
We, therefore, choose $\{N=50, G= 300\times300\}$ for simulations with 10201 sampling points, i.e. in cases where $y$ is not set to 0 within this paper.

Indeed, these parameters result only in roughly comparable computational effort, and better matching could be found if intermediate values were used. However, this is not necessary and is not the aim here, since the computational effort depends on additional simulation parameters and will change for other setups, particularly the properties of the wavefront that is to be decomposed, as well as the sampling grid in the target plane. 
However, our experience showed, that the given choice was consistently resulting in a comparable computational effort for all simulations compared within this paper.

Finally, we can compare the elapsed real time and the CPU time, and thereby the parallelization of both methods.
For comparable parameters, i.e., $\{N=50, G= 400\times400\}$ (\cref{mem_computational_effort}, \cref{gbd_computational_effort}) and  $\{N=50, G= 300\times300\}$ (\cref{mem_computational_effort_10201} and \cref{gbd_computational_effort_10201}), the CPU time of GBD is longer. This shows that the GBD is more strongly parallelized than the MEM in the used IfoCAD version.

\section{Method Comparison} \label{se:4} 
In this section, we compare the performance of the MEM and GBD directly for various scenarios, which include non-clipped and clipped Gaussian beams in free space (\cref{sec:GBs-free-space}), aberrated wavefronts (\cref{se:5}), and reflection from optical components (\cref{se:6}). 

\subsection{Non-clipped Gaussian beams and clipped Gaussian beams in free space}\label{sec:GBs-free-space}
In every comparison within this subsection, we compute the introduced errors to evaluate the quality of each method. Unfortunately, this requires the electric field $E$ to be known in every target plane, which strongly restricts the number of possible test cases. We, therefore, test in \cref{sec:Non-clipped_GBs} the performance of the MEM and GBD for non-clipped circular and general astigmatic Gaussian beams for which the analytic representation of the electric field is widely known. In \cref{sec:Clipped_GBs}, we further investigate the case of circular symmetric clipped Gaussian Beams in the near-, far-, and extreme far-field, for which we can use again the analytic representations provided by Campbell \cite{1987Fresnel} and Tanaka et al.\ \cite{1985Field} in the Fresnel and the Fraunhofer region. With extreme far-field, we refer to propagation distances of a few million kilometers, which occur in space gravitational wave detectors such as LISA \cite{2006LISA} and Taiji \cite{2017The}.
\subsubsection{Non-clipped Gaussian beams}\label{sec:Non-clipped_GBs}
\subparagraph{Circular Gaussian beam}\label{sec:Non-clipped-circular-GBs} 
The simplest case to compare the MEM and GBD is using non-clipped Gaussian beams, for which the electric field is analytically known in any propagation distance. We, therefore perform a first comparison of the MEM and GBD on the example of a non-clipped circular-symmetric Gaussian beam with the parameters listed in \cref{tnoclipping}. 
\begin{table}[htbp]
	\caption{Parameters list for non-clipped circular Gaussian beam.}\label{tnoclipping}
	\footnotesize
	%\centering
	%\resizebox{\textwidth}{!}
	{
		\begin{tabular}{@{}lll}
			\hline
			\textbf{parameters} & \textbf{description} & \textbf{value}  \\ 
			\hline
			$\lambda$          & wavelength  & \SI{1064}{\nm} \\ 
			$P_0$          & beam power  & \SI{1}{W} \\  
			$w_0$         & beam waist  &  \SI{1}{\mm}     \\ 
			$z_0$        & distance from the waist           &   \SI{0}{\mm}      \\ 
			$R_a$          & aperture radius       &  \SI{4}{\mm}     \\ 
			$N$          & mode order of the MEM    &   50     \\ 
			$w_{0d}$          & waist of the modes used in the MEM  & \SI{0.8}{mm}    \\ 
			$G$          &grid size of the GBD             &  400$\times$400      \\ 
			$L$          & window size of the GBD             &   \SI{8}{\mm}     \\ 
			$f_{\rm ws}$        &waist scaling factor of the GBD             &  10/3  \\
			$w_{0g}$          & grid beam waist of the GBD  & \SI{0.0333}{mm}    \\ 
			grid shape          &grid shape of the GBD             &   square     \\ 
			$d$                 &propagation distance & $0.001z_r$, $z_r$, $1000z_r$, \SI{3}{Gm}\\
			$X$                 &number of sampling points & 3001 \\
			\hline
		\end{tabular}
	}
\end{table}
\normalsize
It is noteworthy that we still define an aperture here. This is due to the IfoCAD version used here, which requires an aperture to be defined for both methods. 
However, with a radius being 4 times larger than the Gaussian waist radius, it is effectively not clipping the beam, given that the clipped power is approximately \SI{1.3 e-12}{\%} of the full beam power. 
Instead, the aperture radius effectively defines the lateral range used in the numerical evaluation of the integral in \cref{eq:amn}. In the GBD, we are then using the very same aperture radius to decompose the exact same input field as in the MEM case. We chose not to overscale the window further and set the window size (which is a full width) to equal the aperture diameter, to not place unnecessarily many grid beams in regions without field amplitude. 

\begin{figure}[htbp]
	\centering
	\includegraphics[]{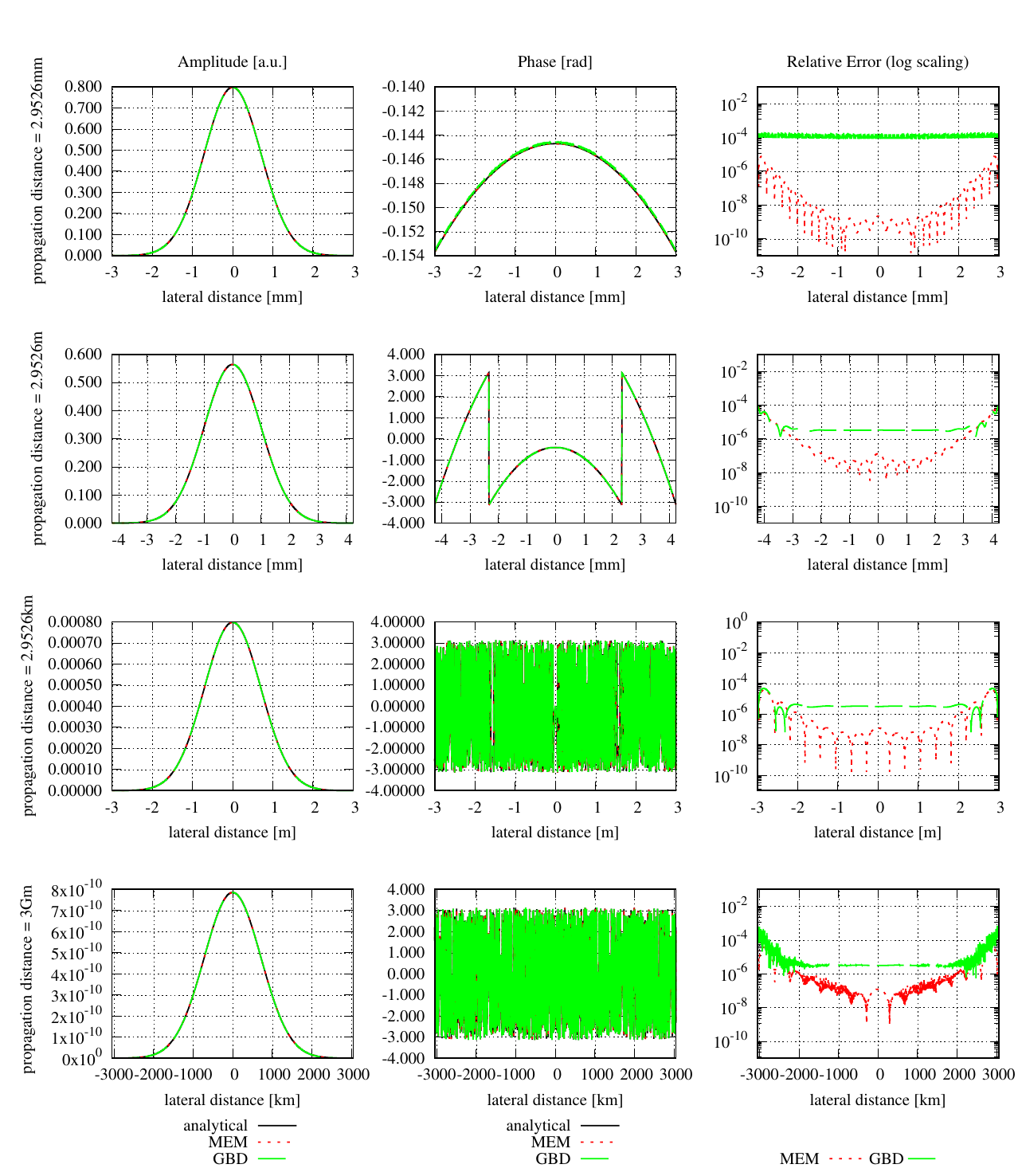}
	\caption{The amplitude (absolute value), phase, and relative error distributions at propagation distances, $z_r/1000$ , $z_r$ (near-field), $1000\,z_r$ (far-field) and 3 million kilometers (extreme far-field). The simulation parameters are listed in \cref{tnoclipping}. The analytical results refer to the complex electric field of the non-clipped Gaussian beam in this case. It can be seen that MEM behaves better than GBD. }
	\label{noclipping} 
\end{figure}

\begin{figure}[htbp]
	\begin{center}
		\includegraphics[]{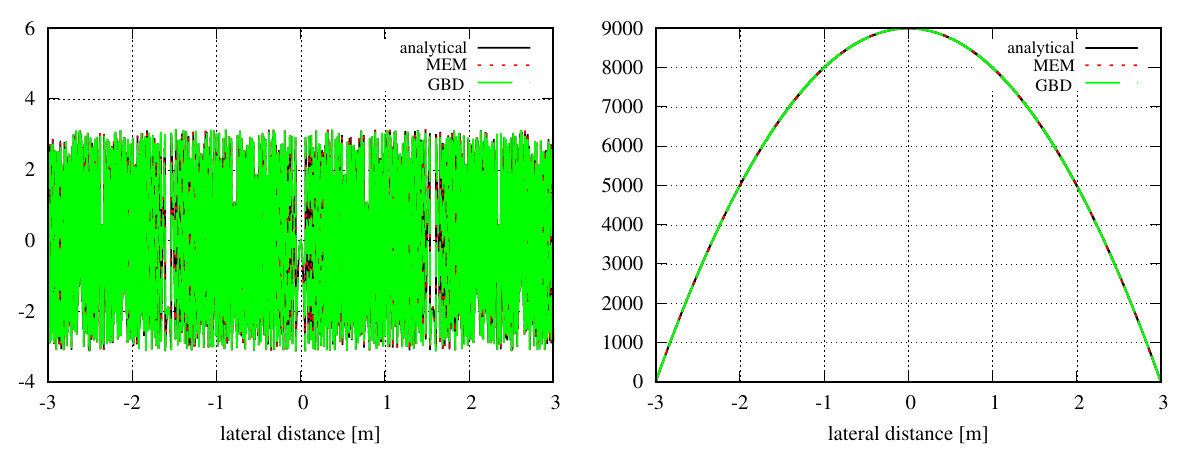} 
		
		\caption{The wrapped and unwrapped phase of propagation distance \SI{2.9526}{km} for the non-clipped circular Gaussian beam. }
		\label{unwrap1}
	\end{center}
\end{figure}

Concerning the optical setup, we define a circular Gaussian beam with a waist radius of \SI{1}{\mm} which is centered in this aperture. The mode order, grid size, and sampling points are chosen according to \cref{se:3}. The waist $w_{0d}$ of modes used in the MEM is calculated from \cref{borghi}; the GBD grid beam waist $w_{0g}$ is calculated correspondingly by \cref{fws}. After the decomposition, the MEM beam and GBD beam propagate for $z_r/1000$, $z_r$, $1000\,z_r$ and 3 million kilometers (\SI{3}{Gm}), with $z_r = \SI{2.9526}{m}$, where $z_r$ refers to the Rayleigh range of the incident Gaussian beam. 
We then speak of the far-field, if $z\gg z_{r}$\cite{siegman1986lasers}. The resulting amplitude, phase, and relative error are shown in \cref{noclipping}, 	%
and the corresponding discretized NMSE, and the summed relative errors are summarized in \cref{noclippingError}.
\begin{table}[htbp]
	\caption{The discretized NMSE and the summed relative error of the MEM and GBD respectively for different  propagation distances. The number of sampling points $X$ is 3001.}
	\footnotesize 
	%\resizebox{\textwidth}{!}
	%\centering
	{
		\begin{tabular}{@{}llll}
			\hline
			\textbf{propagation distance} & \textbf{method}  & \textbf{\makecell[c]{DNMSE \\$\varepsilon^\text{DNMSE}_\circ$}} & \textbf{\makecell[c]{the summed relative error \\$\varepsilon_{\sum}^\text{rel}$}} \\ 
			\hline
			\multirow{2}{*}{\SI{2.9526}{mm}}&	MEM &  1.3118$\times10^{-17}$ &5.0045$\times10^{-5}$\\
			&	GBD & 1.3786$\times10^{-8}$ &	7.0568$\times10^{-3}$\\  \hline
			\multirow{2}{*}{\SI{2.9526}{m}}&	MEM &  4.1645$\times10^{-15}$ &0.0011\\
			&	GBD &1.0192$\times10^{-11}$ &0.0019\\  \hline
			\multirow{2}{*}{\SI{2.9526}{km}}&	MEM & 5.1083$\times10^{-15}$ &371.2160\\
			&	GBD & 1.0202$\times10^{-11}$ &518.2059\\ \hline
			\multirow{2}{*}{\SI{3}{Gm}}&	MEM & 1.3516$\times10^{-14}$ &3.8923$\times10^{14}$ \\
			&	GBD & 1.0223$\times10^{-11}$ &	2.2340$\times10^{15}$\\
			\hline
		\end{tabular}
	}
	
	\label{noclippingError}
\end{table}
\normalsize
The chosen lateral ranges of 3 times the local spot sizes cause all amplitudes shapes to appear identically. This also holds for the shape of the phase profile. However, this is not immediately visible due to phase wrapping particularly in the far- and extreme far-field, which can be resolved by using a phase-tracking algorithm. \Cref{unwrap1} shows the unwrapped phase for the propagation distance of about \SI{3}{km}, i.e. 1000\,$z_r$.

From \cref{noclipping}, it can be seen clearly that both MEM and GBD can accurately represent the circular Gaussian beam. However, in all shown propagation distances, the MEM is more accurate than the GBD (right hand side column of graphs), which can also be seen from the discretized NMSE $\varepsilon^\text{DNMSE}$ (column 3) and summed relative error $\varepsilon_{\sum}^\text{rel}$ (column 4) in \cref{noclippingError}. 
The DNMSE error of the MEM is notably small, considering that the waist of the modes used in the MEM determined through \cref{borghi} is 0.8 mm, which closely aligns with the waist of the non-clipped Gaussian beam. While it is possible to obtain a 0 error by choosing the waist of the modes equal to the non-clipped beam waist of 1 mm, such a comparison would be meaningless in this particular scenario. However, this agrees with the analytically calculated NMSE (\cref{eq:ierror}) which is 1.3989$\times10^{-14}$. The discretized NMSE of the MEM result is found to be even slightly smaller than this, with a residual propagation distance dependency. Both of these properties originate from the finite radial range used in the decomposition and error computation. 

In conclusion, we find that in this example both methods accurately resolve the incident wavefront, and that the MEM has exceptional precision and is, therefore, more accurate than the GBD.

A particular challenge in this simulation was the ultra-large propagation distance of 3 million kilometers. 
This gives natural rise to numerical precision problems and was a major concern and one of the initial reasons for why we tested this extreme far field, which is relevant for space-based gravitational wave detectors. 
The given example shows no apparent signs of numerical limitation. 
This was achieved by a separation of the optical pathlength (i.e. the $i k z$-term in the Gaussian beam), from the residual phase contributions\cite{2012Methods}. 

Additionally, it was expected that the GBD could not propagate the beam into this extreme far-field without a re-decomposition in an intermediate plane. This expectation originated from the chosen small grid in the original decomposition plane, compared to the very large spot size in the target plane. After all, in the given example, the spot size of the total beam is 1\,mm in the decomposition plane, for which we chose a window size of 8\,mm. Since the beam is not re-decomposed, this is also being used in the target plane, where the Gaussian beam radius has increased to 1.0160$\times10^3$\,km (cf \cref{noclipping}). Yet, despite that the grid is unfit for the target plane, it is clearly visible in \cref{noclipping} that the beam is well represented. A re-decomposition was not needed in the given example.  

\begin{figure}[htbp]
	\begin{center}
		
		\includegraphics[]{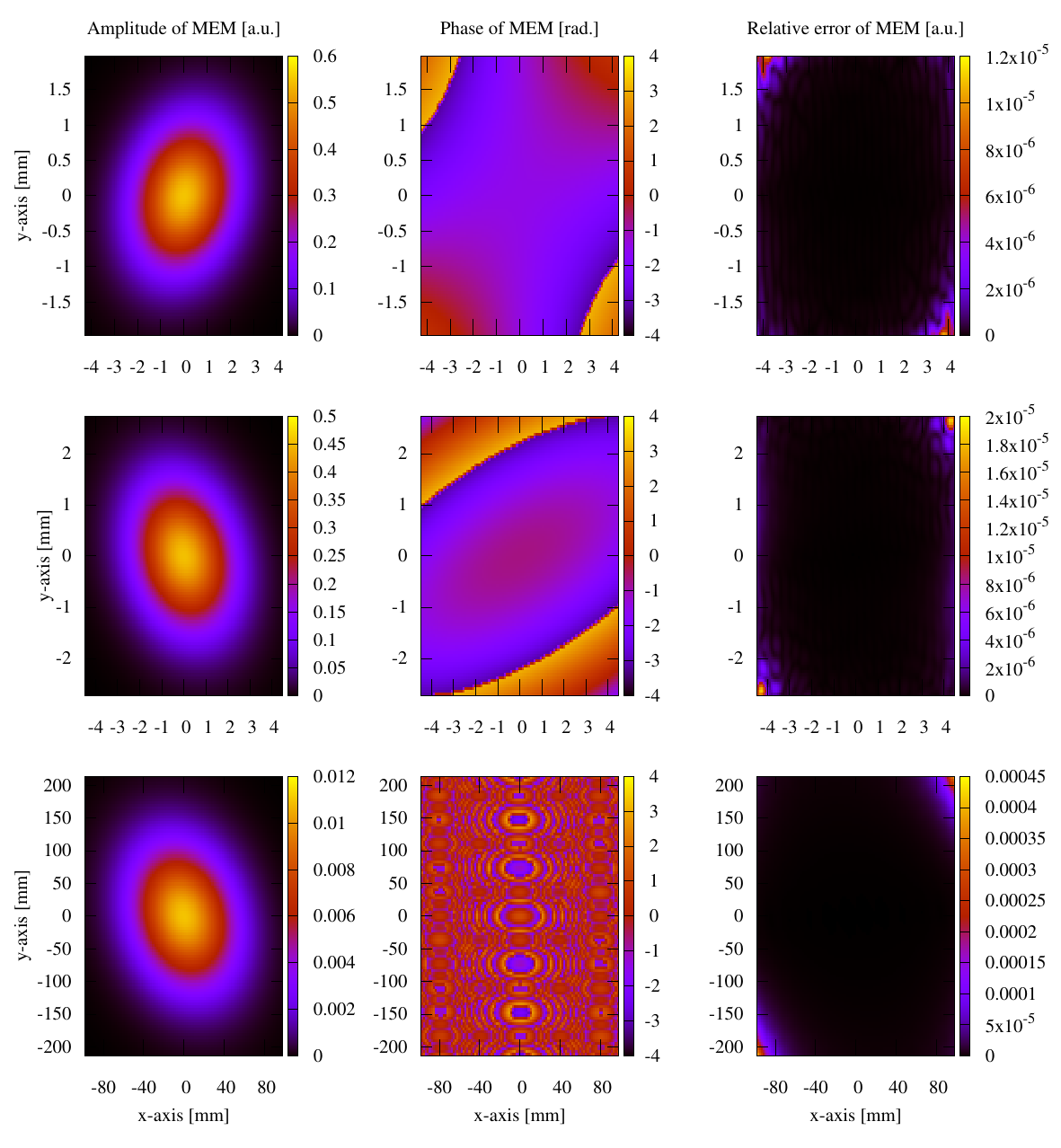} 
		
		\caption{MEM representation of a general astigmatic Gaussian beam at different propagation distances. From left to right: amplitude (absolute value), phase and relative error distribution of the MEM. From top to bottom: propagation distance \SI{29.526}{\mm}, \SI{2.9526}{\m} and \SI{295.26}{\m} respectively. All the simulation parameters are listed in \cref{tgagb}. It can be seen from the third column of this figure that MEM shows a good agreement with the initial general astigmatic Gaussian beam.}
		\label{gagbMEM}
	\end{center}
\end{figure}

\begin{figure}[htbp]
	\begin{center}
		\includegraphics[]{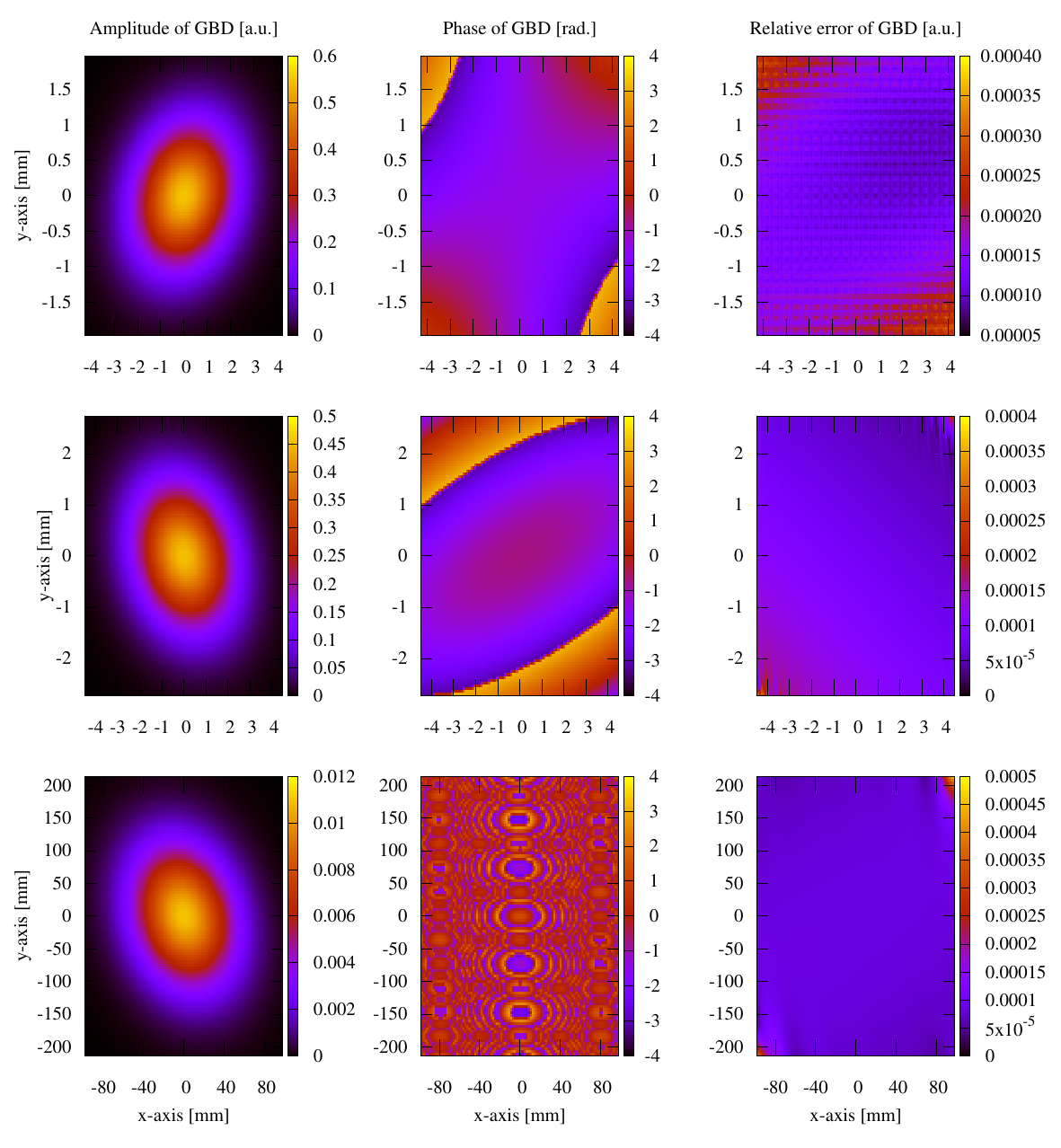} 
		
		\caption{GBD representation of a general astigmatic Gaussian beam at different distances. From left to right: amplitude (absolute value), phase, and relative error distribution of the GBD. From top to bottom: propagation distance \SI{29.526}{\mm}, \SI{2.9526}{\m} and \SI{295.26}{\m} respectively. All the simulation parameters are listed in \cref{tgagb}. It can be seen from the third column of this figure that GBD shows a good agreement with the initial general astigmatic Gaussian beam.}
		\label{gagbGBD}
	\end{center}
\end{figure}

\subparagraph{General astigmatic Gaussian beam}	
After testing the performance in the case of a non-clipped circular symmetric Gaussian beam, we now test the performance of these methods on a general astigmatic Gaussian beam. In that case, two beam waists and a complex angular orientation $\theta$ need to be defined \cite{2015Stigmatic}. The parameters used for the performance comparison are listed in \cref{tgagb}. 
\begin{table}[htbp]
	\caption{Parameters list for non-clipped general astigmatic Gaussian Beam.}
	\footnotesize
	%\centering
	%\resizebox{\textwidth}{!}
	{
		\begin{tabular}{@{}lll}
			\hline
			\textbf{parameters} & \textbf{description} & \textbf{value}  \\ 
			\hline
			$\lambda$          & wavelength  & \SI{1064}{\nm} \\ 
			$P_0$          & beam power  & \SI{1}{W} \\  
			$w_{01}$         & beam waist in $XZ$ plane & \SI{1}{\mm}      \\ 
			$z_{01}$        & distance from the waist in $XZ$ plane           &    \SI{0}{\mm}      \\ 
			$w_{02}$         & beam waist in $YZ$ plane &  \SI{2}{\mm}       \\ 
			$z_{02}$        & distance from the waist in $YZ$ plane          &  \SI{0}{\mm}     \\ 
			$\theta$    &  tilt angle     &   0.1+0.2i    \\ 
			$R_a$          & aperture radius       &   \SI{8}{\mm}     \\ 
			$N$          & mode order of the MEM          &  50      \\ 
			$w_{0d}$          & waist of the modes used in the MEM         &   \SI{1.6}{\mm}    \\ 
			$G$          &grid size of the GBD             &   300$\times$300    \\ 
			$L$          & window size of the GBD             &   \SI{16}{\mm}     \\ 
			$f_{\rm ws}$        &waist scaling factor of the GBD          &   10/3   \\ 
			$w_{0g}$          & grid beam waist of the GBD  & \SI{0.0444}{mm}    \\ 
			grid shape          &grid shape of the GBD             &    square     \\
			$d$                & propagation distance               & $0.01\,z_{r1}$, $z_{r1}$, $100\,z_{r1}$    \\
			$X$               & number of sampling points                    & $101\times101 = 10201$ \\
			\hline
		\end{tabular}
	}
	\label{tgagb}
	
\end{table}
\normalsize
In this case, the lateral parameter $y$ is not set to 0. As shown in \cref{faircomparison}, we therefore use a mode order of $N=50$ and compare this with a GBD grid size of $300\times300$. The resulting electric field profiles of the MEM are plotted in \cref{gagbMEM} and for the GBD in \cref{gagbGBD} for three different propagation distances: $z_{r1}/100, z_{r1}, 100 z_{r1}$, with $z_{r1} = \SI{2.9526}{m}$ being the XZ-plane Rayleigh range. 
We do not show the results of \SI{3}{Gm} in this case because the sampling points of $101\times 101$ is too low for such an extreme-far distance and we have already shown that both MEM and GBD are accurate at \SI{3}{Gm} for the non-clipped circular Gaussian beam. We have computed the electric field in lateral distances $(x,y)$ out to 2 times the spot size. This choice results in good visibility of the amplitude profiles, however, it slightly masks the magnitude of ellipticity in the resulting images. 

\begin{figure}[htbp]
	\begin{center}
		\includegraphics[]{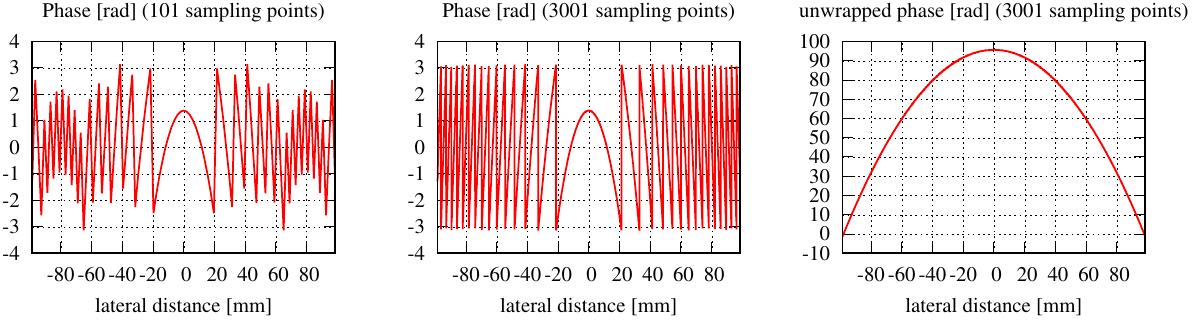} 	
		\caption{Cross-section of the phase obtained with the MEM along $y=0$ and $z=\SI{295.26}{m}$. The left-hand side image shows the phase using 101 sampling points, the centered image for 3001 sampling points, and the right-hand side image again for 3001 sampling points after a phase-tracking algorithm unwrapped the data. One sees that the small side-maxima, which are seen as circles in \cref{gagbMEM} and \cref{gagbGBD}, disappear when a higher sampling rate is used, and indeed a smooth but strongly curved phase profile is restored in the far field. }  
		\label{unwrap2}
	\end{center}
\end{figure}

In the first column of both figures, one sees the known characteristics of a general astigmatic Gaussian beam: its elliptical amplitude pattern which has a rotating orientation during propagation. Both methods equally well resolve this characteristic.
The second column, showing the phase of the general astigmatic beam, is mostly smooth at propagation distances less than \SI{3}{m} (top and central row), except of two lines of phase jumps. In the far field  (lowest row), there is a pattern of circular shapes. This pattern is an aliasing effect originating from the high wavefront curvature, the resulting high number of phase jumps in combination with the low sampling rate. We demonstrate this effect and how it can be resolved for a cross-section of the phase profile in \cref{unwrap2}. 
Finally, the third column in \cref{gagbMEM} and \cref{gagbGBD} shows the accuracy of each method. The DNMSE and summed relative error of the MEM and GBD at different propagation distances are shown in \cref{noclippingError_ga}. 
\begin{table}[htbp]
	\caption{The DNMSE and the summed relative error of the MEM and GBD for non-clipped general astigmatic Gaussian beam at different propagation distances. The number of sampling points $X$ is 10201.}
	\footnotesize
	%\centering
	%\resizebox{\textwidth}{!}
	{ 
		\begin{tabular}{@{}llll}
			\hline
			\textbf{propagation distance} & \textbf{method}  & \textbf{\makecell[c]{DNMSE \\$\varepsilon^\text{DNMSE}_\Box$}} & \textbf{\makecell[c]{the summed relative error \\$\varepsilon_{\sum}^\text{rel}$}} \\ 
			\hline
			\multirow{2}{*}{\SI{29.526}{mm}}&	MEM  &6.2454$\times10^{-17}$&  1.9562$\times10^{-5}$\\
			&	GBD &6.7856$\times10^{-9}$ & 0.0050\\  \hline
			\multirow{2}{*}{\SI{2.9526}{m}}&	MEM  &2.3443$\times10^{-16}$& 3.5741$\times10^{-5}$\\
			&	GBD  & 4.7449$\times10^{-12}$& 2.6332$\times10^{-4}$\\  \hline
			\multirow{2}{*}{\SI{295.26}{m}}&	MEM  &1.1112$\times10^{-14}$&  0.4653\\
			&	GBD &  4.7519$\times10^{-12}$&	0.6285\\ 
			\hline
		\end{tabular}
	}
	\label{noclippingError_ga}
	
\end{table}
\normalsize

It can be seen from this table, as well as from the third column of both \cref{gagbMEM} and \cref{gagbGBD} that the MEM is again consistently more accurate than the GBD. Additionally, one can see that both the DNMSE and the summed relative error of the MEM increase with the propagation distances, while for the GBD, the DNMSE nonmonotonically decrease with the propagation distances, and the summed relative error again changes inconsistently.

\subsubsection{Clipped Gaussian beam} \label{sec:Clipped_GBs}  	
\begin{figure}[htbp]
	\begin{center}
		\includegraphics[width=0.95\textwidth]{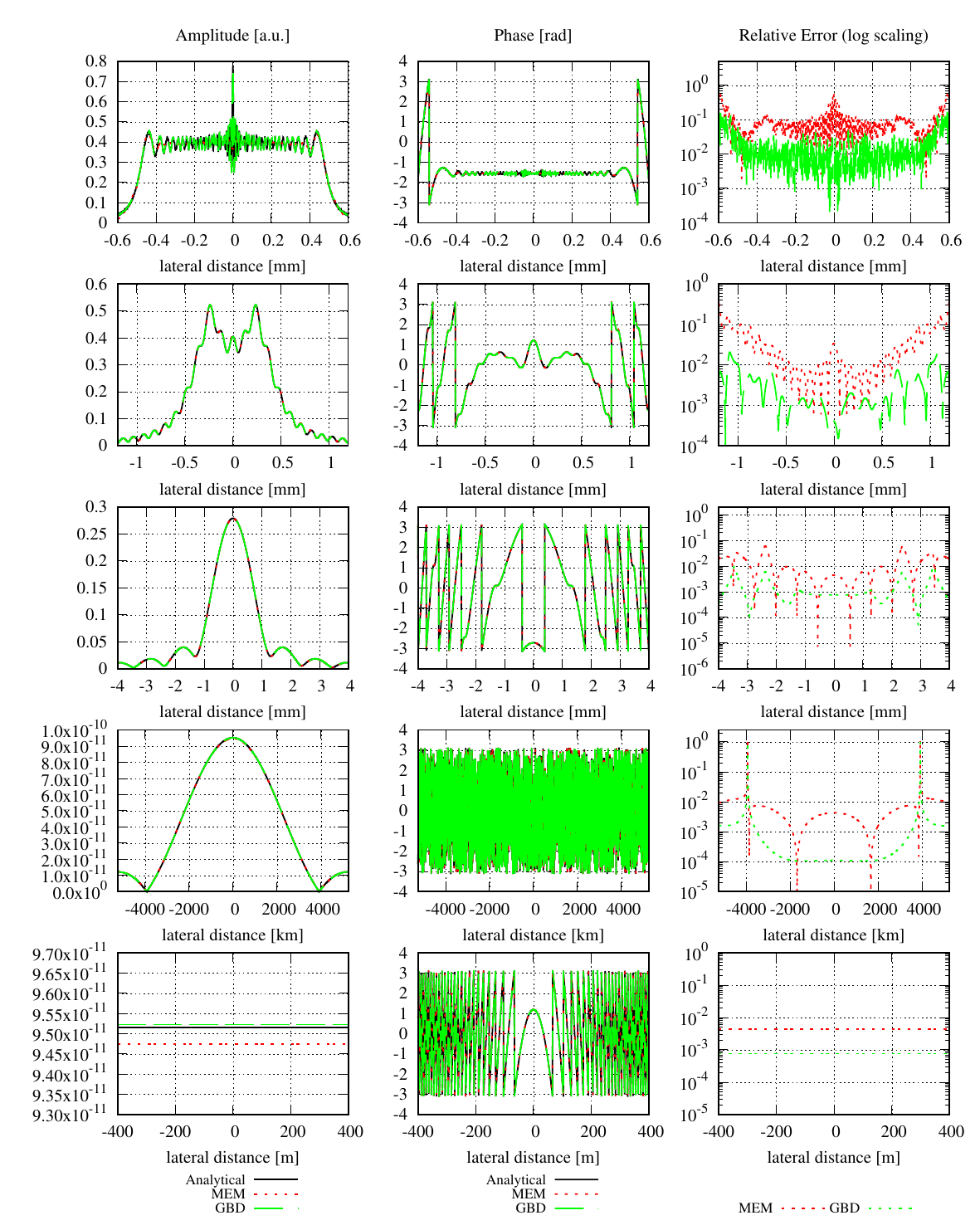}
	\end{center}
	\caption{Amplitude (absolute value), phase and relative error distribution of a clipped circular Gaussian beam with propagation distances \SI{5}{mm}, and \SI{100}{mm} in the near field, \SI{1000}{mm} in the far field, and \SI{3}{Gm} in the extreme far field (from top to bottom). The lateral range in the fourth row is 2 times the local spot size. The simulation parameters are listed in \cref{tclippedgauss}. The analytical methods for the near and far field are Campbell \cite{1987Fresnel} and Tanaka et al.\ \cite{1985Field}, respectively. The graphs in the third column show the relative error, which indicates that the GBD performs better than the MEM.}
	\label{clippedgauss}
\end{figure}

In this subsection, we directly compare the performance of the MEM and GBD for a circular Gaussian beam clipped by a circular aperture. The parameter settings are summarized in \cref{tclippedgauss}, for convenience.
\begin{table}[htbp]
	\caption{Parameters list for circular Gaussian beam clipped by a circular aperture centered in the beam waist.}
	\footnotesize
	%\centering
	%\resizebox{\textwidth}{!}
	{
		\begin{tabular}{@{}lll}
			\hline
			\textbf{parameters} & \textbf{description} & \textbf{value}  \\ 
			\hline
			$\lambda$          & wavelength  & \SI{1064}{\nm} \\ 
			$P_0$          & beam power  & \SI{1}{W} \\  
			$w_0$         & beam waist  &  \SI{2}{\mm}     \\ 
			$z_0$        & distance from the waist           &   \SI{0}{mm}      \\ 
			$R_a$          & aperture radius       &  \SI{0.5}{\mm}     \\ 
			$N$          & mode order of the MEM    &   50     \\ 
			$w_{0d}$          & waist of the modes used in the MEM         &  \SI{0.1}{\mm}      \\ 
			$G$          &grid size of the GBD             &  400$\times$400      \\ 
			$L$          & window size of the GBD             &   \SI{1.5}{\mm}     \\ 
			$f_{\rm ws}$        &waist scaling factor of the GBD             &  3/2\\ 
			$w_{0g}$       &grid beam waist of the GBD     & \SI{0.0029}{\mm}\\
			grid shape          &grid shape of the GBD             &   square     \\ 
			$d$          & propagation distance            &   \SI{5}{mm}, \SI{100}{mm}, \SI{1000}{mm}, \SI{3}{Gm}   \\ 
			$X$                 & number of sampling points          & 3001 \\
			\hline
		\end{tabular}
	}
	\label{tclippedgauss}
\end{table}
\normalsize
In this case, the diffracted beam propagates \SI{5}{mm}, and \SI{100}{mm} in the near field, \SI{1000}{mm} in the far field, and \SI{3}{Gm} in the extreme far field, with Fresnel numbers of 46.9925, 2.3496, 0.2350 and $7.83208\times 10^{-11}$, respectively. 
The resulting amplitude, phase and relative error profiles are depicted in \cref{clippedgauss},
and the corresponding errors are summarized in \cref{clippedError}.
\begin{table}[htbp]
	\caption{The DNMSE and the summed relative error of the MEM and GBD respectively for different  propagation distances. The number of sampling points is 3001.}
	\footnotesize 
	%\resizebox{\textwidth}{!}
	
	{
		\begin{tabular}{@{}llll}
			\hline
			\textbf{propagation distance} & \textbf{method} & \textbf{\makecell[c]{discretized NMSE \\$\varepsilon^\text{DNMSE}_\circ$}} & \textbf{\makecell[c]{the summed relative error \\$\varepsilon_{\sum}^\text{rel}$}} \\ 
			\hline
			\multirow{2}{*}{\SI{5}{mm}}&	MEM &  0.0057 &0.2465 \\
			&	GBD & $9.5762 \times10^{-4}$ &  0.1318 \\  \hline
			\multirow{2}{*}{\SI{100}{mm}}&	MEM &  $1.9470\times 10^{-4}$&0.5380\\
			&	GBD &$6.9726\times10^{-6}$&   0.0857 \\  \hline
			\multirow{2}{*}{\SI{1000}{mm}}&	MEM & $2.5676\times 10^{-5}$ &1.67784	 \\
			&	GBD & $6.6764\times10^{-7}$ &0.3279\\ \hline
			\multirow{2}{*}{\SI{3}{Gm} (2 times spot size)}&	MEM & $5.5759\times 10^{-6}$ &	8.7671$\times10^{17}$  \\
			&	GBD & $7.5033\times10^{-8}$ &  3.0557$\times10^{17}$	\\ \hline
			\multirow{2}{*}{\SI{3}{Gm} (\SI{400}{m})}&	MEM & $7.2154\times 10^{-13}$ &	 4.3396 $\times10^{9}$ \\
			&	GBD & $ 2.4326\times10^{-15}$ & 2.5197$\times10^{8}$  \\ 
			\hline
		\end{tabular}
	}
	
	\label{clippedError}
\end{table}
\normalsize
We have chosen again lateral ranges that allow good visibility of the amplitude profile. For \SI{3}{Gm}, we provide two lateral ranges: approximately 2 times the spot size and a smaller lateral range of \SI{400}{m}. In fact, we did not attempt to compute the exact spot size of the clipped Gaussian beam, but simply estimate it from two boundary cases: the spot size of a top hat and Gaussian beam. The spot size of a top hat beam with \SI{0.5}{mm} radius can be estimated using \cite[Eq.(8)]{Drege:00} resulting in \SI{2625.3}{km} after \SI{3}{Gm}. For a Gaussian beam with \SI{0.5}{mm} waist, the spot size at a propagation distance of \SI{3}{Gm} would be \SI{2032.1}{km}, and the spot size of the clipped Gaussian beam is expected to be between these two values. We use here the spot size of the top-hat for the lateral ranges for simplicity. 

In \cref{clippedgauss}, it can be seen that both methods describe the clipped Gaussian beam well, however, in all shown propagation distances, the GBD behaves better than the MEM. Quantitatively, we can also reach the same conclusion from the discretized NMSE and the summed relative errors listed in \cref{clippedError}. Especially in the near field with propagation distances \SI{5}{\mm} and \SI{100}{\mm}, the differences in the errors of the MEM and GBD is significant. This is the same finding as in \cref{se:exampleMEM}, that the MEM is insufficiently resolving the high-frequency spatial oscillation in the very near field. With the increase of the propagation distance, at \SI{1000}{\mm} (third row), the differences between both methods become smaller, and in the extreme far-field \SI{3}{Gm} (fourth and lowest row), they narrow further. We can also see that both methods become more accurate with increasing propagation distances. Like before, the extreme far-field electric field was computed with the GBD in one step and did not require a re-decomposition in an intermediate plane.

\begin{figure*}[htbp]
	\begin{center}
		\mbox{\includegraphics[width=\textwidth]{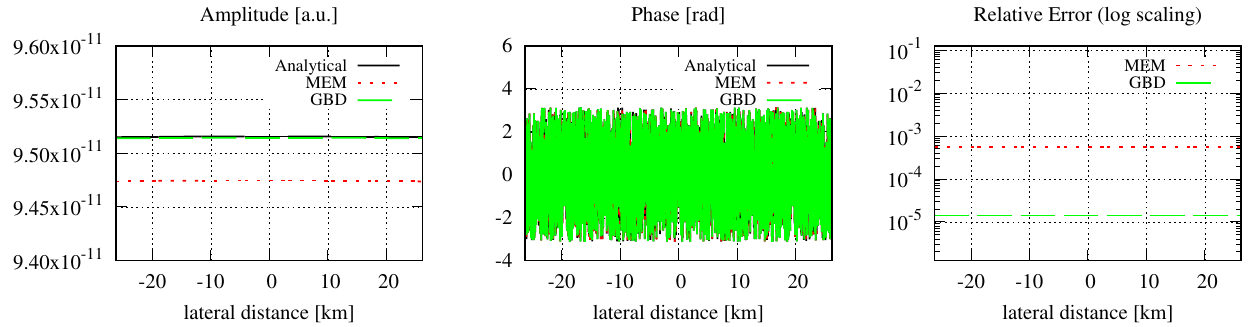}}
	\end{center}
	\caption{Amplitude (absolute value), phase and relative error distribution of a clipped circular Gaussian beam at propagation distances \SI{3}{Gm}. Both the beam waist diameter and aperture diameter are \SI{30}{cm}. The right-hand side image shows the relative error, which indicates that the GBD performs better than the MEM.}
	\label{clippedgauss1}
\end{figure*}

The \SI{0.5}{mm} aperture radius is a typical value in laboratory experiments and, therefore, fits well with the shown near-field propagation distances. However, it is not a realistic value for the extreme far-field simulation case originating from space gravitational-wave detectors. The aperture diameter in LISA-like missions is usually between \SI{20}{cm} and \SI{40}{cm}. Therefore, here we show another example, where both the beam waist diameter and aperture diameter are \SI{30}{cm}, for propagation distance \SI{3}{Gm}. Other parameters are chosen from \cref{tclippedgauss}, except the window size, which is \SI {35}{cm} in this case, and the lateral range is chosen 3 times the spot size estimated by \cite[Eq.(8)]{Drege:00}. The resulting amplitude, phase, and relative error are shown in \cref{clippedgauss1}.

In this figure, it can be seen that the order of magnitude of the amplitude is $10^{-8}$ (left-hand side image), and the relative error shown in the right-hand side image indicates that the GBD is again more accurate than the MEM.

\subsection{Aberrated wavefronts} \label{se:5}

Aberration is a typical phenomenon in optics and a case in which the beam cannot be propagated analytically through the setup.
In LISA-like missions, aberration occurs for instance when the beam propagates through the telescope. The beam that is launched towards the remote spacecraft is, therefore, not a perfect clipped Gaussian beam but has additional wavefront distortions. Likewise, the received beam is not a perfect top hat beam. These aberrations affect the readout noise, and therefore need to be studied.

We study here how the MEM and GBD decompose and propagate aberrated wavefronts. Unfortunately, we do not have an analytic solution to compare the results with. We can, therefore, only qualitatively compare the results of both methods without being directly able to judge which one is more accurate. Instead, we test whether the methods generate qualitatively agreeing results. 

Mathematically, the wavefront aberration can be described by adding an additional phase term $\varOmega_a$ in the complex electric field of the original beam, i.e.:
\begin{equation}
	E_a(r; 0)=E(r; 0)\exp\left(ik\varOmega_a\right)\,,
\end{equation}
where $E_a$ is the complex electric field of the beam with aberration, $E$ is the complex electric field of the original beam, and $\varOmega_a$ is the additional phase distribution caused by aberration. Such a phase term is usually described by Zernike polynomials \cite{Mahajan:94}:
\begin{equation}
	\varOmega_a(x,y)=\sum_{n=0}^{N}\sum_{m=-n}^{n} c_{n}^{m}\,Z_{n}^{m}(x,y),
\end{equation}
where $ c_{n}^{m}$ represents coefficients and $Z_{n}^{m}$ are Zernike Polynomals, where $n-m\geq0$ is an even number. 

To demonstrate that both the MEM and GBD can describe wavefront aberrations, we show their results for the well-known effects that are caused by individual Zernike polynomials up to fourth order (cf. \cite[Fig.\,4]{Masalehdan:10}, \cite[Fig.\,3]{Vera2012}, and \cite[Fig.\,9.8]{bornwolf}). The effect of optical aberration is often represented by calculating the point spread function (PSF)  \cite{Masalehdan:10}, i.e. the intensity profile at a distance $z$ which is usually computed by Fourier transformation:	%
\begin{equation}
	\text{PSF}(r;z)= | \text{FT}_{z}(E_a(r; 0)) | ^2\,,
	\label{eq:psf}
\end{equation}	
which is also known as the response of the pupil function after Fourier transform (FT) under a certain distance $z$ \cite{goodman1996introduction}. In the far field, the beam source can usually be regarded as a point source compared to the propagation distance, and the PSF equals then the intensity profile computed by Fraunhofer diffraction. In this section, we use the MEM and GBD to represent the effect of optical aberration, by calculating the amplitude profile of a diffracted Gaussian beam with aberration at the Fraunhofer region, instead of directly calculating the PSF.

In this example, we calculate amplitude profiles of the first 4 orders of Zernike polynomials by the MEM and GBD for an aperture with \SI{1}{mm} radius at $\lambda = \SI{1064}{nm}$, the coefficient of Zernike polynomials $c_{n}^{m}$ for all $m$ and $n$ is set to 10.
Parameter settings of the example are listed in \cref{AberrationSetting1}. The propagation distance is \SI{5}{km} ($F = 1.8797\times10^{-4}$), therefore, it is in the Fraunhofer region.
\begin{table}[htbp]
	\caption{Parameters list of the wavefront aberration.}
	\footnotesize
	%\centering
	%\resizebox{\textwidth}{!}
	{
		\begin{tabular}{@{}lll}
			\hline
			\textbf{parameters} & \textbf{description} & \textbf{value}  \\ 
			\hline
			$\lambda$          & wavelength  & \SI{1064}{\nm} \\ 
			$P_0$          & beam power  & \SI{1}{W} \\
			$w_0$         & beam waist  &  \SI{1}{\mm}     \\ 
			$z_0$        & distance from the waist           &   0      \\ 
			$R_a$          & aperture radius       &  \SI{1}{\mm}     \\ 
			$N$          & mode order of the MEM    &   50     \\ 
			$w_{0d}$          & waist of the modes used in the MEM    &   \SI{0.2}{\mm}     \\ 
			$G$          & grid size of the GBD    &   $150\times150$     \\ 
			$L$          & window size of the GBD    &  \SI{3}{\mm}     \\ 
			$f_{\rm ws}$          & waist scaling factor of the GBD    &  8/3\\ 
			$w_{0g}$          & grid beam waist of the GBD    &   \SI{0.0267}{\mm}   \\ 
			grid shape 				 & 	grid shape of the GBD & square \\
			$c_{n}^{m}$          & coefficient of Zernike Polynomals    &   10     \\
			$d$          & propagation distance  &  \SI{5}{\km}     \\ 
			$X$         & number of sampling points & $201\times201 = 40401$ 	\\  
			\hline
		\end{tabular}
	}
	\label{AberrationSetting1}
\end{table}

\begin{figure*}[htbp]
	\begin{center}
		\includegraphics[width=\textwidth]{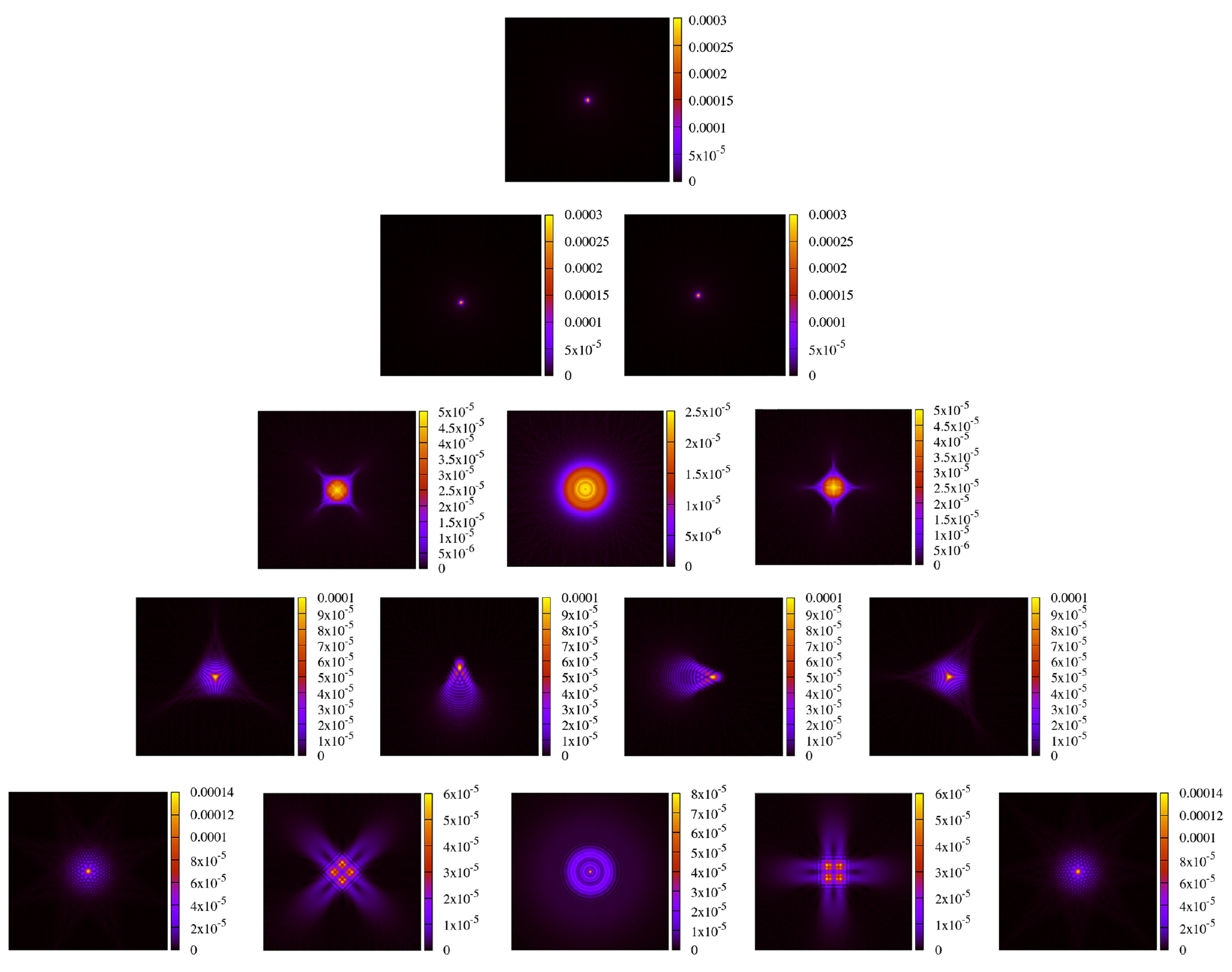}
	\end{center}
	\caption{Amplitude profile of the first 4 orders of wavefront aberrations described by Zernike polynomials, which is generated by the GBD. Aberrations of order n=1 are called Tilt, n=2 are Astigmatism and Defocus, n=3 are Coma and Trefoil, and n=4 are Tetrafoil, 2nd Astigmatism, and Spherical Aberration. The shown images cover $200\,m \times200\,m$ each.}
	\label{Aberration}
\end{figure*}

For a good resolution of our results, we used $201\times201 = 40401$ sampling points and chose for the MEM a mode order of $N = 50$ and for the GBD a grid size $G = 150\times150$. The results computed by the GBD are shown in \cref{Aberration}, and show the expected amplitude profiles (compare e.g. (cf.\cite[Fig.\,4]{Masalehdan:10}, \cite[Fig.\,3]{Vera2012}, and \cite[Fig.\,9.8]{bornwolf}). 

\begin{figure*}[htbp]
	\begin{center}
		\includegraphics[width=\textwidth]{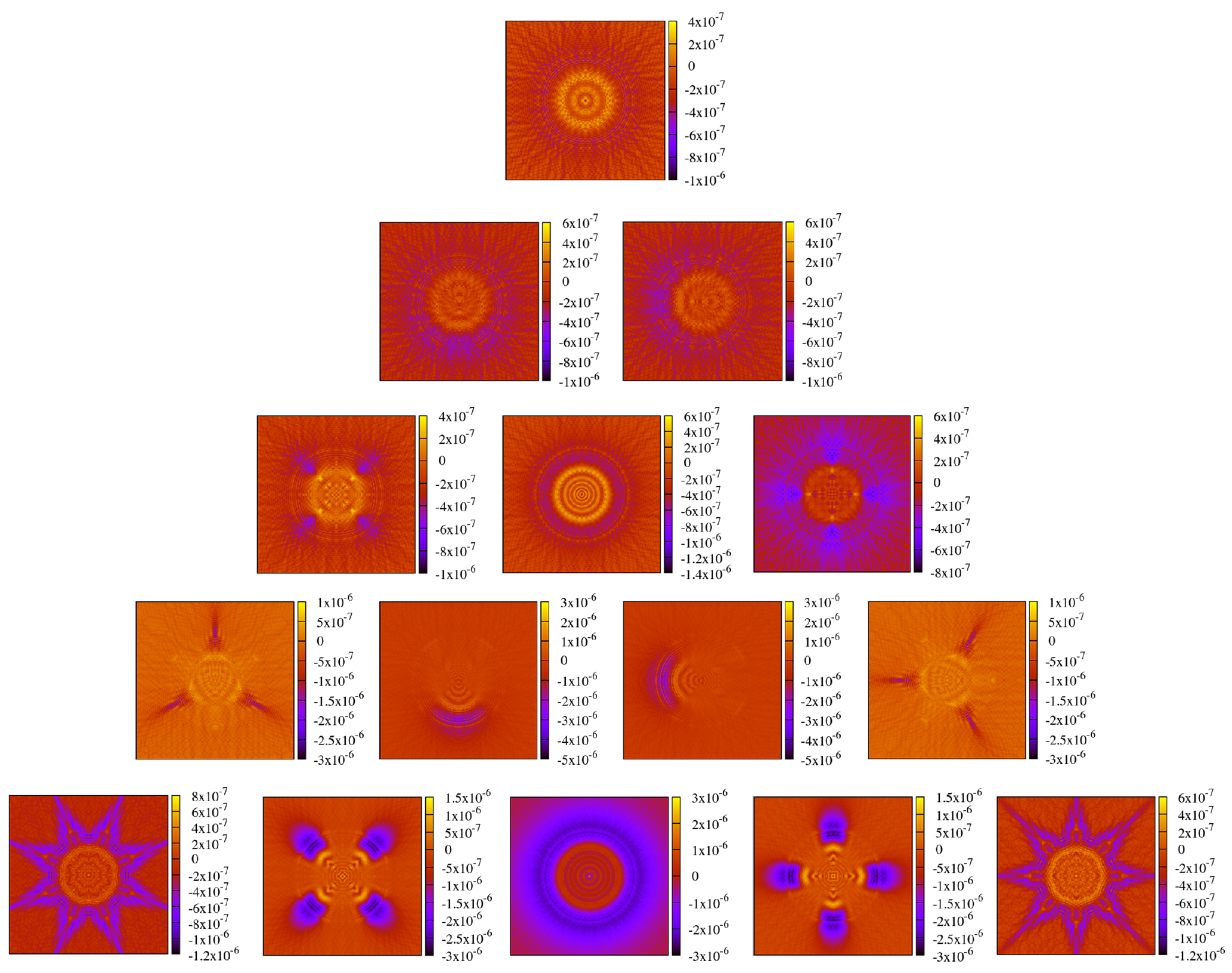}
	\end{center}
	\caption{Amplitude difference profile between the MEM and GBD of the first 4 orders of  wavefront aberrations described by Zernike polynomials. The shown images cover $200\,m \times200\,m$ each.}
	\label{Aberration_diff}
\end{figure*}

The MEM results are nearly identical and are not shown here to avoid unnecessary duplications. Instead, we show the difference between the MEM and GBD results in \cref{Aberration_diff}.
It can be seen from this figure that the difference between  MEM and the GBD is in the order of $10^{-7}$ to $10^{-6}$. If this is related to the individual amplitudes of approximately $10^{-5}$ to $10^{-4}$, we would speak of relative deviations in the order of a few percent. Even though we cannot judge which of the two methods is more accurate, we can conclude that either method can generally be used for decomposing and propagating aberrated wavefronts. Please note, we intentionally plot here only the difference between the results, and perform only a qualitative relative deviation, because the computation of a relative difference like $|E_\text{GBD}- E_\text{MEM}|/|E_\text{MEM}|$ causes divisions by zero with the shown lateral ranges.

\subsection{Reflection from optical components} \label{se:6}

\begin{figure}[htbp]
	\begin{center}
		\includegraphics[width=\textwidth]{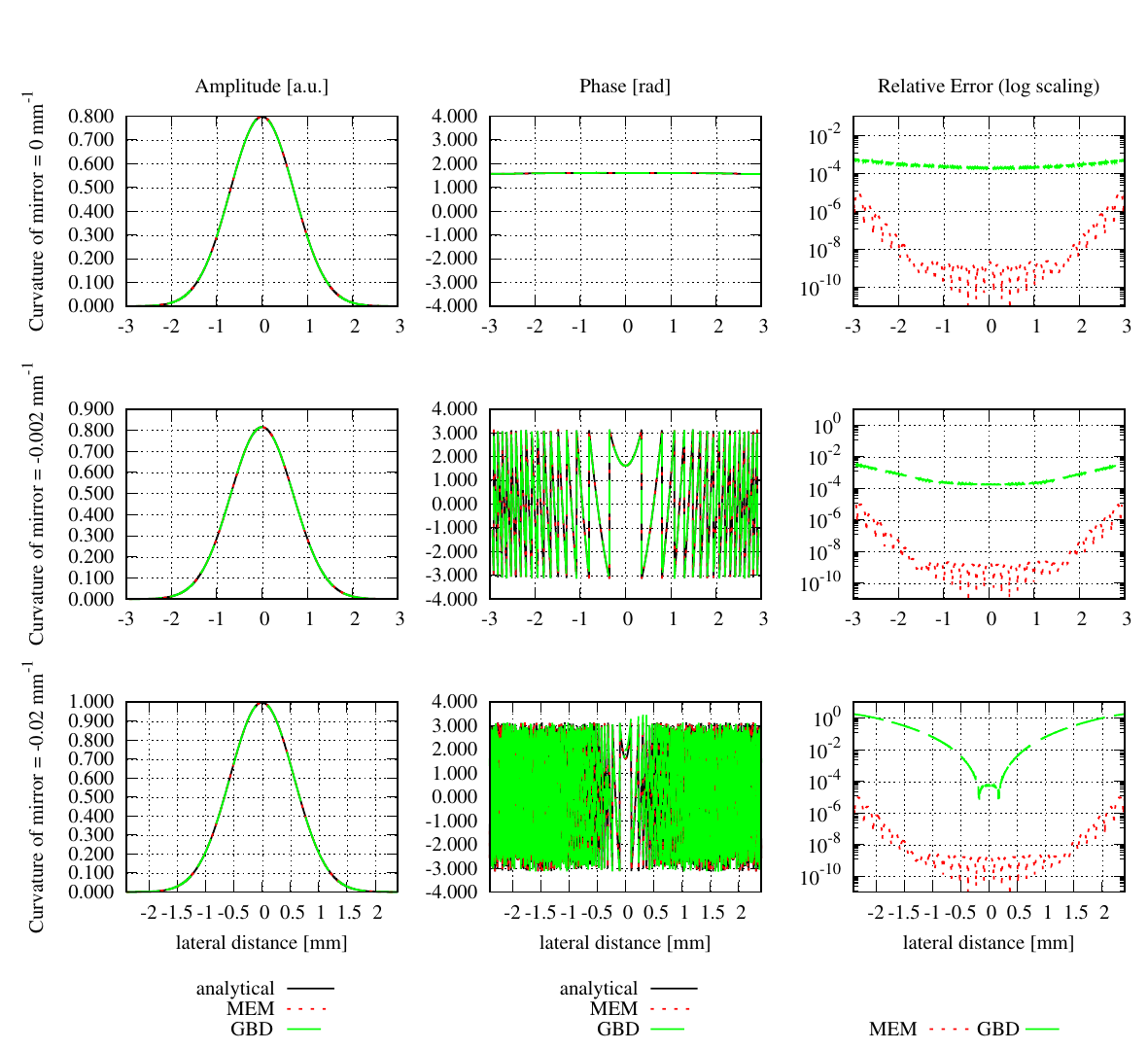} 
	\end{center}
	\caption{Amplitude, phase distribution and relative error after the beam reflected from mirrors with different curvatures. The graphs in each row represent the amplitude, phase and relative error respectively from left to right, and each row represents the different curvature of mirrors \SI{0}{mm}$^{-1}$,\SI{-0.002}{mm}$^{-1}$, and \SI{-0.02}{mm}$^{-1}$  from top to bottom. The simulation parameters are listed in \cref{mirrorparameters}. The relative deviation of the GBD from the Gaussian beam and MEM increases when the curvature of the mirror increases.}
	\label{ReflectonGBDCMEM}
\end{figure}

\begin{figure}[htbp]
	\begin{center}
		\includegraphics[width=0.8\textwidth]{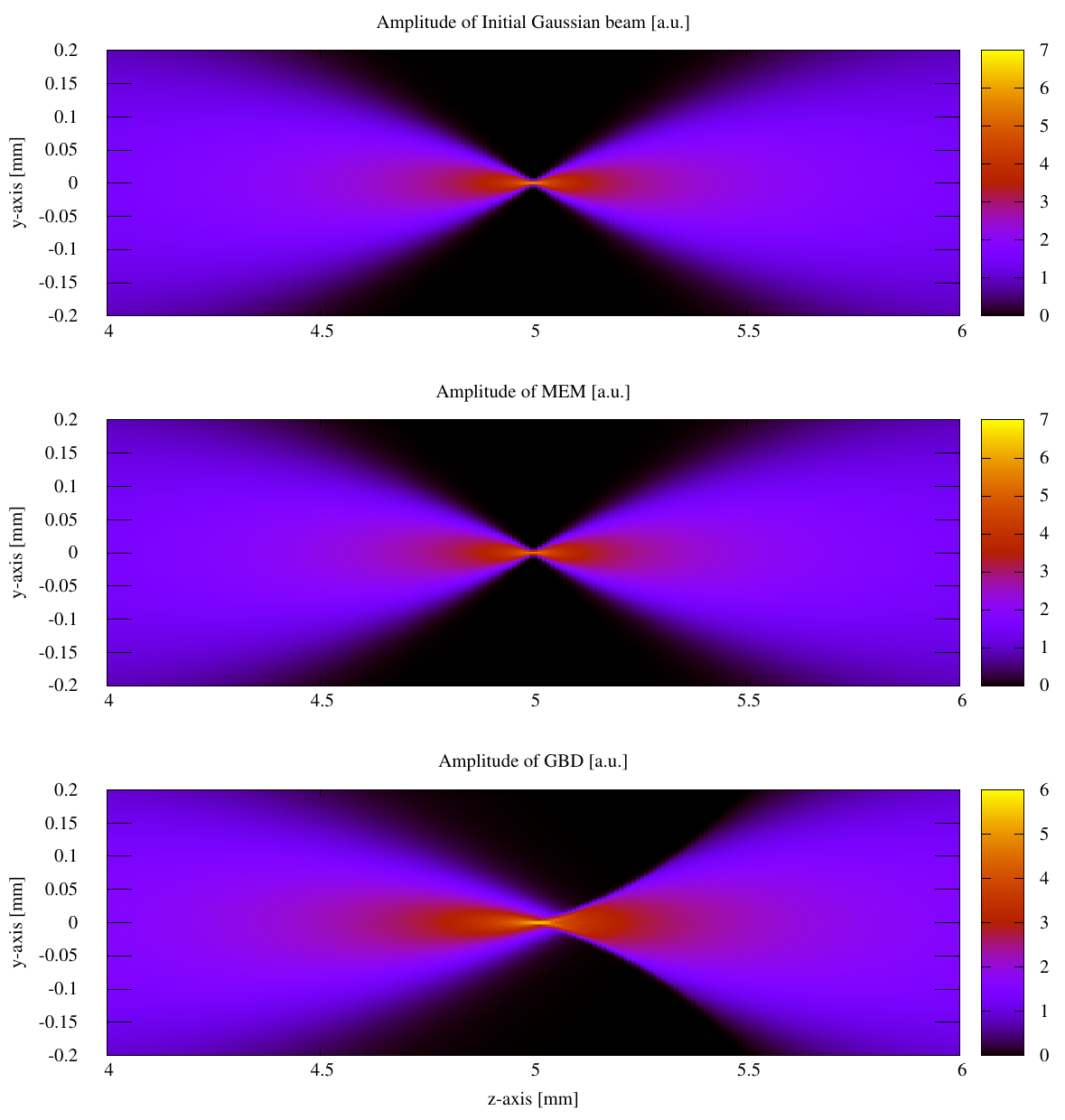} 
	\end{center}
	\caption{Amplitude profile of the initial Gaussian Beam, MEM and GBD through longitudinal
		sections near the focal point after reflected from a concaved spherical mirror with the curvature of \SI{-0.1}{mm}$^{-1}$. The Amplitude is
		scaled by log(Amplitude+1). Only the GBD shows signs of spherical aberration.}
	\label{concaveSpherical}
\end{figure}

In the previous subsections, the behavior of the MEM and GBD in free space propagation have been compared. However, a major difference between both methods arises when the decomposed beams are propagated through an optical setup. This is an important test case because of the differences in the decomposition methods: while all MEM modes share the very same axis, the GBD grid beams are distributed on a grid. This means when an MEM beam interacts with a surface (i.e. reflects or refracts), it effectively interacts only with the intersection point and its local curvature, while the GBD beam probes the surface in multiple intersection points.
We illustrate this difference for a simple test case: the reflection of a Gaussian beam from a spherically curved mirror with varying curvature. We then expect the GBD to show increasing levels of spherical aberration with increasing mirror curvature, while the MEM is expected not to resolve the occurring spherical aberration. This means we decompose a Gaussian beam with an MEM and a GBD and reflect the original Gaussian as well as the MEM and GBD representations of this Gaussian from a spherically curved mirror. We then expect for increasing mirror curvature an increasing level of deviation between the GBD beam and the reference Gaussian, while the MEM is not expected to show this behavior. The results of this simple test are illustrated in \cref{ReflectonGBDCMEM} and for a stronger curvature case in \cref{concaveSpherical}, which confirm the expected behaviour.

For this simulation, we assumed the Gaussian beam with a 1\,mm waist radius located in its origin which propagated by $z=\SI{10}{mm}$ before it impinged orthogonally and centered onto the spherically curved mirror which had a diameter of x\,mm. The mirror was, therefore, sufficiently oversized to reflect the full beam. After reflection, the complex electric field was calculated at an observation plane, which was $\SI{5}{mm}$ away from the mirror and orthogonal to the beam. All simulation parameters are summarized in \cref{mirrorparameters}.

Please note, we use the wording `relative error' in \cref{ReflectonGBDCMEM} for consistency with the previous sections. However, this should be rather seen as a relative deviation, given that the Gaussian beam, which serves as a reference, cannot be trusted to be physically correct in this example because it is itself insensitive to spherical aberration.

\begin{table}[htbp]
	\caption{Parameters list for reflection from optical components. Like in \cref{sec:Non-clipped-circular-GBs}, we define here an aperture only for implementation reasons. However, the beam is effectively not being clipped. Therefore, the complex electric fields of the MEM and GBD in the observation plane can be directly compared with the results for a fundamental Gaussian beam.} 
	\footnotesize
	%\centering
	\begin{tabular}{@{}lll}
		\hline
		\textbf{parameters} & \textbf{description} & \textbf{value}  \\ 
		\hline
		$\lambda$          & wavelength  & \SI{1064}{\nm} \\
		$P_0$          & beam power  & \SI{1}{W} \\   
		$w_0$         & beam waist  &  \SI{1}{\mm}     \\ 
		$z_0$        & distance from the waist           &   0      \\ 
		$R_a$          & aperture radius       &  \SI{4}{\mm}     \\ 
		$N$          & mode order of the MEM    &   50     \\ 
		$w_{0d}$          & waist of the modes used in the MEM  & \SI{0.8}{mm}    	\\ 
		$G$          &grid size of the GBD             &  400$\times$400      \\ 
		$L$          & window size of the GBD             &   \SI{8}{\mm}     \\ 
		$f_{\rm ws}$        &waist scaling factor of the GBD             &  	10/3      \\ 
		$w_{0g}$        &waist of the grid beam used in the GBD            &  \SI{0.0333}{mm}     \\ 
		grid shape          &grid shape of the GBD             &   square     \\ 	 
		$C$                 & curvatures of the mirror       & \SI{0}{mm}$^{-1}$,\SI{-0.002}{mm}$^{-1}$, \SI{-0.02}{mm}$^{-1}$ , \SI{-0.1}{mm}$^{-1}$ \\
		size & the diameter of the mirror & \SI{1}{cm}$\times$\SI{1}{cm} \\
		$X$                 &number of sampling points       & 3001\\ 
		\hline
	\end{tabular}
	\label{mirrorparameters}
\end{table}

%

%%%%%%%%%%%%%%%%%%%%%%%%%%%%%%%%%%%%%%%%%%%%%	
\section{Summary and Conclusion} \label{se:7}
In this paper, we have compared the two wavefront decomposition methods MEM and GBD for different test cases. To judge the performance of either of the methods and to allow a direct comparison of both, several different types of error estimates have been introduced: the normalized mean square error NMSE, its discrete analog DNMSE, a relative error, and its sum. The properties of all these errors were discussed and compared. % error definitions 

%findings for the different DNMSE and NMSE
We found for the MEM that even though the well-known NMSE is propagation distance independent, the DNMSE is usually not so because the lateral ranges chosen in the target plane are too small. To achieve the propagation independency of the DNMSE, the lateral ranges in the given example were so large that the MEM with mode orders up to 50 was not able to resolve the necessary lateral range due to the finite spot sizes of the involved modes.

% findings for the relative error and its sum
While the NMSE and its discretized analog are commonly known and used errors, they do not allow visualizing the error distribution over the cross-section of the field of interest. For this, the relative error can be used. Finally, the summed relative error is a useful addition to the relative error, quantifying the graphical findings. 

% fair compaison
To allow a direct comparison, we have tested the simulation runtime for different settings in a typical test case. We showed that the GBD runtime significantly depends on the number of sampling points in the target plane, unlike the MEM. A direct comparison of the precision of these methods for comparable runtime, therefore, depends significantly on the number of sampling points chosen in the target plane. Naturally, particularly this finding depends on the chosen software tool and the implementation of the methods. All simulations performed throughout this paper were performed using the software library IfoCAD (runtime tests with Version of 2022/10, git commit adf19a5b). However, the introduced method of finding settings that allow a fair comparison is independent of computer systems, software choices, and implementation details. Additionally, the findings underline that the GBD method in any software implementation should not only be optimized in the decomposition but also for the superposition in the target plane.  % 

% findings for the MEM 
For the individual performance of the MEM, we found that it is not ideally resolving the high-frequency spatial oscillations in the near-field, but resolves the far field wavefronts accurately even with small mode orders. The accuracy of the MEM naturally improves with higher mode orders, but in the usual ranges of interest also with the propagation distance.

% findings for the GBD
For the individual performance of the GBD, we likewise found that the high-frequency spatial oscillations in the near field are insufficiently resolved with typical settings. The comparably smooth far fields, in comparison, are resolved with higher accuracy. Naturally, the accuracy of the GBD becomes better with increasing grid sizes. However, this can quickly result in grid beam waists that are so small that they violate the paraxial approximation. 
We intentionally tested the performance of the GBD in such an imperfect case and found that the precision was not impaired by the non-ideal, extremely small waist sizes. That means the relative error and its sum decreased for GBDs with increasing grid sizes, despite the use of smaller and smaller waist sizes that violated the paraxial approximation. 

%direct comparison of MEM and GBD for cases we analytically know
We have directly compared the MEM and GBD for cases where the electric field amplitude and phase are analytically known in different propagation distances. We have performed this test for typical propagation distances in the near and far-field and additionally in the extreme far-field of millions of kilometers, as needed for space gravitational wave detectors. We showed that both methods can resolve the field at this extreme far distance without the need for a re-decomposition at an intermediate distance. 
The direct method comparison showed a better performance of the MEM for the decomposition and free-beam propagation of non-clipped circular and general astigmatic Gaussian beams. In the cases of clipped circular Gaussian beams, the GBD showed higher accuracy. However, these findings might well depend on the settings, the used software with its implementation of both methods, and the computer, operation system, and compilers. %

%aberration and reflection from mirrors
Additionally, we compared the MEM and GBD representation of aberration and showed a qualitative agreement between the results. Finally, we showed in one example that the GBD is a superior method for propagation through an optical setup, where interactions with surfaces occur. While the MEM decomposes the initial field with modes that all share the very same beam axis, the GBD decomposes into fundamental Gaussian beams on a grid. Consequently, the MEM beam probes the curvature of a surface only at one intersection point between its beam axis and the surface, while the grid beams of a GBD beam probe the surface curvature in a grid of intersection points. We have shown this difference by a qualitative comparison of the electric fields of a Gaussian beam and its MEM and GBD representations after reflection from a curved mirror. We have shown that the GBD beam showed the expected spherical aberration unlike the Gaussian or MEM beam.

%final conclusions
We can generally conclude that both methods are useful for decomposing and propagating non-Gaussian beams. Once the fields are decomposed, the propagation in free space or through an optical setup is computationally trivial. Which method is more accurate depends on the test case and simulation settings. However, for the propagation through optical layouts, where the beam interacts with surfaces, and particularly if non-spherical surfaces exist in the setup, the GBD with its grid of beams has a clear advantage over the MEM.

\section*{Acknowledgement}
	% The funding of Mengyuan Zhao
	Firstly, we express our sincere gratitude to the National Key R\&D Program of China (2020YFC2200100) for providing funding that supported the contributions made by Mengyuan Zhao in this work.
	% LEGACY members
	Secondly, this work has been supported by the Chinese Academy of Sciences (CAS) and the Max Planck Society (MPG) in the framework of the LEGACY cooperation on low-frequency gravitational wave astronomy (M.IF.A.QOP18098, CAS’s Strategic Pioneer Program on Space Science XDA1502110201). 
	%GW -> DLR + QF 
	Likewise, we gratefully acknowledge the German Space Agency, DLR and support by the Federal Ministry for Economic Affairs and Energy based on a resolution of the German Bundestag (FKZ 50OQ1801) as well as the Deutsche Forschungsgemeinschaft (DFG) funding the Cluster of Excellence QuantumFrontiers (EXC 2123, Project ID 390837967) for funding the work contributions by Gudrun Wanner. 
	% geoQ for Sönke, TerraQ for Kevin
	We gratefully acknowledge DFG for funding the Collaborative Research Centres CRC 1128: geo-Q - Relativistic Geodesy and Gravimetry with Quantum Sensors, project A05 and all work contributions to this paper made by Sönke Schuster, as well as CRC 1464: TerraQ – Relativistic and Quantum-based Geodesy, Project B03 and Project-ID 434617780, for all contributions made by Kevin Weber.
	% PhoenixD + QF 
	We gratefully acknowledge DFG for funding the Clusters of Excellence PhoenixD (EXC 2122, Project ID 390833453) and QuantumFrontiers (EXC 2123, Project ID 390837967) which offer an excellent scientific exchange on optical simulations.

\bibliographystyle{elsarticle-num}
\bibliography{paperbib}

\begin{thebibliography}{10}
\expandafter\ifx\csname url\endcsname\relax
  \def\url#1{\texttt{#1}}\fi
\expandafter\ifx\csname urlprefix\endcsname\relax\def\urlprefix{URL }\fi
\expandafter\ifx\csname href\endcsname\relax
  \def\href#1#2{#2} \def\path#1{#1}\fi

\bibitem{hecht2001optics}
E.~Hecht, Optics 4th edition, Optics 4th edition by Eugene Hecht Reading.

\bibitem{ghatak1989optics}
A.~Ghatak, Optics, Tata Mcgraw hill publishing company, 1989.

\bibitem{bea1991fundamentals1}
S.~BEA, M.~Teich, Fundamentals of photonics, Wiley (1991) 127.

\bibitem{siegman1986lasers}
A.~Siegman, \href{https://books.google.com/books?id=1829MgEACAAJ}{Lasers},
  University Science Bks, University Science Books, 1986.
\newline\urlprefix\url{https://books.google.com/books?id=1829MgEACAAJ}

\bibitem{born2013principles}
M.~Born, E.~Wolf, Principles of optics: electromagnetic theory of propagation,
  interference and diffraction of light, Elsevier, 2013.

\bibitem{2006LISA}
T.~A. Prince, P.~Binetruy, J.~Centrella, L.~S. Finn, L.~Team, Lisa: Probing the
  universe with gravitational waves, New York, N.Y. : American Institute of
  Physics.

\bibitem{2017The}
W.~R. Hu, Y.~L. Wu, The taiji program in space for gravitational wave physics
  and the nature of gravity, National Science Review 4~(5) (2017) 2.

\bibitem{amaro2017laser}
P.~Amaro-Seoane, H.~Audley, S.~Babak, J.~Baker, E.~Barausse, P.~Bender,
  E.~Berti, P.~Binetruy, M.~Born, D.~Bortoluzzi, et~al., Laser interferometer
  space antenna, arXiv preprint arXiv:1702.00786.

\bibitem{bea1991fundamentals2}
S.~BEA, M.~Teich, Fundamentals of photonics, Wiley (1991) 26--31.

\bibitem{goubau1961}
G.~Goubau, F.~Schwering, On the guided propagation of electromagnetic wave
  beams, IRE Transactions on Antennas and Propagation 9~(3) (1961) 248--256.
\newblock \href {http://dx.doi.org/10.1109/TAP.1961.1144999}
  {\path{doi:10.1109/TAP.1961.1144999}}.

\bibitem{Tanaka:88}
K.~Tanaka, K.~Yoshida, M.~Taguchi, Analytical and experimental investigations
  of the diffraction field of a gaussian beam through a sequence of apertures:
  applicability of the beam mode expansion method, Appl. Opt. 27~(7) (1988)
  1310--1312.
\newblock \href {http://dx.doi.org/10.1364/AO.27.001310}
  {\path{doi:10.1364/AO.27.001310}}.

\bibitem{NS2003Modeling}
N.~Petrović, A.~Rakić, Modeling diffraction in free-space optical
  interconnects by the mode expansion method, Applied Optics 42~(26) (2003)
  5308.

\bibitem{2007Modeling}
J.~J. Snyder, Modeling laser beam diffraction and propagation by the
  mode-expansion method, Applied Optics 46~(22) (2007) 5056.

\bibitem{2010Interferometer}
A.~Freise, K.~Strain, Interferometer techniques for gravitational-wave
  detection, Living Reviews in Relativity 13~(1) (2010) 1.

\bibitem{2010Optimum}
F.~Ghasemi, K.~Mehrany, Optimum waist of localized basis functions in truncated
  series employed in some optical applications, Applied Optics 49~(8) (2010)
  1210.

\bibitem{Borghi1996Optimization}
R.~Borghi, F.~Gori, M.~Santarsiero, Optimization of laguerre-gauss truncated
  series, Optics Communications 125~(4-6) (1996) 197--203.

\bibitem{2014Simulated}
R.~J. Mahon, J.~A. Murphy, Simulated propagation of ultrashort pulses modulated
  by low-fresnel-number lenses using truncated series expansions, Applied
  Optics 53~(25) (2014) 5701--11.

\bibitem{xiao2019laguerre}
Y.~Xiao, X.~Tang, C.~Wan, Y.~Qin, H.~Peng, C.~Hu, B.~Qin, Laguerre-gaussian
  mode expansion for arbitrary optical fields using a subspace projection
  method, Optics letters 44~(7) (2019) 1615--1618.

\bibitem{Yongxin2006Truncated}
Y.~Liu, B.~Lü, Truncated hermite–gauss series expansion and its application,
  Optik International Journal for Light \& Electron Optics.

\bibitem{popov1982new}
M.~M. Popov, A new method of computation of wave fields using gaussian beams,
  Wave motion 4~(1) (1982) 85--97.

\bibitem{2014Fat}
A.~W. Greynolds, Fat rays revisited: a synthesis of physical and geometrical
  optics with gau$\rm beta$lets, in: International Optical Design Conference,
  2014.

\bibitem{1985Propagation}
A.~W. Greynolds, Propagation of generally astigmatic gaussian beams along skew
  ray paths, in: Proc Spie, 1985.

\bibitem{narayananl2004gaussian}
G.~Narayananl, Gaussian beam analysis of relay optics for the sequoia focal
  plane array, in: 15th International Symposium on Space Terahert Technology,
  Vol.~8, 2004.

\bibitem{ashcraft2020open}
J.~N. Ashcraft, E.~S. Douglas, An open-source gaussian beamlet decomposition
  tool for modeling astronomical telescopes, in: Modeling, Systems Engineering,
  and Project Management for Astronomy IX, Vol. 11450, SPIE, 2020, pp.
  354--366.

\bibitem{white1987some}
B.~White, A.~Norris, A.~Bayliss, R.~Burridge, Some remarks on the gaussian beam
  summation method, Geophysical Journal International 89~(2) (1987) 579--636.

\bibitem{leye2016gaussian}
P.~O. Leye, A.~Khenchaf, P.~Pouliguen, et~al., The gaussian beam summation and
  the gaussian launching methods in scattering problem, Journal of
  Electromagnetic Analysis and Applications 8~(10) (2016) 219.

\bibitem{SPIES2000155}
M.~Spies,
  \href{https://www.sciencedirect.com/science/article/pii/S0963869599000365}{Modeling
  of transducer fields in inhomogeneous anisotropic materials using gaussian
  beam superposition}, NDT \& E International 33~(3) (2000) 155--162.
\newblock \href
  {http://dx.doi.org/https://doi.org/10.1016/S0963-8695(99)00036-5}
  {\path{doi:https://doi.org/10.1016/S0963-8695(99)00036-5}}.
\newline\urlprefix\url{https://www.sciencedirect.com/science/article/pii/S0963869599000365}

\bibitem{Alonso:02}
M.~Alonso, G.~Forbes,
  \href{https://opg.optica.org/oe/abstract.cfm?URI=oe-10-16-728}{Stable
  aggregates of flexible elements give a stronger link between rays and waves},
  Opt. Express 10~(16) (2002) 728--739.
\newblock \href {http://dx.doi.org/10.1364/OE.10.000728}
  {\path{doi:10.1364/OE.10.000728}}.
\newline\urlprefix\url{https://opg.optica.org/oe/abstract.cfm?URI=oe-10-16-728}

\bibitem{kong2013design}
H.~B. Kong, D.~J. Cho, Design and analysis of infrared diffractive optical
  systems using beam synthesis propagation, Korean Journal of Optics and
  Photonics 24~(4) (2013) 189--195.

\bibitem{Alonso:13}
M.~A. Alonso,
  \href{https://opg.optica.org/josaa/abstract.cfm?URI=josaa-30-6-1223}{Ray-based
  diffraction calculations using stable aggregates of flexible elements}, J.
  Opt. Soc. Am. A 30~(6) (2013) 1223--1235.
\newblock \href {http://dx.doi.org/10.1364/JOSAA.30.001223}
  {\path{doi:10.1364/JOSAA.30.001223}}.
\newline\urlprefix\url{https://opg.optica.org/josaa/abstract.cfm?URI=josaa-30-6-1223}

\bibitem{tanushev2009gaussian}
N.~M. Tanushev, B.~Engquist, R.~Tsai, Gaussian beam decomposition of high
  frequency wave fields, Journal of Computational Physics 228~(23) (2009)
  8856--8871.

\bibitem{Sahin:13}
E.~\c{S}ahin, L.~Onural,
  \href{https://opg.optica.org/josaa/abstract.cfm?URI=josaa-30-3-527}{Calculation
  of the scalar diffraction field from curved surfaces by decomposing the
  three-dimensional field into a sum of gaussian beams}, J. Opt. Soc. Am. A
  30~(3) (2013) 527--536.
\newblock \href {http://dx.doi.org/10.1364/JOSAA.30.000527}
  {\path{doi:10.1364/JOSAA.30.000527}}.
\newline\urlprefix\url{https://opg.optica.org/josaa/abstract.cfm?URI=josaa-30-3-527}

\bibitem{worku2017}
N.~Worku, H.~Gross, \href{https://doi.org/10.1117/12.2273919}{{Vectorial field
  propagation through high NA objectives using polarized Gaussian beam
  decomposition}}, in: K.~Dholakia, G.~C. Spalding (Eds.), Optical Trapping and
  Optical Micromanipulation XIV, Vol. 10347, International Society for Optics
  and Photonics, SPIE, 2017, p. 103470W.
\newblock \href {http://dx.doi.org/10.1117/12.2273919}
  {\path{doi:10.1117/12.2273919}}.
\newline\urlprefix\url{https://doi.org/10.1117/12.2273919}

\bibitem{2018Decomposition}
N.~G. Worku, H.~Ralf, G.~Herbert, Decomposition of a field with smooth
  wavefront into a set of gaussian beams with non-zero curvatures, Journal of
  the Optical Society of America A 35~(7) (2018) 1091--.

\bibitem{2019Propagation}
N.~G. Worku, H.~Gross, Propagation of truncated gaussian beams and their
  application in modeling sharp-edge diffraction, Journal of the Optical
  Society of America A 36~(5) (2019) 859.

\bibitem{YANRong2006Application}
Y.~Rong, M.~Shan-jun, L.~Bai-da, Application of laguerre-gauss truncated series
  expansion, High Power Laser and Particle Beams 18~(6) (2006) 931--934.

\bibitem{2015Modeling}
J.~E. Harvey, R.~G. Irvin, R.~N. Pfisterer, Modeling physical optics phenomena
  by complex ray tracing, Optical Engineering 54~(3) (2015) 035105.

\bibitem{ifocad}
Ifocad, \url{https://www.aei.mpg.de/ifocad}.

\bibitem{kochkina2013simulating}
E.~Kochkina, G.~Heinzel, G.~Wanner, V.~M{\"u}ller, C.~Mahrdt, B.~Sheard,
  S.~Schuster, K.~Danzmann, Simulating and optimizing laser interferometers,
  in: 9th LISA Symposium, 2013, pp. 291--292.

\bibitem{Kimel1993}
I.~Kimel, L.~Elias, Relations between hermite and laguerre gaussian modes,
  {IEEE} Journal of Quantum Electronics 29~(9) (1993) 2562--2567.
\newblock \href {http://dx.doi.org/10.1109/3.247715}
  {\path{doi:10.1109/3.247715}}.

\bibitem{ONEIL200035}
A.~T. O'Neil, J.~Courtial,
  \href{https://www.sciencedirect.com/science/article/pii/S0030401800007367}{Mode
  transformations in terms of the constituent hermite–gaussian or
  laguerre–gaussian modes and the variable-phase mode converter}, Optics
  Communications 181~(1) (2000) 35--45.
\newblock \href
  {http://dx.doi.org/https://doi.org/10.1016/S0030-4018(00)00736-7}
  {\path{doi:https://doi.org/10.1016/S0030-4018(00)00736-7}}.
\newline\urlprefix\url{https://www.sciencedirect.com/science/article/pii/S0030401800007367}

\bibitem{item_1660139}
C.~Mahrdt, {Laser Link Acquisition for the GRACE Follow-On Laser Ranging
  Interferometer}, Ph.D. thesis (2014).

\bibitem{1987Fresnel}
C.~Campbell, Fresnel diffraction of gaussian laser beams by circular apertures,
  Optical Engineering 26~(3) (1987) 270--275.

\bibitem{1985Field}
K.~Tanaka, N.~Saga, H.~Mizokami, Field spread of a diffracted gaussian beam
  through a circular aperture, Appl Opt 24~(8) (1985) 1102.

\bibitem{Drege:00}
E.~M. Dr\`{e}ge, N.~G. Skinner, D.~M. Byrne,
  \href{http://opg.optica.org/ao/abstract.cfm?URI=ao-39-27-4918}{Analytical
  far-field divergence angle of a truncated gaussian beam}, Appl. Opt. 39~(27)
  (2000) 4918--4925.
\newblock \href {http://dx.doi.org/10.1364/AO.39.004918}
  {\path{doi:10.1364/AO.39.004918}}.
\newline\urlprefix\url{http://opg.optica.org/ao/abstract.cfm?URI=ao-39-27-4918}

\bibitem{1980spot}
W.~H. Carter, Spot size and divergence for hermite gaussian beams of any order,
  Appl Opt 19~(7) (1980) 1027--1029.

\bibitem{2012Methods}
G.~Wanner, G.~Heinzel, E.~Kochkina, C.~Mahrdt, B.~S. Sheard, S.~Schuster,
  K.~Danzmann, Methods for simulating the readout of lengths and angles in
  laser interferometers with gaussian beams, Optics Communications 285~(24)
  (2012) 4831--4839.

\bibitem{2015Stigmatic}
E.~Kochkina, Stigmatic and astigmatic gaussian beams in fundamental mode, Ph.D.
  thesis (2015).

\bibitem{Mahajan:94}
V.~N. Mahajan,
  \href{http://opg.optica.org/ao/abstract.cfm?URI=ao-33-34-8121}{Zernike circle
  polynomials and optical aberrations of systems with circular pupils}, Appl.
  Opt. 33~(34) (1994) 8121--8124.
\newblock \href {http://dx.doi.org/10.1364/AO.33.008121}
  {\path{doi:10.1364/AO.33.008121}}.
\newline\urlprefix\url{http://opg.optica.org/ao/abstract.cfm?URI=ao-33-34-8121}

\bibitem{Masalehdan:10}
H.~Masalehdan, E.~Lotfi, A.~Lotfi, K.~Jamshidi-Ghaleh,
  \href{http://opg.optica.org/abstract.cfm?URI=BIOMED-2010-JMA98}{Modeling of
  zernike optical aberrations by mtf and psf}, in: Biomedical Optics and 3-D
  Imaging, Optica Publishing Group, 2010, p. JMA98.
\newblock \href {http://dx.doi.org/10.1364/BIOMED.2010.JMA98}
  {\path{doi:10.1364/BIOMED.2010.JMA98}}.
\newline\urlprefix\url{http://opg.optica.org/abstract.cfm?URI=BIOMED-2010-JMA98}

\bibitem{Vera2012}
F.~A. Vera-Daz, N.~Doble, The human eye and adaptive optics, InTech.

\bibitem{bornwolf}
M.~Born, E.~Wolf, A.~B. Bhatia, P.~C. Clemmow, D.~Gabor, A.~R. Stokes, A.~M.
  Taylor, P.~A. Wayman, W.~L. Wilcock, Principles of Optics: Electromagnetic
  Theory of Propagation, Interference and Diffraction of Light, 7th Edition,
  Cambridge University Press, 1999.
\newblock \href {http://dx.doi.org/10.1017/CBO9781139644181}
  {\path{doi:10.1017/CBO9781139644181}}.

\bibitem{goodman1996introduction}
J.~Goodman, P.~Sutton, Introduction to fourier optics, Quantum and
  Semiclassical Optics-Journal of the European Optical Society Part B 8~(5)
  (1996) 1095.

\end{thebibliography}


\end{document}